\documentclass[article,11pt,leqno, fleqn]{article}
\usepackage{Other/Templates/paper}

\AtEveryBibitem{%
  \clearfield{note}
  \clearfield{addendum}
  \clearfield{annotation}
  \clearfield{annote}
}

\usepackage[]{hyperref}
\hypersetup{
     colorlinks=true,          
     linkcolor=blue,           
     citecolor=blue,
}
\usepackage{ifthen}
\newboolean{anon}
\setboolean{anon}{false}  

\newcommand{\anon}[1]{%
  \ifthenelse{\boolean{anon}}%
    {[Author Information Redacted]}%
    {#1}%
}

\usepackage{setspace}
\newcommand{\titleLong}{\vspace{-2cm}
Beyond Parents? Prediction Gaps in University Completion Using Population-Scale Networks and Flexible Machine Learning
}

\hypersetup{pdftitle={\titleLong}}

\usepackage{soul} 

\begin{document}

\title{\titleLong}


\author[1, 2]{Javier Garcia-Bernardo}
\author[3]{Eva Jaspers}
\author[3]{Weverthon Machado}
\author[4,5]{Samuel Plach}
\author[6]{Erik Jan van Leeuwen}

\affil[1]{ODISSEI Social Data Science Team, Department of Methodology and Statistics, Utrecht University, Utrecht, the Netherlands}
\affil[2]{Centre for Complex Systems Studies, Utrecht University, Utrecht, the Netherlands}
\affil[3]{Department of Sociology, Utrecht University, Utrecht, the Netherlands}
\affil[4]{Department of Social and Political Sciences, Università Commerciale Luigi Bocconi, 20136 Milan, Italy}
\affil[5]{Carlo F. Dondena Centre for Research on Social Dynamics and Public Policy, Università Commerciale Luigi Bocconi, 20136 Milan, Italy}
\affil[6]{Department of Information and Computing Sciences, Utrecht University, Utrecht, the Netherlands}

\date{}

\begin{titlepage}
\maketitle


    


\begin{abstract}

How predictable is children's educational attainment from the wider social contexts in which they grow up, once parental background is known? Sociological research places households, schools, neighborhoods, and extended kin central to intergenerational reproduction, yet their predictive value beyond parental background is rarely tested. 
This matters because the contribution of social contexts helps distinguish whether they operate as independent sources of inequality or as channels through which parental advantage is reproduced. 
Using Dutch population-scale administrative data, we construct a network linking a full cohort of children aged 11--12 to parents, extended kin, classmates, household members, and neighbors, and predict university completion at ages 24--25. We compare logistic regression and gradient boosting---which use individual-level aggregates of these contexts---with graph neural networks (GNNs) operating directly on the network. We refer to differences in out-of-sample performance as \textit{prediction gaps}. 
We find small overall prediction gaps because parental socioeconomic background captures most predictable variation; even GNNs add little once parents are known. Prediction gaps are largest among children without a registered father, especially girls and those with less-educated mothers. Methodologically, we argue that prediction gaps can support sociological theory-building: small gaps show where current theory-based models already explain what can be measured; large gaps identify where targeted mechanism-focused research is warranted.

\end{abstract}

    
\end{titlepage}

\section{Introduction}\label{s:introduction}

\begin{figure}[ht]
    \centering
        \includegraphics[width=\linewidth]{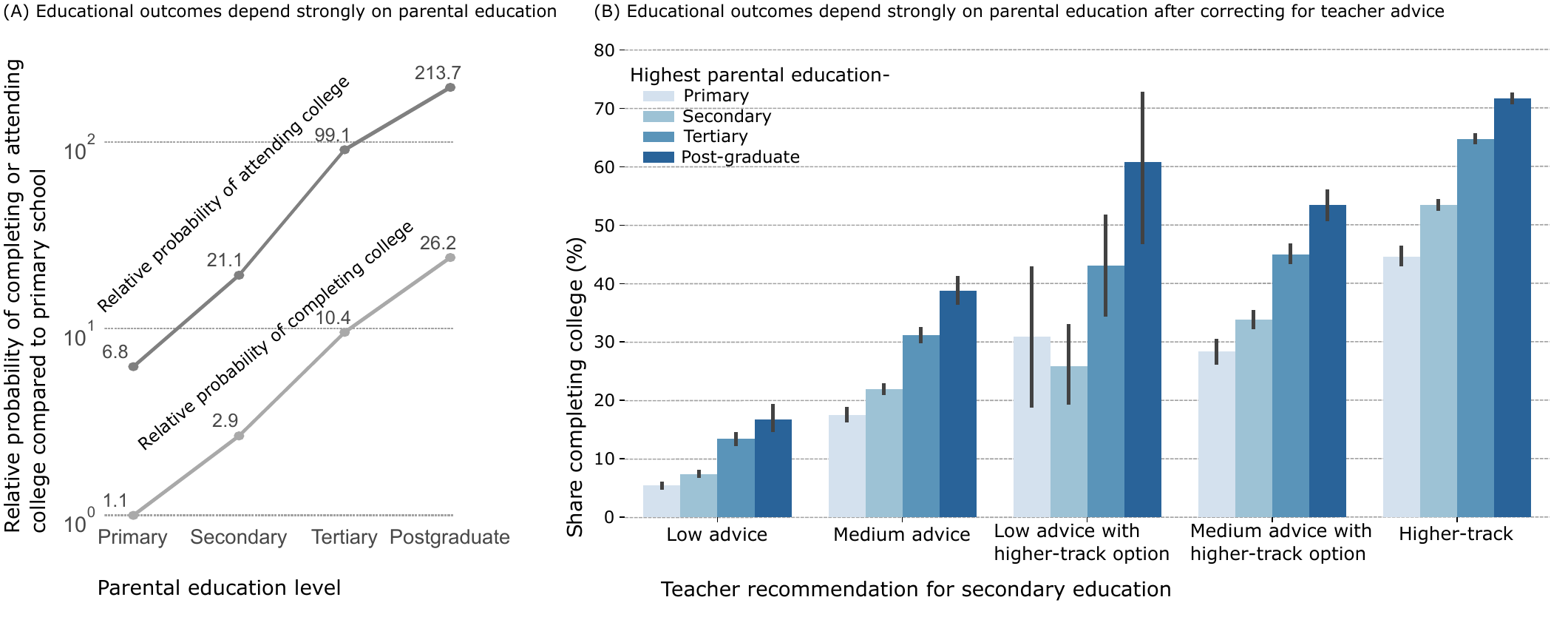}
    \caption{\textbf{Educational outcomes are deeply unequal.} For our cohort of students transitioning to secondary school in 2010 (A) Probability of completing (dark gray) or attending (light gray) university compared with completing only primary school by 2023. (B)  Share of students completing university by 2023 as a function of the teacher’s track recommendation and the highest education achieved by the parent. Low advice corresponds to practical education and lower and intermediate vocational education; medium advice to advanced vocational education; higher-track to tracks preparing students for university. Categories with higher-track options allow students to choose higher-track secondary schools. Whiskers show 95\% bootstrap confidence intervals.}
    \label{fig:context_inequality}
\end{figure}

Educational attainment strongly predicts individuals’ future opportunities, including income, job security, and health \citep{torche_is_2011, parolin_intergenerational_2024, raghupathi_influence_2020}. Yet education outcomes remain profoundly unequal. Children’s chances of completing university are still strongly shaped by the education and income of their parents. In the United States, high-achieving students from low-income families are less likely to complete a bachelor’s degree than low-achieving students from wealthy families \citep{frank_success_2016}. We observe similar inequality in the Netherlands, where 64\% of children with at least one highly educated parent complete university by age 24, compared with only 19\% of children whose parents completed only primary school (Fig. \ref{fig:context_inequality}A).
These gaps persist even when children perform equally well in school (Figs. \ref{fig:context_inequality}B and \ref{fig:context_inequality_income}) or have a similar genetic pool \citep{belsky_genetic_2018}, reinforcing a large literature placing parental background at the center of educational stratification \citep{bourdieu_reproduction_1990, engzell_its_2020, erikson_how_2019, blanden_educational_2023}.

\begin{figure}[ht!]
    \centering
    \includegraphics[width=1\linewidth]{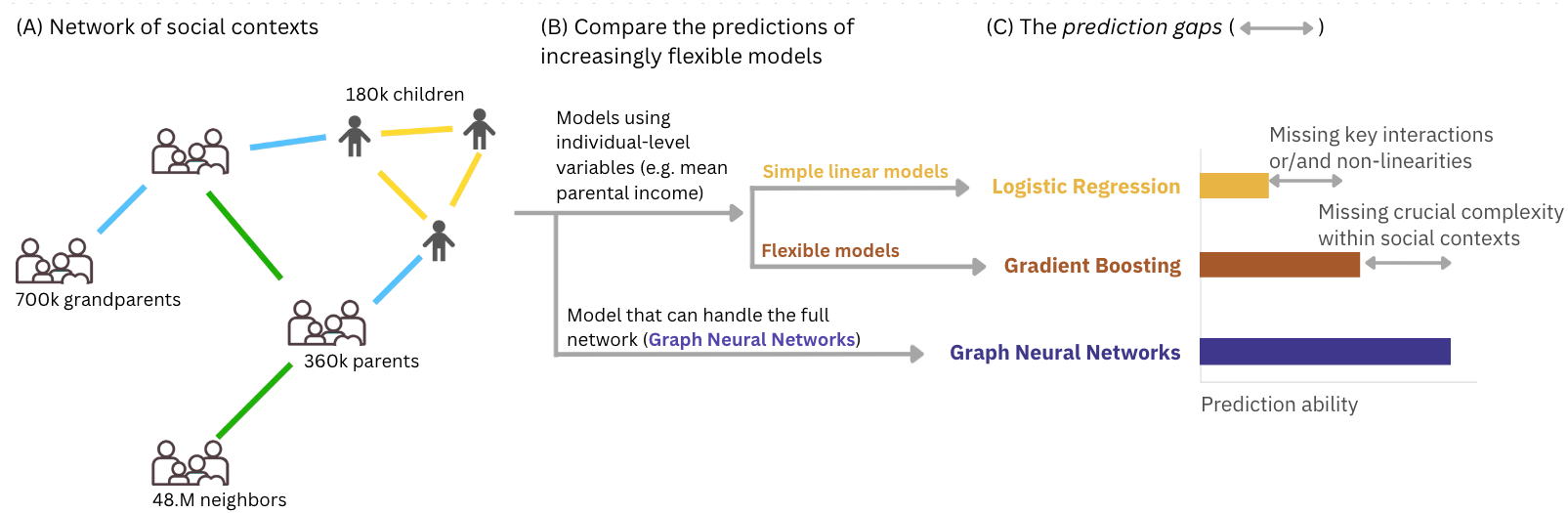}
    \caption{\textbf{Conceptual overview of prediction gaps.} (A) A population‐scale network linking children (gold), parents (blue), grandparents (blue), and neighbors (green), illustrating the multiple overlapping social contexts that shape educational trajectories. Numbers show the sample size of this study. (B) Three statistical models of increasing flexibility: logistic regression and gradient boosting using only aggregated individual‑level covariates, and graph neural networks that explicitly leverage the full network structure. (C) Expected out‑of‑sample performance,
    where horizontal arrows denote prediction gaps---between logistic regression and gradient boosting to reveal missing interactions/non‐linearities, and between gradient boosting and GNNs to expose deeper social‐context complexity---highlighting the unequal association between social contexts and educational outcomes.}
    \label{fig:conceptual}
\end{figure}

Decades of sociological research have argued that this inequality is shaped not only by parents but also by an overlapping set of social contexts (Fig.~\ref{fig:conceptual}A), including extended kin \citep{hallsten_grand_2017}, schools \citep{coleman_equality_1966, owens_neighborhoods_2010, hermansen_long-term_2020}, or neighborhoods \citep{putnam_bowling_2000, massey_american_1990, ainsworth_why_2002, sharkey_where_2014, hedefalk_social_2020}. Several mechanisms have been proposed to link these contexts to educational success, over and above the parental selection into them. Extended kin may compensate for limited parental resources, or they may multiply existing parental advantages so that children from already advantaged families benefit most \citep{jaegerExtendedFamilyChildren2012, mareMultigenerationalViewInequality2011}. Neighborhoods may differ in their collective capacity to monitor children and enforce shared expectations \citep{sampsonNeighborhoodsViolentCrime1997}, in the adult role models they provide \citep{wilson_truly_1987}, and in the institutional resources available to families \citep{leventhalNeighborhoodsTheyLive2000, smallUnanticipatedGainsOrigins2009}. Schools may affect educational success in two distinct ways. First, they differ in quality: some schools have more effective teachers \citep{chettyMeasuringImpactsTeachers2014, rivkinTeachersSchoolsAcademic2005}, hold higher expectations of their students \citep{jussimTeacherExpectationsSelfFulfilling2005, rosenthal_pygmalion_1968}, provide more emotional support \citep{hamreCanInstructionalEmotional2005}, or have more financial resources to spend on enriching educational activities \citep{jacksonEffectsSchoolSpending2016}. Second, schools may bring together different sets of peers and, indirectly, of peers' parents \citep{coleman_equality_1966}. 

Whether the wider contexts add predictive information beyond parental background is substantively important. If they do, then children’s educational trajectories depend on resources, constraints, and institutional environments not reducible to parental background. If they do not, then social contexts may still matter causally, but add little predictive information because they are strongly stratified by parental background and act mostly as channels through which parental advantage is reproduced.

Estimating the added contribution of social contexts is difficult because individuals are embedded in complex networks of these social contexts (Fig.~\ref{fig:conceptual}A), which interact not only with individual characteristics \citep{torche_is_2011, lee_success_2014} but also with other social contexts \citep{owens_neighborhoods_2010, hermansen_long-term_2020, solon_correlations_2000, page_correlations_2003, altonji_role_2011, kuyvenhoven_neighbourhood_2021} to hinder or support educational outcomes. For instance, the effect of an absent father on children’s educational attainment interacts not only with their own gender \citep{chung_peers_2020}, but also with their parents' social class \citep{bernardi_understanding_2016, gratz_when_2015, lang_does_2001}, their household's composition \citep{monserud_household_2011}, and even their peers’ parents \citep{chung_peers_2020}. Standard additive models may therefore understate the role of social contexts if their contribution lies in nonlinearities, interactions, or relational structure rather than in simple average differences.

This motivates a central question: how much predictable variation in university completion is contained in children's wider early-life social contexts after accounting for parental background? We answer this question by comparing models of increasing flexibility trained on the same population. We define a \textit{prediction gap} as the difference in out-of-sample performance between a simpler, theory-based model and a more flexible model capable of capturing complexity within and between social contexts\footnote{Out-of-sample performance measures how well a model predicts outcomes that were set aside during model fitting. If a complex model merely memorizes patterns specific to the training data (if it \textit{overfits} the data), it will not predict these held-out outcomes better than a simpler model. }. These gaps reveal how much predictive information remains beyond parental background. When prediction gaps vary across subpopulations, they identify groups for whom social contexts contain additional predictive information about university completion that is not captured by the simpler model. 

This use of prediction is not meant to replace explanation, but to support theory-building. Related work has shown how predictive models can clarify theoretical claims across different domains: neural word embeddings have been formalized as models of how meanings are acquired and organized in cultural sociology \citep{arseniev-koehler_machine_2022}; predictive models have been used to uncover regularities in human decision-making and judgment in cognitive psychology \citep{peterson_using_2021}; and, in inequality research, prediction has helped show that grandparents contribute little to status transmission once both parents are well measured \citep{engzell_its_2020} and helped explain long-term changes in income mobility \citep{engzell_understanding_2023}. In the same spirit, we use prediction gaps as diagnostics: they show where parent-centered accounts of attainment capture most of the measurable structure in the outcome, and where they appear incomplete.

More specifically, our analysis speaks to the role of social capital, broadly conceived, in school attainment. Many of the mechanisms linking wider social contexts to school outcomes can be understood as forms of social capital, that is, resources that children and their parents can access through their social relationships \citep{bourdieu1986forms, colemanSocialCapitalCreation1988, linSocialCapitalTheory2001}. Support from extended kin, the resources and role models provided by schoolmates' parents, and neighborhood institutions can be understood as what \citet{bourdieu1986forms, linSocialCapitalTheory2001} refer to as the resources embedded in social networks. Collective monitoring by neighbors and by the parents of schoolmates, as well as peer norms regarding studiousness, correspond to \citet{colemanSocialCapitalCreation1988} argument that closed networks facilitate the enforcement of shared norms and expectations. The same relationships can also have negative consequences, for instance through obligations toward kin or peer norms that discourage academic effort \citep{portesSocialCapitalIts1998}. Moreover, access to social capital is itself unequally distributed \citep{bourdieu1986forms} and dependent on parental resources, which is one reason why social contexts may simply reproduce parental (dis)advantage rather than independently affect school outcomes. Although we do not aim to directly test these mechanisms, prediction gaps allow us to identify the added contribution of extended contexts to children’s educational attainment, which may in turn inform us where social capital theory can most fruitfully be expanded.

Methodologically, we argue that prediction gaps are informative whether they are large or small. If flexible models add little predictive power once parental characteristics are included (and therefore prediction gaps are small), this suggests that parental education and income already account for most of the predictable variation in university completion. This does not imply that households, extended kin, schools, or neighborhoods are causally irrelevant; it means that, for this cohort and outcome, knowing attributes of the parents is enough to predict the outcome almost as well as possible. Part of what appears as parental influence may operate through these other channels, but they add little independent predictive information. Conversely, a large prediction gap after accounting for parents would indicate that parent-centered accounts leave substantial predictable variation unexplained. For example, if growing up without a father disproportionately affects already disadvantaged children \citep{bernardi_understanding_2016, gratz_when_2015, lang_does_2001}, we expect models that can automatically identify such interactions---like gradient boosting---to outperform simpler models such as logistic regression unless we explicitly include the necessary interaction terms. Similarly, if we observe a substantive gap between traditional models that use aggregated measures and graph neural networks---which directly model an individual's embeddedness in social contexts---it indicates that important effects are not well-captured by our current operationalization of social contexts, such as average education level of peers' parents or average neighborhood education levels. 

We answer these questions using population-scale administrative data from Statistics Netherlands (CBS). We observe a full cohort of children at the end of primary school (ages 11--12), reconstruct their nuclear and extended families, households, schools, and neighborhoods, and predict whether they complete university in early adulthood (ages 24--25). The Dutch case is particularly interesting because students are sorted into differentiated educational tracks around age 12, making early-life family and school contexts highly consequential \citep{knigge_delayed_2022, atav_persistent_2026, keskiner_is_2015, van_de_werfhorst_ethnicity_2007}. Models using only parental characteristics provide the substantive benchmark: they define how much university completion is already predictable from family background. The empirical question is how much more can be predicted once early-life social contexts are added and modeled flexibly. We therefore compare logistic regression, gradient boosting, and graph neural networks (GNNs). Logistic regression provides the additive, theory-guided specification. Gradient boosting tests whether the same tabular inputs contain nonlinearities or interactions that the additive model misses. By modeling recorded ties, GNNs retain relational structure that is lost when schools and neighborhoods are represented only by aggregate characteristics. GNNs test whether population-scale relational structure contains information not captured by aggregate covariates (Fig.~\ref{fig:conceptual}B--C). However, like tabular models, they cannot represent unmeasured properties of schools, neighborhoods, or communities.

We find that parental socioeconomic background captures most predictable variation in university completion. Once parental background is known, early-life social contexts---including extended family, schools, and neighborhoods---add little predictive power, even when measured with rich administrative data and modeled using flexible graph neural networks. It is possible that the early-life social contexts of children are not independently predictive for university completion because they are largely stratified by parental background and act as channels through which parental advantage is reproduced. Alternatively, both contexts and parental characteristics might reflect shared unobserved factors, or the relevant contextual processes might simply not be captured in administrative data.

Importantly, prediction gaps do exist for subpopulations. They are largest among children with absent fathers, especially girls and those with less-educated mothers. For these groups, early-life social contexts contribute substantially to the predictability of university completion. Possible explanations include gendered coping strategies, compensatory ties beyond the nuclear family, differential institutional responses, or unmeasured family and community processes. We therefore treat this pattern as a candidate puzzle for future research.

This article makes important contributions to three bodies of literature. First, we speak to the intergenerational mobility literature by confirming that even population-scale network data and flexible network models leave parental socioeconomic status as the dominant predictor of educational success. Second, we contribute to the promising literature investigating the relationship between social capital and mobility. Our findings that social contexts do not add much predictability in the pooled sample suggest that social contexts either operate as mediators for the persistence of intergenerational advantage, or there are common unmeasured factors impacting both parental characteristics and social contexts. Finally, we contribute to an emerging literature showing that machine learning prediction methods can support sociological theory-building rather than undermine it. 
When prediction gaps appear, as for girls with absent fathers, they identify subpopulations meriting mechanism-focused inquiry. When they do not appear, as elsewhere in our analysis, they strengthen the validity of the modeled theories of educational attainment.

\section{Prediction Gaps as Abductive Diagnostics}\label{s:prediction}
Quantitative social science is often described as deductive: researchers start with a theory, derive hypotheses, and then test these using statistical models and empirical data \citep{pankowska2024potential}. In practice, research is often iterative. Theories rarely generate predictions precise enough to be cleanly confirmed or refuted. Instead, evidence accumulates through revision, extension, and boundary-setting. \citet{engzell_understanding_2023} describe this as a ``scaffolding'' view of theory, in which theory organizes heterogeneous evidence and is evaluated by asking where, when, and for whom it applies. Abductive inference  \citep{brandt_abductive_2021} operates on this ``scaffolding'' view: it begins from an anomaly relative to current theoretical expectations and proceeds to the best available explanation of it.

The prediction gap is built around this logic. The baseline model encodes the conventional, theory-based specification of the field: a linear, additive function of inputs that the literature treats as the relevant determinants of attainment. The flexible comparison model is given the same inputs but is free to use them in any way: nonlinearities, high-order interactions, or, in the case of GNNs, network embeddedness. A prediction gap is the difference in out-of-sample performance between the flexible model and the baseline. It is an anomaly relative to the standard specification, and it is theoretically motivated precisely because the baseline carries theoretical commitments about which variables and which functional forms should matter.

These comparisons are diagnostic, not explanatory. A prediction gap can arise because a theory-relevant interaction is omitted, because a social context is poorly operationalized, or because measurement and modeling choices leave structure that flexible models can exploit. Thus, prediction gaps locate puzzles rather than resolve them. We highlight two productive uses of the prediction gap, both substantive.

\subsection{When prediction gaps are small: the strength of existing accounts}

A small prediction gap is not a failed result. It means that, for a given outcome, population, and measurement strategy, a more flexible model extracts little additional predictive information beyond the simpler benchmark. In our case, this is substantively important because educational stratification research gives strong reasons to expect parental background to dominate attainment, while also giving strong reasons to expect schools, neighborhoods, households, and kin networks to matter. If models that incorporate wider social contexts and greater flexibility add little predictive power after parental education and income are included, this suggests that measured social contexts contain little additional information once parental background is known.

One possibility is that contextual measures are strongly correlated with parental background: children from advantaged families are more likely to live in advantaged neighborhoods, attend advantaged schools, and be embedded in advantaged kin networks. Another is that the relevant contextual processes operate in ways not captured by administrative measures. In either case, identifying small prediction gaps is useful because they narrow the empirical puzzle. They shift attention away from whether wider contexts contain large unmodeled predictive structure and toward how parental background, contextual sorting, and social reproduction are linked. 

\subsection{When prediction gaps are large: anomalies as puzzles}
When a flexible model outperforms the additive baseline, the standard specification is leaving predictive information in the data unused. This matters because it suggests that the baseline omits variables relevant to educational attainment, whether in the form of nonlinearities, interactions, relational patterns, or poorly measured contexts. The gap does not identify the mechanism responsible for the improvement. Instead, it tells researchers where to look. Large gaps identify populations, contexts, or specifications for which existing models are least complete and where targeted confirmatory research may be most valuable. In this study, we treat  subgroup gaps in this way: as candidate anomalies rather than resolved explanations.

Prediction gaps between the models in this study have three main sources. Every statistical model modeling an outcome variable (e.g. completing university) can be written with the formula $y  = f(X) + \epsilon$, where $y$ is the outcome variable, $X$ is a vector of predictor variables, $f(\cdot)$ is a function that maps inputs $X$ to a prediction of $y$, and $\epsilon$ is the random error term. This function $f(X)$ may be simple (as in linear regression) or complex (as in a neural network), but in all cases it expresses the model's best guess at how inputs relate to outcomes. A prediction gap can arise because the model is given different information ($X$), because it uses the same information differently ($f()$), or because social contexts are represented in a different form (e.g., a table or a network).

\subsubsection{Features: adding early-life social contexts}

The first source of prediction gaps is the set of features, or variables, made available to the model. A model using only demographic and parental characteristics asks how much university completion is predictable from individual characteristics and family background. Adding household, extended-family, school, and neighborhood measures asks whether early-life social contexts contain additional predictive information beyond parents. A gap between these feature sets would indicate that measured contexts contribute independently to prediction once parental background is known. Conversely, a small gap would suggest that parental background already absorbs much of the predictive information contained in these measured contexts, or that the relevant contextual processes are not captured by the available administrative measures.

\subsubsection{Model specification: Interactions and non-linearities}

Here, we compare two different specifications of $f(X)$: a simple logistic regression model, which assumes linear effects and no interactions, and a gradient boosting model, which flexibly learns interactions and nonlinearities from the data.
Both models use the same individual-level variables (e.g., parental income and education, or migration background) to predict the probability of completing university. In this case, a substantive prediction gap between the models unveils important interactions or nonlinearities that the linear model fails to capture (see also \citet{hindman_building_2015, verhagen_incorporating_2024}). While social scientists generally accept that the world is theoretically nonlinear, they often rely on linear models for their simplicity. Prediction gaps show when such simplifications overlook important patterns that should not be ignored. Thus, prediction gaps are a useful tool for abductive inquiry: when the interactions and non-linearities underlying prediction gaps align with theoretical expectations, they provide evidence in favor of specific theories and help us improve model specification. When they are unexpected, they suggest areas where theory might need to be revised.

\subsubsection{The operationalization of social contexts}
Every predictive model makes assumptions not only about the function $f$---how inputs relate to outcomes---but also about the structure of the social contexts $X$. In most traditional models, $X$ includes individual-level variables like parental income or educational background. To incorporate social context, researchers often aggregate features from the environment, such as the average income in a neighborhood or the share of highly educated parents in a school. These are then included as additional covariates to statistical models, preserving the individual-level structure of the data.

However, this approach treats social contexts as isolated, indendent properties. In reality, social contexts interact with each other and with individual characteristics, producing complex and layered effects. Traditional models struggle to incorporate this complexity, as they require variables to be defined at the level of the individual observation.

Graph Neural Networks (GNNs) offer a solution by fundamentally changing how both $X$ and $f$ are defined. Introduced by \citet{scarselli_graph_2009}, GNNs are a class of models specifically designed to operate over data structured as graphs---i.e., networks. In our case, the graph represents the population-scale network linking individuals to others in their households, extended families, schools, and neighborhoods. At the same time, the function $f(\cdot)$ becomes a more complex process: GNNs predict outcomes not just from a person’s own characteristics, but by recursively aggregating information from their social contexts (see Section~\ref{s:methods}). 

Here, we compare a gradient boosting model---where $X$ consists of social contexts aggregated at the individual-level (e.g., mean neighborhood education, share of educated peers)---to a GNN that directly uses the raw network structure $X$ and a relational function $f$ to capture complex network interactions. This distinction is key: boosting relies on pre-specified aggregate covariates, whereas GNNs learn directly from the relational graph. A prediction gap between these models suggests that conventional operationalizations (e.g., including mean neighborhood education as a covariate) fail to fully capture how social contexts are associated with the outcome. For example, segmented assimilation theory suggests that migrant children may do better in school when they live near highly educated migrant neighbors who offer support and act as role models \citep{portes_new_1993}. Boosting models may miss this kind of pattern unless we explicitly include it---for instance, by adding the share of highly-educated migrant neighbors. In contrast, GNNs can pick up these connections automatically, which would result in a prediction gap. These gaps again provide a basis for abductive inference: they reveal that social mechanisms are either underspecified or entirely overlooked in traditional models, offering new directions for sociological theory.

\section{Data and Methods}\label{s:methods}
\subsection{Data}
We use population-scale administrative data from Statistics Netherlands (CBS). Access to these data is facilitated by ODISSEI (Open Data Infrastructure for Social Science and Economic Innovations). These longitudinal microdata cover all residents of the Netherlands and are linkable across registers (e.g., tax authority, registrations, education) using anonymized personal identifiers. This enables us to jointly analyze how household, extended family, school, and neighborhood contexts are associated with long-term educational outcomes.

\subsubsection{Cohort Construction}
Our study cohort includes all children who completed the final year of primary school (grade 8) in 2010---when the social contexts data start---and transitioned to secondary school in 2011, when children were approximately age 11--12. This age provides a theoretically consequential baseline immediately before Dutch pupils enter differentiated secondary-school tracks. It captures family, primary-school, and neighborhood contexts before track placement reorganizes peer networks and constrains subsequent educational pathways (see Appendix Section~\ref{s:educational_system}). We use demographic characteristics and information on social contexts (nuclear and extended family, schools, and neighborhoods) observed in 2010. All contextual predictors used in this study are measured at this baseline. 

The outcome is whether an individual had completed university by 2023 (approximately age 24--25). This window allows several years for delayed entry or completion, although it excludes later returners and longer educational trajectories. We excluded 4,734 individuals with missing educational attainment (this happens for example when the person dies) and 1,272 individuals who were absent from the Netherlands for more than 1 year of the 13-year period studied. The final sample size is 188,011 individuals.

We then reconstructed children's early-life social environments by linking them to their nuclear and extended family members, school classmates, and neighborhood residents using CBS population registries. This is possible through the network files \citep{van_der_laan_whole_2022, bokanyi_anatomy_2023}, which connect individuals to household members, extended family members, schoolmates, and neighbors. We do not impute individuals or network ties; relations are taken directly from CBS administrative registers. For address-based relations, CBS validation studies found that 96--97\% of registered individuals lived at their recorded address between 2016--2024 \citep{Adreskwaliteit2024}.

\subsubsection{Variables included} \label{s:m:variables}
For the logistic regression and gradient boosting models, the input data are structured as a tabular dataset where each row corresponds to an individual (student), and each column corresponds to a variable. Since we are mostly interested in the prediction ability of the models, we include a large number of variables across six thematic domains: individual, nuclear family, household, extended family, school, and neighborhood. Network variables are created using the open-source Python library \texttt{netCBS} \citep{garcia-bernardo_netcbs_2024}. The nuclear-family domain captures parental background directly: whether each parent is known in the registry, each parent's university completion, and income rank, as well as the number of siblings and the educational weight used for school funding which depends only on parental education. The full list of variables is shown in Table~\ref{tab:desc_stats}. The motivation for inclusion and datasets are detailed in Appendix Section~\ref{a:s:var_selection} and the correlation between variables in Appendix Section \ref{a:s:correlations}.

\begin{table}[htbp]
\centering
\renewcommand{\arraystretch}{1.2}
\resizebox{\textwidth}{!}{%
\begin{tabular}{p{5.5cm}rrrrrrrrr}
\toprule
Variable & \# values & N & Mean & Std & 1\% & 10\% & 50\% & 90\% & 99\% \\
\midrule
I: Migration generation & 3 & 188001 & 0.3879 & 0.7752 & 0 & 0 & 0 & 2 & 2 \\
I: Registered gender (male) & 2 & 188001 & 0.5047 & 0.5 & 0 & 0 & 1 & 1 & 1 \\
I: Birth year & 8 & 188001 & 1,999 & 0.5589 & 1,997 & 1,998 & 1,999 & 1,999 & 2,000 \\
I: Disability indicator & 2 & 188011 & 0.02227 & 0.1476 & 0 & 0 & 0 & 0 & 1 \\
I: Special education (SBO) & 2 & 188011 & 0.04317 & 0.2032 & 0 & 0 & 0 & 0 & 1 \\
F: WPO weight & 3 & 179912 & 0.08651 & 0.2724 & 0 & 0 & 0 & 0.3 & 1.2 \\
F: Number of siblings & Cont & 188011 & 1.701 & 1.285 & 0 & 1 & 1 & 3 & 6 \\
F: Father univ. degree & 2 & 101182 & 0.4432 & 0.4968 & 0 & 0 & 0 & 1 & 1 \\
F: Mother univ. degree & 2 & 116542 & 0.3487 & 0.4766 & 0 & 0 & 0 & 1 & 1 \\
F: Father's income & Cont & 188011 & 44,910 & 165,800 & -5,861 & 5,437 & 36,530 & 78,580 & 211,500 \\
F: Mother's income & Cont & 188011 & 19,070 & 22,800 & -2,314 & 0 & 16,780 & 37,960 & 86,590 \\
F: Father's income rank & Cont & 188011 & 0.7297 & 0.2765 & 0 & 0.3123 & 0.8266 & 0.9763 & 0.9979 \\
F: Mother's income rank & Cont & 188011 & 0.5055 & 0.2438 & 9.875e-05 & 0.116 & 0.5214 & 0.8407 & 0.9816 \\
F: Father known & 2 & 188011 & 0.9482 & 0.2217 & 0 & 1 & 1 & 1 & 1 \\
F: Mother known & 3 & 188011 & 0.9925 & 0.09381 & 1 & 1 & 1 & 1 & 1 \\
E: Known grandparents & 5 & 188011 & 2.562 & 1.231 & 0 & 1 & 3 & 4 & 4 \\
E: Min grandparent income rank & Cont & 188011 & 0.3652 & 0.1835 & 0 & 0.116 & 0.366 & 0.5693 & 0.8867 \\
E: Max grandparent income rank & Cont & 188011 & 0.6272 & 0.2602 & 0 & 0.3542 & 0.6558 & 0.9367 & 0.9956 \\
E: Min grandparent income & Cont & 188011 & 10,530 & 14,300 & 0 & 0 & 9,445 & 19,130 & 43,890 \\
E: Max grandparent income & Cont & 188011 & 31,670 & 105,000 & 0 & 9,272 & 24,080 & 54,940 & 154,500 \\
E: Grandparents in municipality & 7 & 188011 & 0.4558 & 0.4324 & 0 & 0 & 0.5 & 1 & 1 \\
H: Mean household income & Cont & 187417 & 20,080 & 38,680 & -87 & 6,394 & 16,880 & 34,710 & 79,110 \\
H: Total household income & Cont & 187687 & 64,060 & 177,100 & -272 & 18,890 & 53,800 & 107,800 & 258,400 \\
H: Mean household income rank & Cont & 187417 & 0.4531 & 0.1428 & 0.1479 & 0.2786 & 0.4478 & 0.629 & 0.8765 \\
H: Total household income rank & Cont & 187687 & 0.5 & 0.2887 & 0.01 & 0.09858 & 0.5 & 0.9 & 0.99 \\
H: Number of earners & Cont & 187687 & 3.396 & 5.235 & 1 & 2 & 3 & 5 & 7 \\
S: Class size & Cont & 188011 & 22.99 & 7.88 & 6 & 13 & 24 & 30 & 41 \\
S: Average class size & Cont & 188011 & 21.02 & 5.889 & 7.8 & 13.54 & 21.84 & 25.64 & 33.18 \\
S: School urbanicity & Cont & 188011 & 147,900 & 222,000 & 11,750 & 22,090 & 59,490 & 392,200 & 1,006,000 \\
S: Mean WPO weight (s) & Cont & 179802 & 0.08821 & 0.1493 & 0 & 0 & 0.036 & 0.2196 & 0.7714 \\
S: std. WPO weight (s) & Cont & 179727 & 0.1712 & 0.1634 & 0 & 0 & 0.1122 & 0.4269 & 0.5876 \\
S: Mean income (s) & Cont & 187832 & 63,870 & 46,310 & 26,320 & 39,180 & 60,260 & 86,860 & 153,700 \\
S: std. income (s) & Cont & 187753 & 48,550 & 159,900 & 12,510 & 19,160 & 33,860 & 76,470 & 264,100 \\
S: Mean income rank (s) & Cont & 187832 & 0.517 & 0.13 & 0.1882 & 0.3359 & 0.5309 & 0.6703 & 0.7927 \\
S: std. income rank (s) & Cont & 187753 & 0.261 & 0.04105 & 0.142 & 0.211 & 0.2634 & 0.3086 & 0.3524 \\
S: Share income reported (s) & Cont & 187832 & 0.959 & 0.08784 & 0.7778 & 0.8889 & 0.9531 & 1 & 1.276 \\
N: Mean income (n) & Cont & 187788 & 21,080 & 11,950 & 8,271 & 12,880 & 19,400 & 29,880 & 54,760 \\
N: std. income (n) & Cont & 187788 & 24,260 & 40,570 & 8,987 & 12,410 & 19,160 & 36,380 & 104,000 \\
N: Mean income rank (n) & Cont & 187788 & 0.4914 & 0.07407 & 0.3213 & 0.3996 & 0.4893 & 0.5855 & 0.6801 \\
N: std. income rank (n) & Cont & 187788 & 0.2904 & 0.04454 & 0.177 & 0.2317 & 0.2929 & 0.3455 & 0.3848 \\
N: Mean education (n) & Cont & 187703 & 0.2195 & 0.1537 & 0 & 0.03846 & 0.2 & 0.4211 & 0.6667 \\
N: std. education (n) & Cont & 187515 & 0.3712 & 0.145 & 0 & 0.1961 & 0.414 & 0.5071 & 0.5345 \\
\bottomrule
\end{tabular}
}
\caption{Summary statistics for all model inputs across six domains: individual (I), nuclear family (F), extended family (E), household (H), school (S), and neighborhood (N) (denoted by the letter next to the variable name). For each variable, we show the number of distinct values, sample size (N), mean, standard deviation, and the 1st, 10th, 50th, 90th, and 99th percentiles. Minimum and maximum values cannot be shown to preserve the privacy of Dutch citizens. }
\label{tab:desc_stats}
\end{table}

\paragraph*{Preprocessing}
We supply the same transformed inputs to all three model families so that differences in performance reflect the models rather than their inputs. Scaling is primarily important for stable GNN optimization, and does not substantively affect unpenalized logistic regression or gradient boosting. We therefore used identical inputs to ensure a fair comparison.

Income variables are converted to ranks, before modeling or aggregation following common practice in mobility research \citep{engzell_understanding_2023}. The relevant reference population depends on the measure and is documented in the accompanying code. Ranking reduces the influence of extreme values and accommodates incomes that are zero or negative. Parental income ranks are calculated relative to the full Dutch population, whereas household, school, and neighborhood ranks are calculated within the study sample. These ranks already range from 0 to 1 and are not Z-score standardized.

For count and ordered variables measured on different scales---birth year, migration generation, number of siblings, number of known grandparents, and number of household income earners---we use min--max scaling:
\[
x^{*}=\frac{x-\min(x)}{\max(x)-\min(x)}.
\]
This maps the smallest observed value to 0 and the largest to 1 while preserving their ordering. For continuous school variables---class size, average class size, and log school urbanicity---we use Z-score standardization:
\[
x^{*}=\frac{x-\bar{x}}{s_x}.
\]
After this transformation, 0 represents the sample average and one unit represents one standard deviation. School urbanicity is logged before standardization to reduce its strong right skew.

Table~\ref{tab:desc_stats} provides the complete variable list. The preprocessing code and the aggregated data needed to reproduce the figures are available in the repository listed in the Code and data availability statement. Several variables had a few missing values (see Table~\ref{tab:desc_stats}) and two variables had a large share of missing values. These are parental education (father and mother), where information is largely missing for older cohorts (Section~\ref{a:s:miss_education} in Appendix), and the educational weight used for school funding (WPO weight), which is missing for students attending special education. Since our goal is to maximize prediction, we created for parental education a corresponding binary indicator variable to flag the original missingness. WPO weights are missing for students attending special education. Since we also include this variable in the model, we replace missing WPO weights with zero without loss of information. This strategy allows the models to use information contained in the missingness pattern. \citep{salganikIntroductionSpecialCollection2019, sperrinMissingDataShould2020}. Since our models include highly correlated variables and missingness indicators, we do not interpret the resulting coefficients.

\paragraph*{Graph Neural Network Variables} \label{s:gnn_variables}
For the Graph Neural Network (GNN), we encode the same information into the heterogeneous network structure representing the embeddedness of individuals within social contexts. The GNN input consists of two node types: \texttt{child} (students) and \texttt{person} (e.g., parents, neighbors, peers), each with its own set of features:\\
\textit{• Child node features}: Birth year, migration generation, gender (male), disability indicator, special education (SBO) indicator, WPO weight (educational weight used for school funding) and parent-known indicators. Household features included at the child level: mean and total household income; number of known grandparents, indicator for co‑municipal grandparents. School features: class size, average class size, log‑urbanicity, share of parents reporting income. \\
\textit{• Person node features}: Birth year, migration generation, gender (male; imputed to 0.5 if missing), educational attainment (university completed in 2010), missingness indicator for educational attainment, income rank (filled with 0 if missing), missingness indicator for income.

To keep the GNN architecture computationally feasible and comparable to the tabular models, we include some school and household characteristics---such as class size, average household income, and urbanity---directly as input features on the child nodes, rather than including ``school nodes'' in the network and relying on GNN to infer them. We also add binary indicators for whether each parent is known, because the architecture we use (GraphSAGE with mean and max aggregation) does not reliably count the number of connected nodes, such as siblings or household members. Including these features explicitly helps ensure that differences across models reflect modeling flexibility and relational structure rather than unequal access to basic contextual information.









\subsection{Statistical Models and Prediction Approach}
We compare three families of models of increasing complexity chosen for their prominence and complementarity. Logistic regression is the standard tool in sociological research, providing interpretability and direct comparability with prior work. Gradient boosting trees represent the state-of-the-art for flexible tabular data, able to capture non-linearities and interactions without manual specification. Graph neural networks extend this further by leveraging the raw relational structure of social contexts, allowing us to model students’ embeddedness in families, schools, and neighborhoods directly. Taken together, these three models allow us to investigate to what extent the prediction gap is driven by missing interactions and nonlinearities, or by limitations in how we operationalize social contexts (Fig.~\ref{fig:conceptual}).

\subsubsection{Logistic Regression}
Our baseline model is a logistic regression, which models the log-odds of university completion with a linear combination of predictors:
\begin{equation}
\log\left( \frac{P(y = 1 \mid X)}{1 - P(y = 1 \mid X)} \right) = \beta_0 + \beta_1 X_1 + \beta_2 X_2 + \dots + \beta_p X_p
\end{equation}
This model is interpretable and widely used in the social sciences. It assumes additive effects and cannot capture complex interactions or nonlinearities without explicitly specifying them. We use the standard implementation from the \texttt{scikit-learn} Python package \citep{pedregosa_scikit-learn_2011}.

Our predictors consist of individual characteristics (e.g., migration background, gender, education, parental income) and aggregated contextual variables (e.g., average neighborhood income, share of educated neighbors or school peers).  Given our large sample size and our focus on predictive performance and model comparison rather than coefficient interpretation, we included all variables in Table~\ref{tab:desc_stats} simultaneously.

\subsubsection{Gradient Boosting Trees}
Gradient boosting is a machine learning method that builds a sequence of decision trees, each one correcting the errors of the previous. This creates a highly flexible function that can capture complex, high-order interactions and nonlinearities without requiring researchers to specify them manually.

We use a standard implementation (\textit{Histogram-based Gradient Boosting Classification Tree}) from \texttt{scikit-learn}, trained on the same individual-level features as the logistic regression. Formally, the prediction is modeled as:
\begin{equation}
\log\left( \frac{P(y = 1 \mid X)}{1 - P(y = 1 \mid X)} \right) = \sum_{m=1}^{M} \beta_m h_m(X)
\end{equation}
where each $h_m(X)$ is the prediction of a shallow decision tree, and $\beta_m$ is a learned weight.

\subsubsection{Graph Neural Networks (GNNs)}
Graph Neural Networks (GNNs) offer a fundamentally different approach compared to traditional models, both in how they define the function $f(\cdot)$ and how they structure the input space $X$. Rather than treating individuals as isolated units with fixed features, GNNs represent individuals as nodes within a social graph. The GNN input consists of two node types: \texttt{child} (students) and \texttt{person} (e.g., parents, neighbors, peers), each with its own set of features. Each node is connected to others via edges that represent social contexts between the type of nodes. We model five different relationships between the two types of nodes (Fig.~\ref{fig:conceptual}A): (1) shared classrooms (child–child), (2) parent links (child–person), (3) child links (person–child), (4) family links connecting co-parents or siblings (person–person), and (5) neighbor links (person–person).

In this framework, the input $X$ includes not only an individual’s own characteristics, but also the features of their social connections. The function $f(\cdot)$ is redefined as a recursive aggregation process over the network. We use a two-layer \textit{GraphSAGE} architecture \citep{hamilton_inductive_2017} without sampling, from the \texttt{pytorch-geometric} Python package. Each node begins with its own characteristics. At each layer, the model summarizes information from connected individuals separately for each relationship, using either mean or combined mean and maximum aggregation, and combines this information with the node's existing representation through a nonlinear update. Two layers allow information to pass through both direct and indirect connections. The resulting student representation is then used to predict university completion. The complete equations, aggregation and update functions, and implementation details are reported in Appendix Section~\ref{a:s:gnn}.

This GNN-based approach is especially well suited for modeling complex and overlapping social environments. By incorporating the structure of relationships and recursively combining information across direct and indirect connections, the model can capture higher-order patterns---such as information about academically high-achieving peers, or the clustering of disadvantage within schools or neighborhoods---that would need to be explicitly operationalized and included in other types of statistical models.

\subsubsection{Ensuring a fair comparison between models}
To ensure fair comparisons between models, we use the same data (see Section~\ref{s:m:variables}) and apply the same fitting and evaluation procedure to all approaches. Each model is trained to minimize the log-loss (also known as binary cross-entropy loss), which is the standard function in traditional logistic regression. This loss function penalizes incorrect predictions more strongly when the model is more confident in them, encouraging models to produce calibrated probabilities rather than just correct classifications.

The log-loss is defined as%
\begin{equation}
\mathcal{L} = -\frac{1}{N} \sum_{i=1}^N \left[ y_i \log(\hat{y}_i) + (1 - y_i) \log(1 - \hat{y}_i) \right],
\end{equation}
where $y_i \in \{0,1\}$ is the true outcome for individual $i$ (i.e., 0=not completing university, 1=completing university) and $\hat{y}_i$ is the model's predicted probability that the individual completes university. 

Highly flexible models, such as Gradient Boosting and Graph Neural Networks, are prone to overfitting, where a model appears highly accurate simply because it has memorized the data it was trained on, rather than having learned patterns that apply more generally. To avoid this, we evaluate \textit{out-of-sample} prediction ability. We randomly split the dataset into a training set (80\% of the data) and a test set (20\%, 37,603 observations). The models are trained only on the training set, and their performance is evaluated on the held-out test set. The test set provides a clean benchmark for assessing how well the model performs on unseen data, thereby offering a more honest evaluation of predictive accuracy. If a complex model merely fits patterns specific to the training data, including local network patterns, it will not predict these held-out outcomes better. Our approach evaluates generalization to held-out students within the same cohort; it does not establish causal mechanisms or transportability to other cohorts or networks. We construct confidence intervals by performing 1,000 bootstrap resamples of the test dataset.

By keeping the loss function, data inputs, and train/test split constant across models, we can attribute the observed differences in performance to the model’s ability to capture complexity which helps predict university graduation---such as interactions or network structure---rather than differences in data access or overfitting. 
Statistical models, especially highly flexible machine learning approaches such as gradient boosting trees and Graph Neural Networks (GNNs), have several settings or \emph{hyperparameters} that must be specified prior to training. These hyperparameters control how complex and flexible the learned function $f(X)$ can be. Optimally tuning these hyperparameters is crucial: Overly simple models may miss relevant complexity, while overly complex models may overfit the training data, leading to poor generalization to unseen cases. To systematically determine optimal hyperparameters for our GNN, we used an automated optimization procedure via Optuna, a Python package designed for efficient hyperparameter optimization. This is detailed in the Appendix Section \ref{a:s:hyper_tuning}. The outcomes in the test set are not used for model fitting or hyperparameter selection. Hyperparameters are selected using validation partitions or cross-validation within the training set, after which the final models are evaluated on the held-out test outcomes.

\subsubsection{Comparing the Prediction Ability}
Evaluating how well models predict university completion is not straightforward when there are imbalances in university completion. For some groups, success is nearly guaranteed: for instance, over 89\% of children with parents with post-graduate degrees attend university. In these cases, predicting success is easy and uninformative. A good model is one that can also identify the few children who do \emph{not} follow expected trajectories---those who fail despite advantage, or succeed despite disadvantage.

Model fitting and performance evaluation serve different purposes. We minimize log loss because it is differentiable and encourages predicted probabilities that match observed completion rates (calibration). 
Similarly to \citet{savcisens_using_2024}, we use the Matthews Correlation Coefficient (MCC) as our main measure of predictive performance. MCC is threshold-dependent and non-differentiable, so it is not suitable as a common training objective. MCC is a balanced metric that captures how well a model correctly identifies both positive and negative outcomes \citep{chicco_advantages_2020}. Unlike accuracy or F1 score, it remains reliable even when one outcome is much more common than the other. We also report held-out log loss and AUC in Appendix Section~\ref{a:s:auc} and ~\ref{a:s:log_loss}. Mathematically, MCC is defined as:
\begin{equation}
\text{MCC} = \frac{TP \cdot TN - FP \cdot FN}{\sqrt{(TP + FP)(TP + FN)(TN + FP)(TN + FN)}}
\end{equation}
Here, $TP$ refers to true positives, $TN$ to true negatives, $FP$ to false positives, and $FN$ to false negatives. The MCC ranges from $-1$ (perfectly wrong predictions), to $0$ (no better than chance), to $1$ (perfect prediction). Although not intuitive, MCC is mathematically equivalent to the phi coefficient (i.e., Pearson’s correlation for two binary variables). An MCC of zero indicates that predictions are no more associated with outcomes than expected under random classification. 

We retain MCC as our principal classification metric because it distinguishes both completers and non-completers, and interpret it alongside accuracy, and log loss since they directly evaluate absolute predictability on scales where 100\% accuracy and a log loss of 0 indicate perfect held-out predictions.

\section{Results and Discussion}\label{s:results}

\subsection{Parental background absorbs most predictable variation}

We first predict university completion by 2023 (when most students are age 24--25) from demographic characteristics and early-life social contexts (nuclear family, extended family, schools, neighborhoods) observed in 2010, when students were about 11--12. We compare three models of increasing complexity: logistic regression (Linear), gradient boosting (Boost), and graph neural networks (GNNs). We evaluate performance with the Matthews Correlation Coefficient (MCC), a measure robust to outcome imbalance that is equivalent to Pearson's correlation for binary outcomes. The central result is that even when we use increasingly flexible models, parental socioeconomic background accounts for most measurable predictive structure in university completion.

\begin{figure}[ht!]
    \centering
    \includegraphics[width=1\linewidth]{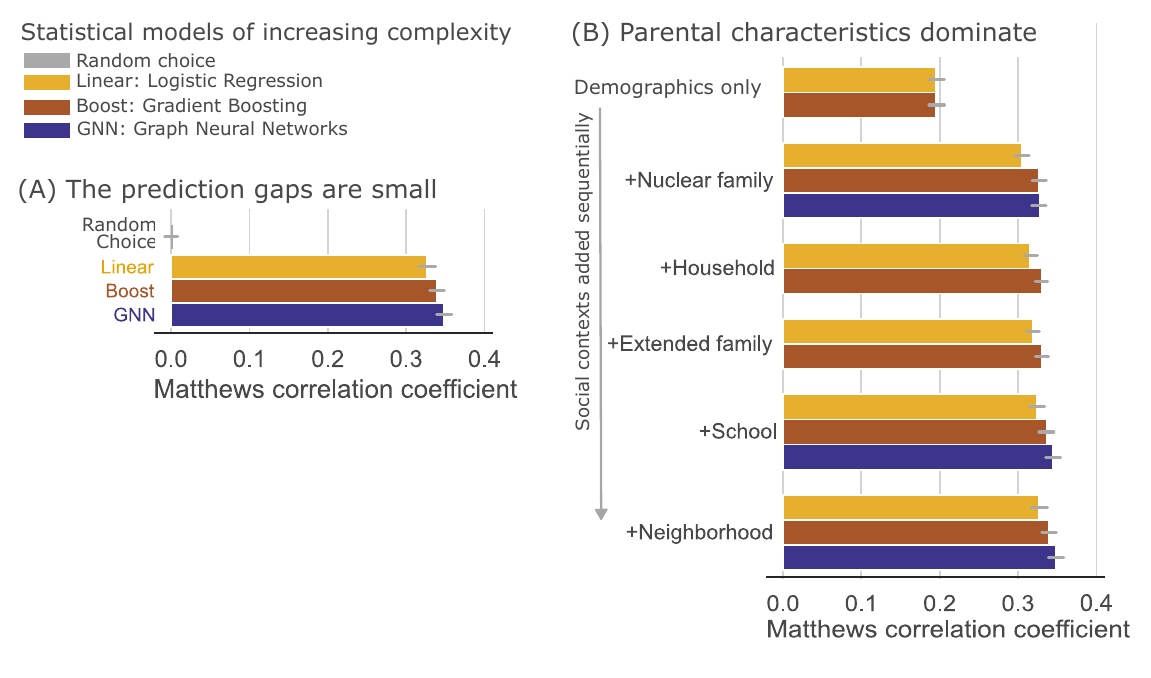}
    \caption{\textbf{Prediction gaps of educational attainment}. Out‐of‐sample predictive performance (Matthews Correlation Coefficient) of three models of increasing flexibility: logistic regression (Linear), gradient boosting (Boost), and graph neural networks (GNN), alongside random guessing for (A) the complete model with all social contexts; (B) models in which social contexts are added sequentially. Note that the confidence intervals (95\% bootstrap confidence intervals) are not independent since they are based on the same data. Results using AUC are shown in Section~\ref{a:s:auc} in the Appendix.}
    \label{fig:pred_gap}
\end{figure}

GNNs outperform both logistic regression and gradient boosting (Fig.~\ref{fig:pred_gap}A), with a 1.24 percentage point MCC gain from Linear to Boost, and an additional 0.84 points with GNNs.\footnote{Equivalent results using AUC-ROC and held-out log loss are shown in Appendix Sections~\ref{a:s:auc} and \ref{a:s:log_loss}, respectively. Log loss, which is also the fitting objective, yields the same model ordering and similarly small prediction gaps. While commonly used, AUC can be misleading in imbalanced datasets \citep{chicco_advantages_2020}.} But these differences are modest---just 247 additional correct predictions on university completion among 37,603 students. Moreover, all three models agree on approximately 90\% of individual predictions, making similar errors and successes (Section~\ref{a:s:conf_matrix} in the Appendix).

Despite drawing on population-scale administrative data covering parental background, extended family, schools, and neighborhoods, the full GNN achieves only moderate predictive performance. Its MCC of 0.348 and AUC of 0.749 indicate that the model captures meaningful differences in university completion, but substantial variation remains unpredictable. The model correctly classifies 70.0\% of students, compared with 61.6\% when always predicting the majority class, and reduces log loss from 0.666 for a prevalence-only model to 0.570. Thus, although the models improve substantially over simple benchmarks, even the most flexible model leaves considerable uncertainty about who completes university (Appendix Sections~\ref{a:s:auc}--\ref{a:s:accuracy}).

The small prediction gaps across models suggest that predictions hinge on strong predictors common to all approaches. To examine this further, we assess how prediction accuracy improves for each model when social contexts are added sequentially. As expected given the extremely dominant role of parental background in shaping educational outcomes \citep{engzell_its_2020}, parental characteristics yield the largest gain in MCC---rising from 19\% to 32\%---when added to a baseline model including only basic demographics (birth year, gender, migration generation, disability status) (Fig.~\ref{fig:pred_gap}B). Adding household, extended family, school, and neighborhood information produces only marginal additional improvements, reaching about 35\%. This suggests that parental background captures most of the measurable variation in educational outcomes. Notably, these smaller incremental gains from additional contexts are consistently modest across all modeling strategies (Fig.~\ref{fig:pred_gap}B), suggesting that interactions and nonlinearities overall do not greatly improve prediction.

Principal Component Analysis of the GNN embeddings confirms this dominance. PCA is a dimensionality reduction technique that is able to combine the information of the embeddings into fewer variables while capturing as much of the remaining variability as possible. Because GNN embeddings represent each student's position within their social context---and serve as input into a logistic regression in the model’s final prediction---examining which dimensions explain the most variation helps us understand what the model has learned. If one dimension accounts for most of the variance in the embeddings and aligns closely with parental background, this suggests that parental background is the most important factor shaping the model’s predictions. That is exactly what we observe. Three components explain nearly 90\% of the variance in the embeddings (67.6\%, 15.5\%, and 6.0\%, respectively). The first dimension correlates strongly with the embedding when only features related to the nuclear family are used (82\% correlation), and with parental education and income (51–53\%). Even in a flexible, network-based model, parental background remains the primary determinant (see Appendix \ref{a:s:embeddings} for full tables and methodology).

The small prediction gaps driven by the dominant role of parental socioeconomic status support social reproduction and status-attainment theories. While the observed lack of additional predictive power from models including social contexts may seem to contradict social capital and neighborhood-effects theories that highlight their importance \citep{colemanSocialCapitalCreation1988, ainsworth_why_2002, sharkey_where_2014}, our findings do not imply that schools and neighborhoods are unimportant. Rather, they suggest that either there is an unobserved confounder, or likely that part of the influence of family background is mediated by school and neighborhood social contexts. Wealthier, better-educated families place children into advantaged contexts \citep{kuyvenhoven_neighbourhood_2021, troost_neighbourhood_2023} which in turn shape children's outcomes. So, once parental characteristics are accounted for, these social contexts add little independent predictive power.

An alternative reason for the small potential gaps is that social contexts are trajectories rather than fixed attributes. The duration, timing, and sequencing of exposure matter: the neighborhood observed at ages 11--12 may represent either a stable childhood context or a recent move, while the same snapshot may conceal cumulative advantage, disadvantage, or important turning points \citep{sharkey_where_2014, chettyImpactsNeighborhoodsIntergenerational2018}. Residential mobility, school transitions, and changes in household composition are socially patterned and impacted by families' socioeconomic backgrounds. Our models collapse these different histories into a single baseline representation. The small prediction gaps consequently show that contexts measured in 2010 add little predictive information beyond parental background; they do not imply that accumulated or changing contextual exposures are unimportant.

\subsection{Prediction gaps as abductive puzzles: unequal patterns of social contexts for girls with absent fathers}
 
\begin{figure}[ht!]
    \centering
    \includegraphics[width=1\linewidth]{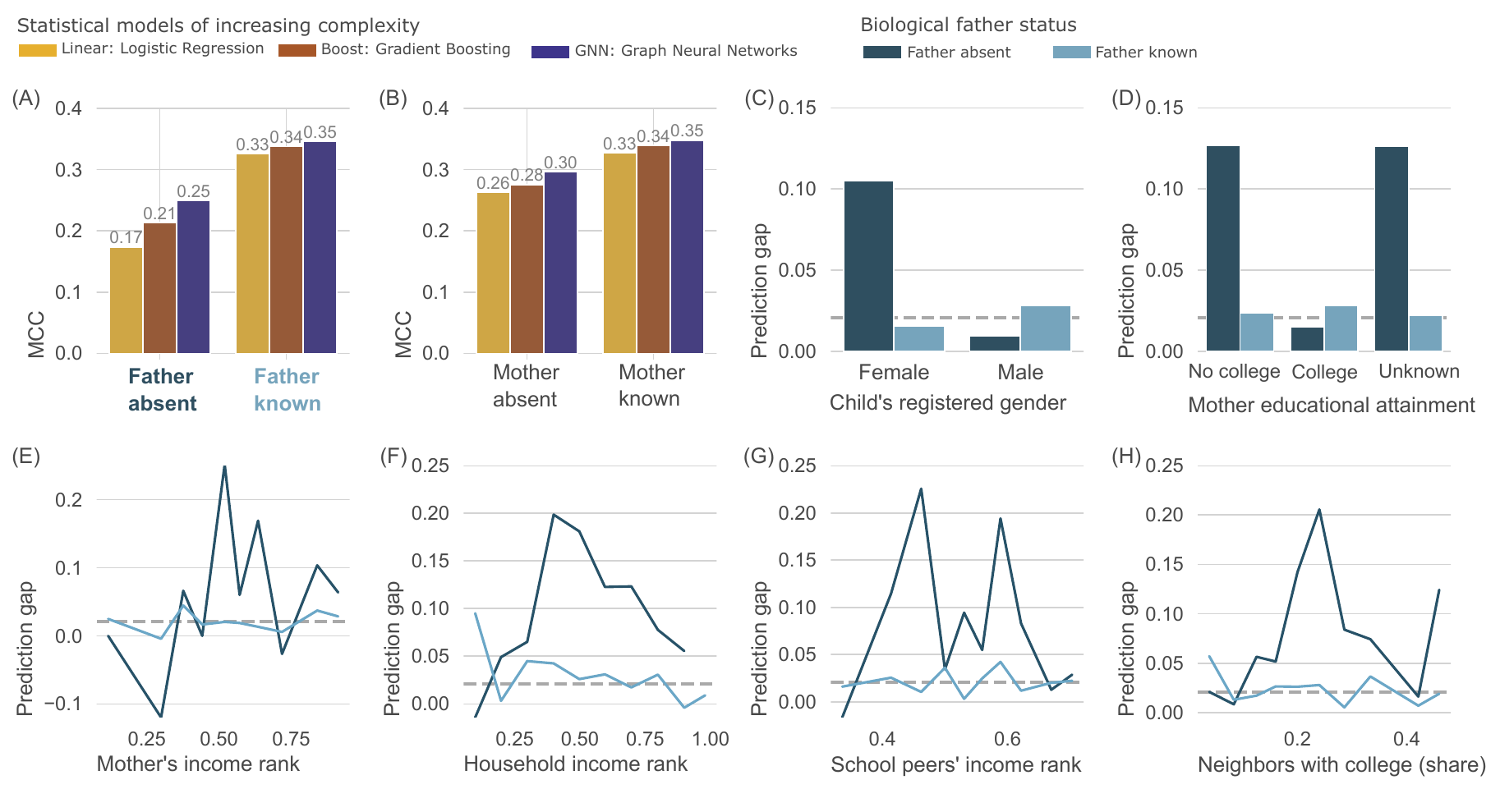}
    \caption{\textbf{Prediction gaps are largest for girls without a father present}. (A--B) Predictive performance (Matthews Correlation Coefficient, MCC) of the linear (gold), boosting (brown) and GNN (purple) models disaggregated by (A) father and (B) mother status. (C--H) Prediction gaps between logistic regression and GNN disaggregated by father status (dark blue, father deceased or not known to the registry; light blue, father known) and (C) child's registered gender (D) mother educational attainment (E) mother's income rank (F) household's income rank (G) school peers' income rank (H) share of neighbors with a college degree. The dashed horizontal line in (C--H) indicates the average gain from using GNNs in the entire population.}
    \label{fig:disaggregation}
\end{figure}

Although population-level prediction gaps are small, they are not uniform across subgroups. The largest prediction gap (between GNN and logistic regression) appears among students whose fathers were deceased or absent from administrative records (Fig.~\ref{fig:disaggregation}A): MCC increases from 0.17 (logistic regression) to 0.21 (gradient boosting) and 0.25 (GNN). No comparable gap appears among those without a registered mother (Fig.~\ref{fig:disaggregation}B), suggesting that social contexts provide additional predictive information when the father is absent. These gains for children with absent fathers are concentrated among girls (Fig.~\ref{fig:disaggregation}C) and among children with less-educated mothers or from middle-income households, schools, and neighborhoods (Fig.~\ref{fig:disaggregation}E–H).\footnote{These gains stem from both the shift to non-linear models (boosting) and the additional value of modeling relational structures with GNNs, as we show in Appendix Section~\ref{a:s:disaggregation_gap}.}

Existing work has documented that family disadvantage and parental disruption disproportionately affect boys' behavioral and educational outcomes on average, while girls are more likely to internalize stress but maintain academic engagement \citep{boertien_gendered_2022, lundberg_father_2017}. Our finding is complementary but distinct. For girls with absent fathers, the prediction gap indicates that attainment depends on combinations of circumstances that are not captured by a simple model centered on parental characteristics, but that are reflected in the measured social context data; for boys with absent fathers it does not. 

One interpretation is that girls compensate more effectively for paternal absence through other social contexts, making their educational trajectories comparatively more predictable for more flexible models. In contrast, boys are typically less sheltered from developing anti-school attitudes and behaviors that align more closely with certain conceptions of masculinity \citep{boertien_gendered_2022}. Whether these attitudes develop and affect educational attainment may depend on unobserved circumstances, such as parenting styles or community support structures. Because these factors are not captured in our data, boys' outcomes may remain harder to predict even with more flexible models.

Which mechanisms generate this structure---compensatory social ties beyond the household, gendered coping strategies, differential institutional response, or something else---is a puzzle our predictive framework identifies but cannot resolve. The point is precisely that: the gap nominates children with absent fathers as a subpopulation for whom mechanism-focused, confirmatory research into the role of wider social contexts is most likely to be informative.


\section{Conclusion}\label{s:conclusion}
This paper examined whether early-life social contexts improve predictions of university completion beyond parental background. Using Dutch population-scale registry data, we compared logistic regression, gradient boosting, and graph neural networks that model students' embeddedness in families, schools, neighborhoods, and extended kin. We find that parental socioeconomic background captures most measurable predictive structure. Even the GNN, our strongest test of whether relational context adds information beyond aggregate covariates, produces only small gains in the pooled population over a simple model with demographics and parental income and education. Moreover, consistent with recent studies of life-outcome prediction \citep{salganikIntroductionSpecialCollection2019,sivak2026births}, the absolute predictability of university completion is modest. The GNN achieves an accuracy of 0.700, compared with 0.616 for a model that always predicts the majority class. The large differences in Figure~\ref{fig:context_inequality} make parental characteristics strong predictors, but considerable variation remains among children with similar backgrounds.

The methodological contribution is to demonstrate how flexible prediction can support, rather than replace, sociological theory. Prediction gaps turn differences in predictive performance into puzzles for abductive inference: they do not explain mechanisms, but they show where theory-guided specifications are more or less complete.
Larger gaps identify places where standard specifications may be missing interactions, relational information, patterned missingness, or unmeasured heterogeneity. In our exploratory analyses, the largest gaps appear among students without a registered father, especially girls. This finding nominates this subgroup as one for which mechanism-focused, confirmatory research on social context and educational attainment is likely to be especially informative. It is possible that social capital embedded in contexts such as extended family, schools and neighborhoods is especially relevant (via, e.g., compensatory mechanisms) for educational attainment in this subgroup. 

The same logic applies when prediction gaps are small. An important contribution of this paper is demonstrating what graph neural networks can and cannot add to sociological inquiry. If population-scale networks contained substantial predictive information not captured by aggregate covariates, the GNN should have been able to exploit it. Instead, the modest pooled gains suggest that the social-context constructs available in the data, such as connections to higher socioeconomic status \citep{chettySocialCapitalMeasurement2022} or the clustering and overlap of social contexts \citep{bokanyi_anatomy_2023}, do not make a large separate predictive contribution once parental background is observed.

Our study has three important limitations, each opening avenues for further research. First, our outcome variable---university completion---compresses a long sequence of educational transitions into one endpoint. This makes it a useful high-stakes summary outcome, but it may also amplify the apparent dominance of parental background, whose influence accumulates across track placement, secondary-school progression, university entry, persistence, and completion.  Future work should examine outcomes less tightly coupled to family background, such as labor market trajectories or adult health.

Second, our analysis concerns one cohort, and its predictor set is a single baseline snapshot from 2010. It captures children's registered family, household, school, and neighborhood contexts at ages 11--12, but not the duration, accumulation, or sequencing of exposure. The interpretation of the prediction gaps is limited to social contexts measured before tracking and completion observed by ages 24--25. Networks formed later in the educational career and degrees completed at older ages fall outside the study. However, residential moves, changes in household composition, and school transitions alter the contexts children encounter and are themselves stratified by parental resources. This may contribute to the small prediction gaps: parental characteristics may partly capture persistent advantage and prior mobility, while a baseline snapshot misses subsequent contextual change. Future work should construct time-varying networks and compare stable, cumulative, and changing exposures across the life course.

Third, important sociological mechanisms are unmeasured in administrative data. Parenting styles, cultural norms, genetics, friendships, and local institutional interactions are absent from our models. This likely contributes to the observed persistent unpredictability, particularly among disadvantaged groups. However, if these mechanisms are indirectly captured through observable behaviors (e.g., parenting styles reflected in enrollment in religious or alternative schools; cultural values visible through patterns of residential mobility), machine learning models could still detect their imprints. 

The central lesson is that machine learning is useful when it is treated as a complement to explanation, not as an explanation itself. Prediction gaps offer a way to turn advances in predictive modeling into opportunities for sociological theory-building. By revealing where existing specifications capture most predictable variation and where they fall short, they identify empirical puzzles that invite theoretical refinement and further investigation. 


\subsection*{Acknowledgments} \label{ack}
\anon{This work would not be possible without ODISSEI (the Open Data Infrastructure for Social Sciences and Economic Innovations), which played a pivotal role in making CBS data more accessible and helping to create network files. ODISSEI also connected researchers, including the authors of this paper. ODISSEI also provided data access for this project (NWO grant number 184.035.014). The initial set-up of this project was enabled by the Utrecht University's Applied Data Science grant (call November 2023) with the research assistance of Bram Hoogendijk. Javier Garcia-Bernardo acknowledges support from the Dutch Research Council (NWO grant VI.Veni.231S.148). The authors would like to thank the participants in the ODISSEI Incubator for helpful discussions, in particular to Tom Emery, Flavio Hafner, Malte Luken, Nicolás Soler, Erik-Jan van Kesteren, and Matt Salganik.}


\subsection*{Code and data availability} 
Administrative data from Statistics Netherlands (CBS) is available for statistical and scientific research under specific conditions. For more details about the data we refer you to \textcite{bokanyi_anatomy_2023, van_der_laan_whole_2022}. For more details about access and usage regulations, please contact \url{microdata@cbs.nl}.

Aggregated data exported from CBS and all analysis scripts can be found at 
\ifthenelse{\boolean{anon}}%
  {\url{https://anonymous.4open.science/r/cbs_gnn_ads-286C}}%
  {\url{https://github.com/jgarciab/cbs_gnn_ads}}%

For further needs, please contact the corresponding author.

\subsection*{Institutional review and consent} 
Data is collected by Statistics Netherlands (CBS) and the National Institute for Public Health and the Environment (RIVM), and made available to researchers for well-defined projects and statistical analysis. Researchers need to be pre-approved before accessing the data, and all data is pseudoanonymized, and available in a secure research environment. The data is safeguarded under the stringent privacy regulations set by the Statistics Netherlands Act (``Wet op het Centraal bureau voor de statistiek'') and the European Union's General Data Protection Regulation, guaranteeing that individual personal information is not revealed during the analysis. All methods were carried out in accordance with relevant guidelines and regulations.

This project has received ethical approval by the Faculty Ethical Review Board of the Faculty of Social and Behavioural Sciences at \anon{Utrecht University (registration number 24-0562).}

\subsection*{Declaration of conflicting interests}

The authors declared no potential conflicts of interest with respect to the research, authorship, and/or publication of this article.

\subsection*{Declaration of generative AI and AI-assisted technologies in the writing process.}
During the preparation of this work the authors used OpenAI's models for copy-editing. The authors reviewed and edited the content as needed and take full responsibility for the content of the published article.

\printbibliography

\newpage
\appendix
\renewcommand{\thesection}{A\arabic{section}}
\renewcommand{\thesubsection}{A\arabic{section}.\arabic{subsection}}
\renewcommand{\thesubsubsection}{A\arabic{section}.\arabic{subsection}.\arabic{subsubsection}}
\setcounter{figure}{0}\renewcommand{\thefigure}{A\arabic{figure}}
\setcounter{table}{0}\renewcommand{\thetable}{A\arabic{table}}

\begin{refsection}
\section*{Appendix}\label{a:appendix1}

\section{The Educational System in the Netherlands}\label{s:educational_system}

\begin{figure}[ht!]
    \centering
    \includegraphics[width=0.9\linewidth]{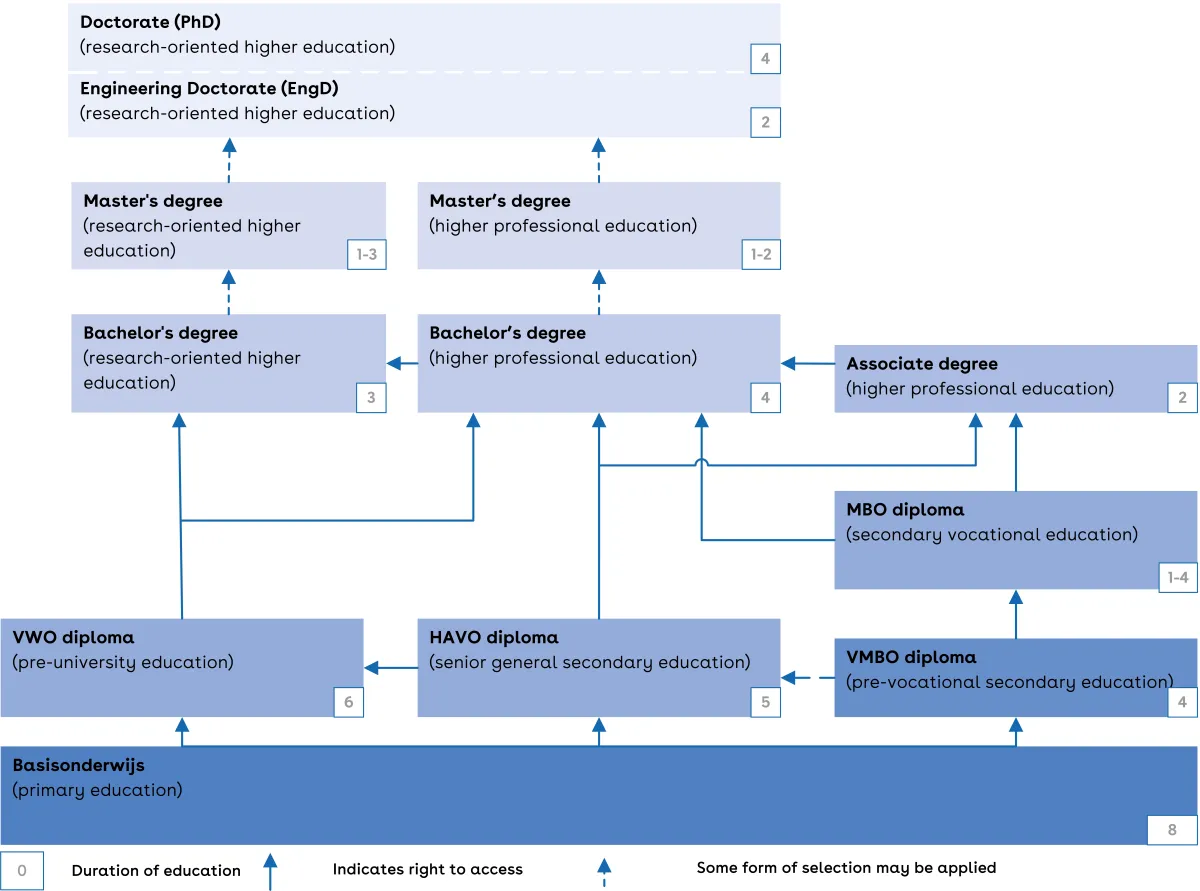}
    \caption{\textbf{Diagram of the Dutch education system}. Progression from primary education through the three secondary tracks (VMBO, HAVO, VWO) into vocational (MBO), professional (HBO), and research (WO) higher education, with typical duration (white boxes) and entry pathways (arrows). Source: Nuffic (the Dutch organisation for internationalisation in education).}
    \label{fig:educ_chart}
\end{figure}

The Dutch educational system separates students into different academic and vocational pathways at an early age (Fig.~\ref{fig:educ_chart}). Primary education continues until age 12, after which students transition to secondary education tracks determined by standardized tests, teacher recommendations, and parental input. The three main secondary tracks are pre-vocational secondary education (VMBO), senior general secondary education (HAVO), and pre-university education (VWO). These tracks significantly shape students' subsequent educational and career opportunities. Students who complete VMBO typically continue to senior secondary vocational education (MBO), HAVO graduates progress to universities of applied sciences (HBO), and VWO graduates have direct access to research universities (WO).

A key feature of the Dutch system is its strong early stratification leading to limited mobility between tracks after initial placement, which exacerbates educational inequalities compared to educational systems with later tracking \citep{knigge_delayed_2022}. By the time students leave primary-school, the upper limits of educational attainment are largely set. While transitions between tracks are possible (Fig.~\ref{fig:educ_chart}), they are uncommon. Social contexts---such as family background, neighborhood characteristics, and school environments---significantly influence these educational trajectories. Dutch students from highly educated or affluent families are disproportionately placed in academic (VWO) tracks and subsequently access university education at higher rates even when controlling for prior achievement (Fig.~\ref{fig:context_inequality}B). Twin evidence points in the same direction: when students are tracked at age 12, the shared family environment explains about a quarter of the variance in their track level beyond genetic influences, compared with almost none when tracking is delayed \citep{knigge_delayed_2022}. Previous research has further demonstrated how parental involvement and cultural capital intensify these outcomes: middle-class families often strategically influence placement decisions and navigate the educational system more effectively, while disadvantaged families may face greater institutional barriers \citep{keskiner_is_2015, van_de_werfhorst_ethnicity_2007}. These inequalities are further reinforced by the socioeconomic composition of neighborhoods and schools: privileged students concentrate in particular schools and neighborhoods \citep{kuyvenhoven_neighbourhood_2021, troost_neighbourhood_2023}.

Ages 11--12 provide a baseline immediately before secondary-school tracking reorganizes students' educational pathways and peer networks. By age 12--13, secondary-school track is already highly predictive of university completion, leaving relatively little room for peer characteristics to add predictive information. Our outcome captures completion of HBO or WO by ages 24--25, allowing several years beyond nominal degree duration while excluding degrees completed later in life.

\section{Missingness in educational outcome}\label{a:s:miss_education}

We define university completion as any degree with a Standard Education Classification (SOI, in Dutch) code starting with 3, which corresponds to programs in higher professional education (HBO) and research universities (WO) under the Dutch Standard Education Classification (SOI).

\begin{figure}[ht!]
    \centering
    \includegraphics[width=0.7\textwidth]{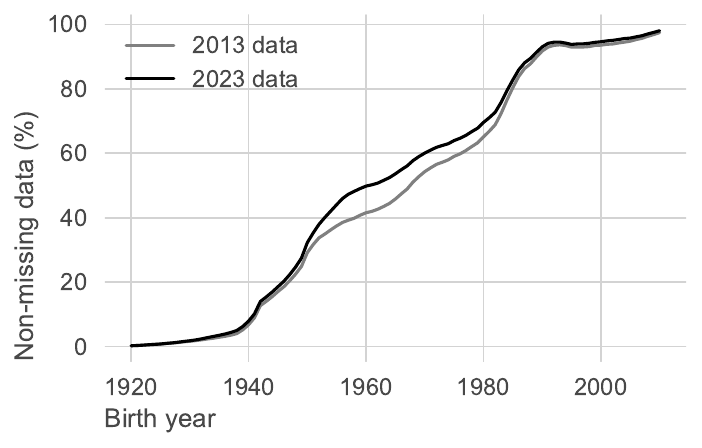}
    \caption{Share of individuals with non-missing educational outcome by birth year, based on two CBS data snapshots: one from 2013 (gray line) and one from 2023 (black line). Note that educational data completeness is much higher in the 2023 data for individuals born 1950--1985.}
    \label{fig:missing_education}
\end{figure}

Educational attainment data at CBS is incomplete for older cohorts, as systematic recording of post-secondary education began only in recent years. Our study cohort consists primarily of individuals born in 1998 and 1999, with a small share born in 1997 (due to grade repetition) or 2000 (due to grade skipping), for whom educational data is nearly complete (Figure~\ref{fig:missing_education}). Missingness is more common among individuals in the social contexts of our cohort---such as parents or neighbors---whose education was completed before registry coverage improved. CBS continues to expand coverage through large-scale representative surveys, increasing data completeness over time.

We address the limitation of unobserved educational attainment in two steps. First, for individuals born after 1980, we use educational attainment recorded in 2013, assuming survey updates outweigh further education. For individuals outside of our cohort (e.g. neighbors) born in 1980 or earlier, we use 2023---the first year educational records were available using the SOI classification.  Second, we include indicators for missing educational attainment. In the linear and boosting models, this applies to parental variables. School and neighborhood summaries are calculated over network members with observed educational attainment; we do not impute missing individuals or ties. In the GNN, each node has a missing-attainment indicator.

\FloatBarrier

\section{Unequal shares of university completion across income groups}

\begin{figure}[ht!]
    \centering
    \includegraphics[width=\linewidth]{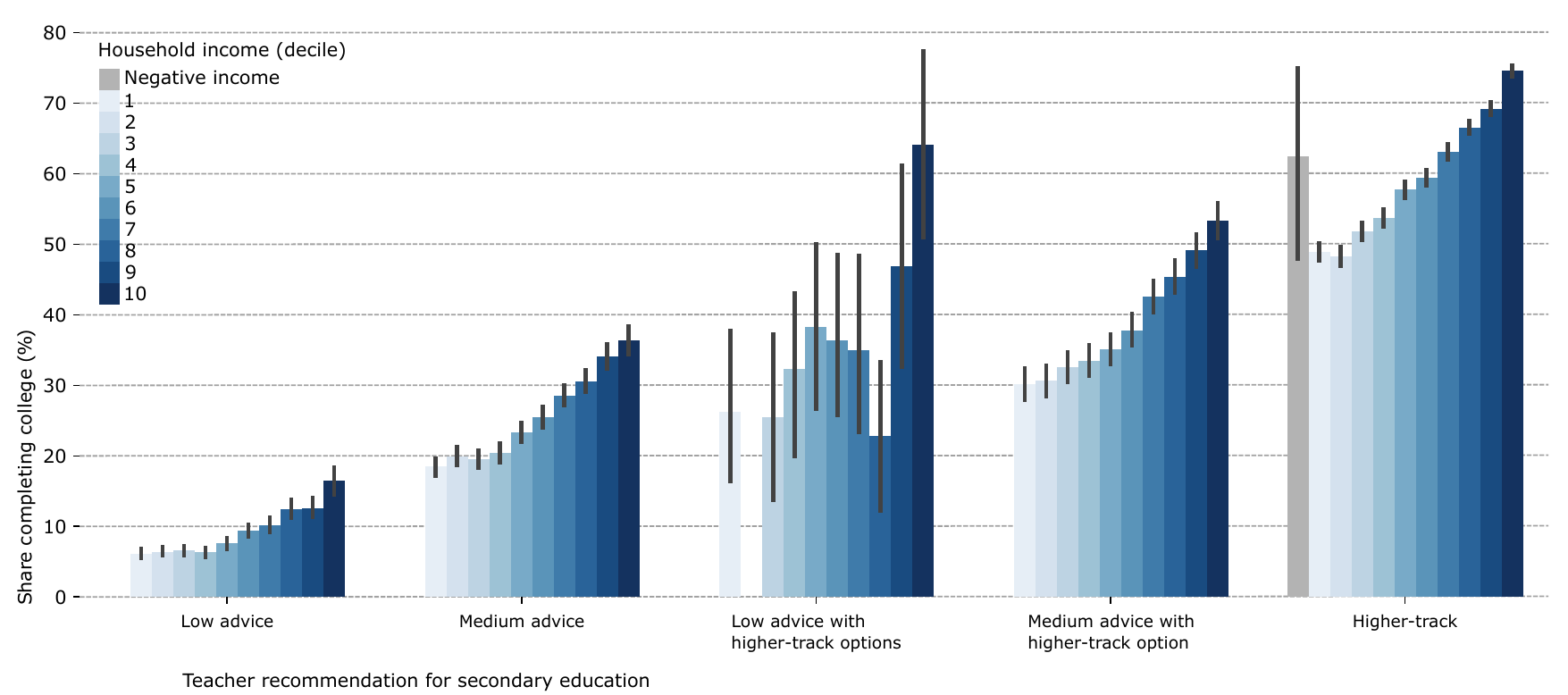}
    \caption{Share of students completing university by teacher recommendation and household income decile. For the description of the teacher recommendations, see Figure \ref{fig:context_inequality}.}
    \label{fig:context_inequality_income}
\end{figure}

\FloatBarrier

\section{Variable selection}\label{a:s:var_selection}

We include a broad set of variables drawn from six key social domains: individual, nuclear family, household, extended family, school, and neighborhood. Our primary goal is to evaluate how well different modeling approaches can predict long-term educational outcomes from early-life social context---rather than estimate effect sizes. We therefore include all available variables that may plausibly reflect structural inequalities, contextual constraints, or resource advantages relevant to educational attainment. These include individual characteristics (e.g., gender, disability), family background (e.g., parental education and income), intergenerational support, peer effects, and neighborhood opportunity structures. Because we use linked administrative records, the scope of features is determined by data availability in our CBS project (e.g., no information on parental occupation or household wealth).

\textit{Datasets used:} population register (\textit{GBAPERSOONTAB}, \citet{cbs_persoonskenmerken_2022}), highest educational attainment (\textit{HOOGSTEOPLTAB}, \citet{cbs_hoogst_2019}), income (\textit{INPATAB}, \citet{cbs_inkomen_2011}), school enrollment (\textit{INSCHRWPOTAB}, \citet{cbs_inschrijvingen_2014}), family networks (\textit{FAMILIENETWERKTAB}, \citet{cbs_familienetwerk_2009}), neighbors (\textit{BURENNETWERKTAB}, \citet{cbs_burennetwerk_2009}), households (\textit{HUISGENOTENNETWERKTAB}, \citet{cbs_huisgenotennetwerk_2009}), classmates (\textit{KLASGENOTENNETWERKTAB}, \citet{cbs_klasgenotennetwerk_2009}), and geographical information (\textit{VSLGTAB}, \citet{cbs_gemeente-_2017}).

\begin{itemize}
    \item \textbf{Individual variables}. Demographics (GBAPERSOONTAB): birth year (\texttt{GEBOORTEJAAR}), migration generation (\texttt{GENERATIE}, first, second, or third+), gender (\texttt{GESLACHT}). Primary school registrations (INSCHRWPOTAB): presence of a disability (\texttt{WPOSOORTIND}), and enrollment in special education (\texttt{WPOTYPEPO}==SBO).    
    
    \item \textbf{Nuclear family}. Parental income and education levels are well-established predictors of children's outcomes.  Previous research has highlighted the importance of measuring different dimensions of socioeconomic background (e.g. education and income), as well as including separate measures for fathers and mothers, for capturing different mechanisms  of intergenerational transmission of (dis)advantage \citep{thaning_end_2020}. Income (INPATAB): We include both parents’ income (\texttt{INPBELI}) separately. Education (HOOGSTEOPLTAB): we included indicators for whether either parent completed university  and binary indicators for missing parental education (\texttt{OPLNIVSOI2021AGG4HBmetNIRWO}). Educational weight (INSCHRWPOTAB): the ``educational weight'' (\texttt{WPOGEWICHT})---used for school funding decisions---summarizes parental educational disadvantage.  Family structure---for example,  growing up in single versus two-parent families---has also been shown to shape several outcomes of children.  Parental presence (GBAPERSOONTAB): We  include indicators for whether each parent is known and alive in 2010 in the dataset. Family structure (FAMILIENETWERKTAB): We also include the number of siblings (relations 306, 307 and 319 in the dataset), to capture potential resource dilution.
    
    \item \textbf{Household variables} (HUISGENOTENNETWERKTAB, relation 401): Household income (\texttt{INPBELI}) and the number of income earners reflect broader financial resources beyond individual parental income, which may influence educational support and expectations.
    
    \item \textbf{Extended family} (FAMILIENETWERKTAB, relation 303): Research suggests intergenerational transmission of advantage often extends beyond the nuclear family. We include the number of known grandparents, share that live nearby (same municipality, variable \texttt{gem2010} in dataset VSLGTAB), and minimum and maximum grandparent income (\texttt{INPBELI}, which includes pensions) to proxy the strength and inequality of extended family support.
    
    \item \textbf{Schools }(KLASGENOTENNETWERKTAB): Peer effects and school-level composition matter greatly for educational expectations and opportunities. We include class size (\texttt{WPOGROEPSGROOTTE}) and the mean and standard deviation of school-level parental income (\texttt{INPBELI}), educational attainment (\texttt{OPLNIVSOI2021AGG4HBmetNIRWO}) and educational weights (\texttt{WPOGEWICHT}). We also include the share of parents reporting income (\texttt{INPPINK}). At the school level, we include average school' class size (\texttt{WPOGROEPSGROOTTE}) and school urbanicity (average municipality size of enrolled students, calculated from the variable \texttt{gem2010}). 
    
    \item \textbf{Neighborhoods} (BURENNETWERKTAB, relation 101): Local socioeconomic context shapes access to resources, exposure to norms, and institutional interactions. We include the mean and standard deviation of neighborhood educational attainment (\texttt{OPLNIVSOI2021AGG4HBmetNIRWO}) and income rank (\texttt{INPBELI}) to capture both average context and heterogeneity.
\end{itemize}

\section{Correlation between variables} \label{a:s:correlations}

\begin{figure}[ht!]
    \centering
    \includegraphics[width=\linewidth]{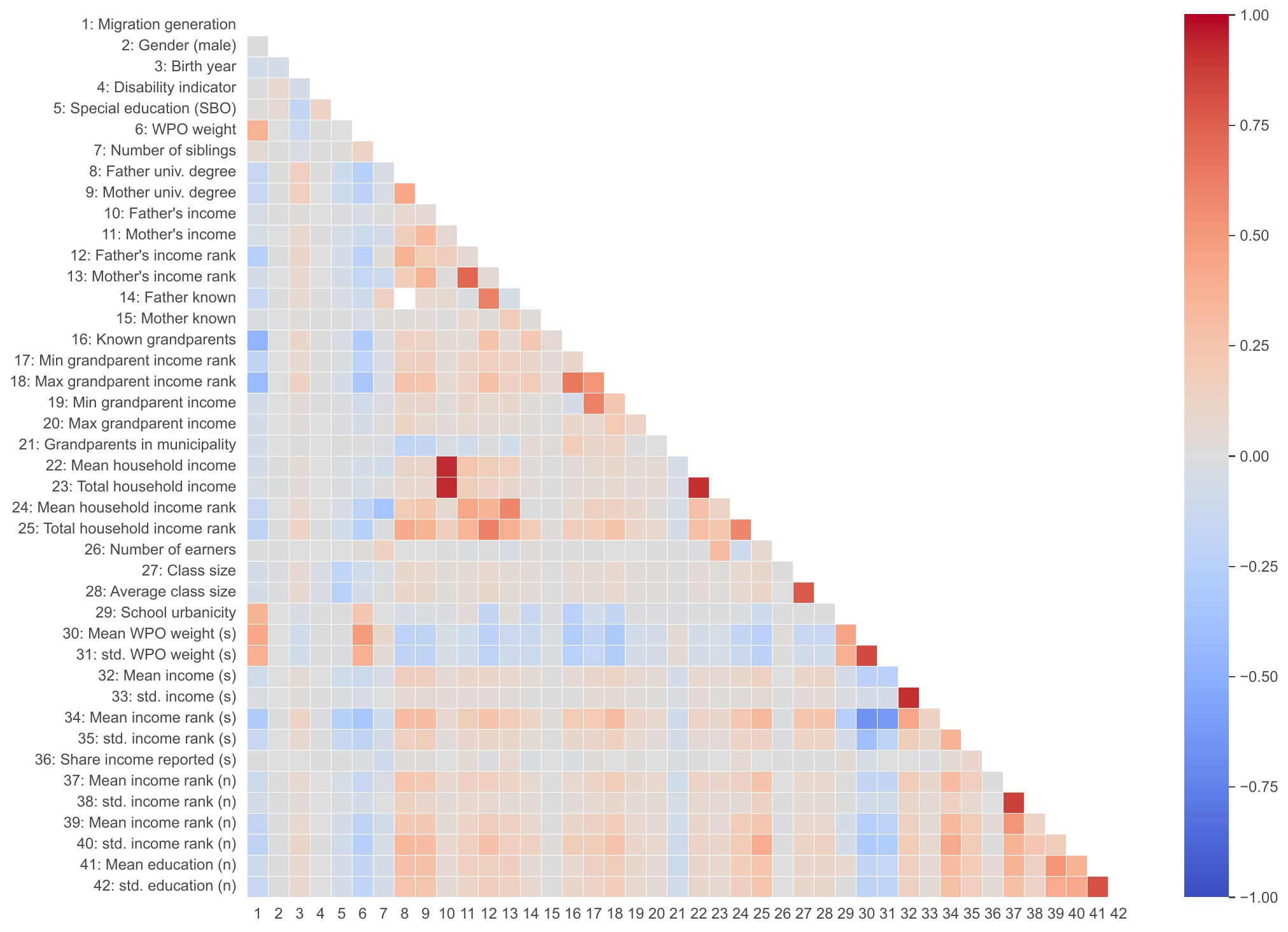}
    \caption{\textbf{Correlation matrix of selected contextual variables.} The color scale indicates the direction and strength of associations (red = positive, blue = negative). Variables referring to schools and neighborhoods are denoted by (s) and (n). Correlations are shown in Table~\ref{tab:desc_corr}.}
    \label{a:fig:heatmap}
\end{figure}

\FloatBarrier

\begin{table}[htbp]
\centering
\renewcommand{\arraystretch}{1.2}
\resizebox{\textwidth}{!}{%
\begin{tabular}{lrrrrrrrrrrrrrrrrrrrrrrrrrrrrrrrrrrrrrrrrrr}
\toprule
 & 1 & 2 & 3 & 4 & 5 & 6 & 7 & 8 & 9 & 10 & 11 & 12 & 13 & 14 & 15 & 16 & 17 & 18 & 19 & 20 & 21 & 22 & 23 & 24 & 25 & 26 & 27 & 28 & 29 & 30 & 31 & 32 & 33 & 34 & 35 & 36 & 37 & 38 & 39 & 40 & 41 & 42 \\
\midrule
1 &  &  &  &  &  &  &  &  &  &  &  &  &  &  &  &  &  &  &  &  &  &  &  &  &  &  &  &  &  &  &  &  &  &  &  &  &  &  &  &  &  &  \\
2 & .00 &  &  &  &  &  &  &  &  &  &  &  &  &  &  &  &  &  &  &  &  &  &  &  &  &  &  &  &  &  &  &  &  &  &  &  &  &  &  &  &  &  \\
3 & -.08 & -.06 &  &  &  &  &  &  &  &  &  &  &  &  &  &  &  &  &  &  &  &  &  &  &  &  &  &  &  &  &  &  &  &  &  &  &  &  &  &  &  &  \\
4 & -.02 & .08 & -.06 &  &  &  &  &  &  &  &  &  &  &  &  &  &  &  &  &  &  &  &  &  &  &  &  &  &  &  &  &  &  &  &  &  &  &  &  &  &  &  \\
5 & .02 & .06 & -.19 & .12 &  &  &  &  &  &  &  &  &  &  &  &  &  &  &  &  &  &  &  &  &  &  &  &  &  &  &  &  &  &  &  &  &  &  &  &  &  &  \\
6 & .37 & -.01 & -.13 & -.01 & -.00 &  &  &  &  &  &  &  &  &  &  &  &  &  &  &  &  &  &  &  &  &  &  &  &  &  &  &  &  &  &  &  &  &  &  &  &  &  \\
7 & .05 & .00 & -.03 & -.01 & .02 & .13 &  &  &  &  &  &  &  &  &  &  &  &  &  &  &  &  &  &  &  &  &  &  &  &  &  &  &  &  &  &  &  &  &  &  &  &  \\
8 & -.16 & .00 & .16 & -.01 & -.11 & -.25 & -.04 &  &  &  &  &  &  &  &  &  &  &  &  &  &  &  &  &  &  &  &  &  &  &  &  &  &  &  &  &  &  &  &  &  &  &  \\
9 & -.15 & .01 & .16 & -.00 & -.11 & -.23 & -.04 & .44 &  &  &  &  &  &  &  &  &  &  &  &  &  &  &  &  &  &  &  &  &  &  &  &  &  &  &  &  &  &  &  &  &  &  \\
10 & -.04 & .00 & .03 & -.00 & -.02 & -.05 & .01 & .08 & .06 &  &  &  &  &  &  &  &  &  &  &  &  &  &  &  &  &  &  &  &  &  &  &  &  &  &  &  &  &  &  &  &  &  \\
11 & -.05 & -.00 & .07 & -.01 & -.05 & -.12 & -.07 & .18 & .31 & .06 &  &  &  &  &  &  &  &  &  &  &  &  &  &  &  &  &  &  &  &  &  &  &  &  &  &  &  &  &  &  &  &  \\
12 & -.25 & .00 & .11 & -.00 & -.08 & -.23 & .02 & .36 & .20 & .17 & .06 &  &  &  &  &  &  &  &  &  &  &  &  &  &  &  &  &  &  &  &  &  &  &  &  &  &  &  &  &  &  &  \\
13 & -.08 & -.00 & .08 & -.01 & -.06 & -.17 & -.12 & .18 & .37 & .02 & .72 & .05 &  &  &  &  &  &  &  &  &  &  &  &  &  &  &  &  &  &  &  &  &  &  &  &  &  &  &  &  &  &  \\
14 & -.15 & .01 & .07 & .00 & -.04 & -.11 & .15 &  & .08 & .06 & -.03 & .62 & -.05 &  &  &  &  &  &  &  &  &  &  &  &  &  &  &  &  &  &  &  &  &  &  &  &  &  &  &  &  &  \\
15 & -.03 & -.00 & .03 & .00 & -.02 & -.03 & .02 & .02 & .03 & .00 & .08 & .02 & .19 & .03 &  &  &  &  &  &  &  &  &  &  &  &  &  &  &  &  &  &  &  &  &  &  &  &  &  &  &  &  \\
16 & -.47 & -.00 & .11 & .01 & -.04 & -.29 & -.02 & .15 & .13 & .04 & .03 & .27 & .06 & .23 & .06 &  &  &  &  &  &  &  &  &  &  &  &  &  &  &  &  &  &  &  &  &  &  &  &  &  &  &  \\
17 & -.22 & -.00 & .07 & .01 & -.03 & -.22 & -.04 & .16 & .17 & .04 & .12 & .16 & .14 & .10 & .03 & .11 &  &  &  &  &  &  &  &  &  &  &  &  &  &  &  &  &  &  &  &  &  &  &  &  &  &  \\
18 & -.42 & -.01 & .13 & .00 & -.06 & -.34 & -.03 & .27 & .25 & .07 & .12 & .28 & .15 & .19 & .06 & .65 & .52 &  &  &  &  &  &  &  &  &  &  &  &  &  &  &  &  &  &  &  &  &  &  &  &  &  \\
19 & -.07 & -.00 & .03 & .00 & -.02 & -.09 & -.02 & .09 & .11 & .02 & .09 & .07 & .09 & .02 & .01 & -.07 & .62 & .24 &  &  &  &  &  &  &  &  &  &  &  &  &  &  &  &  &  &  &  &  &  &  &  &  \\
20 & -.07 & -.00 & .03 & -.00 & -.02 & -.06 & .00 & .13 & .07 & .04 & .07 & .05 & .04 & .03 & .01 & .11 & .09 & .21 & .12 &  &  &  &  &  &  &  &  &  &  &  &  &  &  &  &  &  &  &  &  &  &  &  \\
21 & -.06 & -.00 & -.01 & -.00 & .01 & .01 & -.03 & -.20 & -.18 & -.03 & -.09 & .01 & -.08 & .06 & .02 & .18 & .11 & .13 & .01 & -.00 &  &  &  &  &  &  &  &  &  &  &  &  &  &  &  &  &  &  &  &  &  &  \\
22 & -.06 & .00 & .04 & -.01 & -.03 & -.08 & -.08 & .11 & .12 & .92 & .24 & .17 & .16 & .02 & .00 & .04 & .07 & .09 & .05 & .05 & -.05 &  &  &  &  &  &  &  &  &  &  &  &  &  &  &  &  &  &  &  &  &  \\
23 & -.04 & .00 & .03 & -.01 & -.02 & -.05 & .01 & .08 & .08 & .93 & .17 & .14 & .09 & .03 & .01 & .03 & .05 & .07 & .03 & .04 & -.04 & .91 &  &  &  &  &  &  &  &  &  &  &  &  &  &  &  &  &  &  &  &  \\
24 & -.15 & -.00 & .07 & -.00 & -.05 & -.20 & -.36 & .20 & .25 & .07 & .42 & .35 & .58 & .02 & -.01 & .06 & .15 & .14 & .09 & .03 & -.05 & .28 & .12 &  &  &  &  &  &  &  &  &  &  &  &  &  &  &  &  &  &  &  \\
25 & -.22 & .00 & .13 & -.01 & -.10 & -.25 & -.02 & .42 & .37 & .17 & .35 & .61 & .39 & .18 & .03 & .17 & .19 & .26 & .10 & .07 & -.08 & .29 & .24 & .58 &  &  &  &  &  &  &  &  &  &  &  &  &  &  &  &  &  &  \\
26 & .01 & .00 & -.00 & -.00 & -.00 & .03 & .14 & -.00 & -.00 & .00 & -.02 & .01 & -.04 & .02 & .01 & .00 & -.01 & -.01 & -.01 & .00 & .00 & -.02 & .30 & -.11 & .07 &  &  &  &  &  &  &  &  &  &  &  &  &  &  &  &  &  \\
27 & -.06 & -.02 & .06 & -.04 & -.20 & -.08 & -.03 & .09 & .08 & .02 & .05 & .08 & .05 & .04 & .01 & .05 & .05 & .09 & .03 & .02 & -.01 & .04 & .02 & .07 & .09 & -.01 &  &  &  &  &  &  &  &  &  &  &  &  &  &  &  &  \\
28 & -.06 & -.01 & .07 & -.05 & -.25 & -.08 & -.02 & .10 & .09 & .02 & .05 & .09 & .05 & .04 & .01 & .05 & .05 & .09 & .03 & .02 & -.01 & .04 & .03 & .06 & .10 & -.00 & .78 &  &  &  &  &  &  &  &  &  &  &  &  &  &  &  \\
29 & .35 & -.00 & -.04 & -.02 & .01 & .24 & -.01 & -.04 & -.03 & -.02 & .03 & -.18 & .02 & -.14 & -.02 & -.24 & -.08 & -.19 & -.01 & -.02 & -.02 & -.01 & -.01 & -.03 & -.11 & .01 & -.01 & .01 &  &  &  &  &  &  &  &  &  &  &  &  &  &  \\
30 & .42 & -.01 & -.09 & -.01 & -.01 & .49 & .09 & -.22 & -.21 & -.05 & -.10 & -.23 & -.13 & -.13 & -.03 & -.28 & -.19 & -.31 & -.08 & -.06 & .05 & -.08 & -.05 & -.18 & -.24 & .02 & -.15 & -.14 & .44 &  &  &  &  &  &  &  &  &  &  &  &  &  \\
31 & .38 & -.00 & -.08 & -.01 & -.02 & .39 & .04 & -.21 & -.21 & -.05 & -.09 & -.22 & -.10 & -.13 & -.02 & -.25 & -.17 & -.28 & -.07 & -.06 & .05 & -.07 & -.05 & -.14 & -.23 & .01 & -.09 & -.09 & .38 & .82 &  &  &  &  &  &  &  &  &  &  &  &  \\
32 & -.10 & -.00 & .05 & -.01 & -.10 & -.12 & -.04 & .17 & .16 & .06 & .10 & .10 & .09 & .04 & .01 & .06 & .10 & .13 & .06 & .05 & -.07 & .08 & .06 & .09 & .15 & -.01 & .10 & .11 & -.07 & -.23 & -.24 &  &  &  &  &  &  &  &  &  &  &  \\
33 & -.03 & -.00 & .02 & .00 & -.03 & -.04 & -.01 & .06 & .06 & .03 & .04 & .04 & .03 & .01 & .00 & .01 & .04 & .05 & .03 & .02 & -.03 & .04 & .03 & .02 & .05 & .00 & .04 & .04 & -.02 & -.07 & -.07 & .92 &  &  &  &  &  &  &  &  &  &  \\
34 & -.29 & -.01 & .13 & -.02 & -.25 & -.35 & -.11 & .31 & .30 & .08 & .18 & .26 & .19 & .13 & .03 & .20 & .20 & .30 & .11 & .08 & -.11 & .13 & .09 & .23 & .32 & -.03 & .25 & .26 & -.26 & -.67 & -.63 & .44 & .14 &  &  &  &  &  &  &  &  &  \\
35 & -.14 & -.00 & .07 & -.01 & -.14 & -.21 & -.08 & .16 & .17 & .03 & .10 & .10 & .11 & .05 & .01 & .09 & .11 & .15 & .06 & .04 & -.06 & .06 & .04 & .12 & .14 & -.02 & .12 & .13 & -.06 & -.38 & -.21 & .17 & .09 & .36 &  &  &  &  &  &  &  &  \\
36 & .01 & -.00 & -.00 & .00 & .01 & -.03 & -.10 & .02 & .03 & -.01 & .04 & -.02 & .07 & -.03 & -.00 & -.02 & .02 & -.01 & .02 & -.00 & -.03 & .01 & -.00 & .06 & -.00 & -.01 & -.01 & -.01 & .05 & -.05 & -.01 & -.01 & -.02 & .04 & .10 &  &  &  &  &  &  &  \\
37 & -.12 & .00 & .07 & -.01 & -.05 & -.15 & -.04 & .25 & .21 & .08 & .15 & .16 & .12 & .06 & .01 & .07 & .13 & .17 & .09 & .07 & -.09 & .13 & .10 & .14 & .26 & -.00 & .08 & .09 & -.02 & -.19 & -.18 & .19 & .09 & .30 & .16 & .01 &  &  &  &  &  &  \\
38 & -.06 & .00 & .03 & -.01 & -.03 & -.07 & -.00 & .15 & .11 & .05 & .08 & .08 & .05 & .03 & .00 & .04 & .07 & .09 & .05 & .04 & -.05 & .08 & .06 & .05 & .14 & .01 & .04 & .04 & -.01 & -.09 & -.09 & .11 & .06 & .15 & .08 & -.01 & .86 &  &  &  &  &  \\
39 & -.18 & .00 & .07 & -.00 & -.05 & -.22 & -.09 & .21 & .21 & .05 & .13 & .17 & .14 & .07 & .01 & .11 & .14 & .19 & .08 & .05 & -.08 & .09 & .06 & .19 & .24 & -.02 & .08 & .07 & -.08 & -.26 & -.21 & .14 & .05 & .32 & .19 & .05 & .52 & .13 &  &  &  &  \\
40 & -.22 & .01 & .11 & -.01 & -.09 & -.26 & -.05 & .33 & .30 & .09 & .16 & .28 & .16 & .14 & .02 & .18 & .17 & .27 & .10 & .08 & -.11 & .13 & .10 & .20 & .40 & .00 & .12 & .14 & -.08 & -.30 & -.28 & .20 & .08 & .42 & .19 & .00 & .37 & .24 & .17 &  &  &  \\
41 & -.11 & .00 & .08 & -.01 & -.07 & -.18 & -.07 & .28 & .29 & .06 & .16 & .15 & .17 & .06 & .01 & .07 & .14 & .18 & .09 & .06 & -.12 & .10 & .07 & .18 & .25 & -.00 & .10 & .10 & .08 & -.18 & -.17 & .16 & .06 & .31 & .18 & .06 & .37 & .14 & .52 & .36 &  &  \\
42 & -.14 & .00 & .08 & -.00 & -.07 & -.20 & -.07 & .26 & .25 & .05 & .13 & .17 & .14 & .08 & .01 & .11 & .13 & .18 & .07 & .05 & -.09 & .09 & .07 & .17 & .26 & -.00 & .09 & .09 & -.01 & -.21 & -.20 & .13 & .05 & .30 & .16 & .04 & .29 & .12 & .40 & .41 & .80 &  \\
\bottomrule
\end{tabular}
}
\caption{Correlation statistics for included variables}
\label{tab:desc_corr}
\end{table}

\FloatBarrier

\section{Technical explanation of Graph Neural Networks (GNNs)} \label{a:s:gnn}
Graph Neural Networks (GNNs) offer a fundamentally different approach compared to traditional models, both in how they define the function $f(\cdot)$ and how they structure the input space $X$. Rather than treating individuals as isolated units with fixed features, GNNs represent individuals as nodes within a social graph. The GNN input consists of two node types: \texttt{child} (students) and \texttt{person} (e.g., parents, neighbors, peers), each with its own set of features. Each node is connected to others via edges that represent social contexts between the type of nodes. We model five different relationships between the two types of nodes (Fig.~\ref{fig:conceptual}A): (1) shared classrooms (child–child), (2) parent links (child–person), (3) child links (person–child), (4) family links connecting co-parents or siblings (person–person), and (5) neighbor links (person–person).

In this framework, the input $X$ includes not only an individual’s own characteristics, but also the features of their social connections. The function $f(\cdot)$ is redefined as a recursive aggregation process over the network. We use the \textit{GraphSAGE} architecture \citep{hamilton_inductive_2017} without sampling, from the \texttt{pytorch-geometric} Python package. 

Formally, for each of the five relationships, the process unfolds in recursive steps over $K=2$ convolutional layers. For each node $v$ and each of the two layers, $k$, we compute an embedding $h_v^{(k)}$. We set an initial embedding $h_v^{(0)}$ for each node $v$ to the individual-level variables such as age, gender, or migration background (Section~\ref{s:gnn_variables}).

For a layer $k$, first, each node $v$ aggregates information from its neighbors in a given social relationship $s$ (e.g., schoolmates, neighbors) in a hidden vector $h_v^{(k, s)}$. Let $\mathcal{N}^s(v)$ denote the set of neighbors of node $v$ in relationship $s$. Then set%
\begin{equation}
h_v^{(k, s)} = \text{AGGREGATE}^{(k)} \left( \left\{ h_u^{(k-1)} : u \in \mathcal{N}^s(v) \right\} \right)
\end{equation}
Here, $\text{AGGREGATE}^{(k)}$ is a function that aggregates embeddings from neighbors in relationship $s$. We use the same aggregation function for each GNN layer, which is either the \texttt{mean} or the concatenated \texttt{mean} and \texttt{max}, depending on the relationship (see Appendix Section~\ref{a:s:hyper_tuning}).

Second, the node updates its own embedding by combining the aggregated information received from its neighbors with its prior embedding. This is done separately for each social relationship $s$, and then merged:%
\begin{equation}
h_v^{(k)} = \text{UPDATE}^{(k)} \left( h_v^{(k-1)}, \bigoplus_s h_v^{(k, s)} \right)
\end{equation}
Here, $\bigoplus_s$ is a cross-type aggregation function across social relationships, and $\text{UPDATE}^{(k)}$ is a function that combines the resulting vector with the embedding vector of the previous layer. We use the same standard update function for each GNN layer: a fully connected layer, followed by a ReLU layer providing nonlinearity. The cross-type aggregation function is either \texttt{mean} or \texttt{cat} (concatenation), depending on the social contexts used (see Appendix Section~\ref{a:s:hyper_tuning}).

After two rounds of message passing (aggregation and update), the final node embeddings $h_v^{(K)}$ are used as input to a logistic regression that predicts whether the student completes university:%
\begin{equation}
\log\left( \frac{P(y = 1)}{1 - P(y = 1)} \right) = \beta h^{(K)}.
\end{equation}

\section{Hyperparameter tuning}\label{a:s:hyper_tuning}

\textit{Logistic Regression:} LR has no hyperparameters. In the \texttt{scikit-learn} library, we run standard logistic regression by setting the penalty to \textit{None}.

\textit{Gradient Boosting Trees:}
In contrast, gradient boosting models require tuning several important hyperparameters. We performed systematic optimization using the Optuna package, a Bayesian optimization framework for efficient hyperparameter tuning \citep{akiba_optuna_2019}. We considered the following hyperparameters:

\begin{itemize}
\item \textbf{Learning Rate (0.01–1)}: Determines how aggressively the model updates predictions with each tree.
\item \textbf{Maximum Iterations (50–200)}: The total number of trees built sequentially to form the final model.
\item \textbf{Maximum Leaf Nodes (2–50)}: Controls the complexity of each decision tree.
\item \textbf{Minimum Samples per Leaf (1–100)}: The minimum number of data points required in each leaf node.
\item \textbf{Maximum Tree Depth (1–20)}: The deepest level a tree can grow, constraining complexity.
\end{itemize}

For each configuration suggested by Optuna, the model's performance was evaluated using cross-validated log-loss, guiding the optimization toward the best hyperparameter values. We conducted 100 trials to comprehensively explore this parameter space. We used the visualization tools provided by Optuna to assess optimization history and hyperparameter importance, ensuring that a minimum was reached.

The optimal hyperparameters identified for each prediction outcome were as follows:

\begin{itemize}
\item \textit{Demographics only}: {Learning Rate: 0.664, Maximum Iterations: 105, Maximum Leaf Nodes: 13, Minimum Samples per Leaf: 49, Maximum Tree Depth: 2}
\item \textit{+Nuclear family}: {Learning Rate: 0.059, Maximum Iterations: 167, Maximum Leaf Nodes: 13, Minimum Samples per Leaf: 48, Maximum Tree Depth: 11}
\item \textit{+Household}: {Learning Rate: 0.119, Maximum Iterations: 185, Maximum Leaf Nodes: 10, Minimum Samples per Leaf: 95, Maximum Tree Depth: 4}
\item \textit{+Extended family}: {Learning Rate: 0.111, Maximum Iterations: 187, Maximum Leaf Nodes: 49, Minimum Samples per Leaf: 82, Maximum Tree Depth: 3}
\item \textit{+School}: {Learning Rate: 0.073, Maximum Iterations: 197, Maximum Leaf Nodes: 15, Minimum Samples per Leaf: 45, Maximum Tree Depth: 4}
\item \textit{+Neighborhood}: {Learning Rate: 0.052, Maximum Iterations: 178, Maximum Leaf Nodes: 24, Minimum Samples per Leaf: 56, Maximum Tree Depth: 11}
\end{itemize}

Once the optimal hyperparameters were identified, we retrained the final gradient boosting model on the entire training set.

\textit{Graph Neural Networks (GNNs):} We optimized the hyperparameters of the Graph Neural Network (GNN) models using Optuna. We tuned the following model-specific configurations:

\begin{itemize}
    \item \textbf{Hidden dimension:} The size of the latent embeddings (ranging from 16 to 64). Larger dimensions allow for more expressive representations of social context.
    \item \textbf{Dropout rate:} The proportion of randomly deactivated units during training (between 0.1 and 0.3), helping prevent overfitting.
    \item \textbf{Learning rate:} The step size for parameter updates during training (between 0.01 and 0.1, tested on a logarithmic scale).
    \item \textbf{Edge-type aggregation:} The method for combining messages from the same type of social connection (e.g., parent-child). We tested ``mean,'' ``max,'' ``sum,'' and the concatenations [mean, max] and [mean, max, min].
    \item \textbf{Cross-type aggregation:} The method for aggregating messages from different types of social ties (e.g., parents and classmates). We compared ``mean'' and ``cat'' (concatenation).
    \item \textbf{Normalization:} Whether to apply normalization within GraphSAGE layers for selected edge types.
    \item \textbf{GNN architecture:} We compared \texttt{SAGEConv} (GraphSAGE) and \texttt{GATv2Conv} (Graph Attention Networks). While GATv2 allows different neighbors to exert different influence via attention weights, it is more computationally intensive and prone to overfitting. 
    \item \textbf{Convolutional layers:} We tested two versus three convolutional layers.
\end{itemize}

We first conducted a comprehensive hyperparameter search (40 trials per model) to determine the most promising configurations. GraphSAGE had similar or better performance than GATv2Conv at lower computational cost. Two convolutional layers consistently outperformed three, normalization within GraphSAGE layers improved performance, and aggregation methods such as ``sum'' and [mean, max, min] were consistently outperformed and subsequently excluded.

We then ran 40 optimization trials per model using Optuna and selected the configuration with the lowest validation loss. Final models were retrained on the full training set. Table~\ref{tab:gnn_hyperparameters} summarizes the optimal hyperparameters per GNN model.

\begin{table}[ht!]
\centering
\caption{Final GNN Hyperparameters by Social Context}
\label{tab:gnn_hyperparameters}
\begin{tabular}{@{}p{2cm}p{1.2cm}p{1.2cm}p{1.3cm}p{1.3cm}p{8cm}@{}}
\toprule
\textbf{Model} & \textbf{Cross-type Agg.} & \textbf{Hidden Dim.} & \textbf{Dropout} & \textbf{Learning Rate} & \textbf{Edge-type Aggregation} \\
\midrule
Nuclear family only 
& cat 
& 64 
& 0.2 
& 0.01 
& Concatenated mean and max for all relations \\

Including school 
& mean 
& 32 
& 0.2 
& 0.01 
& Mean for schoolmates and family-type edges; concatenated mean and max for child-parent edges \\

Full context (family, school, neighbors) 
& cat 
& 16 
& 0.1 
& 0.02 
& Mean for schoolmates; concatenated mean and max for child-parent, neighbor, and extended family edges \\
\bottomrule
\end{tabular}
\end{table}

\FloatBarrier

\section{Confusion tables} \label{a:s:conf_matrix}

The confusion tables (Table~\ref{tab:conf_matrix}) compare model predictions to assess how similarly they classify educational outcomes. Panel (A) compares gradient boosting to logistic regression, while panel (B) compares graph neural networks (GNN) to gradient boosting. In both panels, the majority of predictions are in the ``Both correct'' or ``Both wrong'' categories, showing that both models predict the same result (completing or not completing university) for the same observations. The small number of cases where one model is correct and the other is not (e.g., only Boost correct: 1,320; only GNN correct: 1,492) highlights the limited but meaningful instances in which more complex models (Boost or GNN) capture patterns missed by simpler models. These cases reflect situations where interactions or network structure are especially important for prediction.

\begin{table}[htbp]
\centering
 (A)
 \begin{tabular}{rrrrr}
\toprule
Both agreeing & Both correct & Both wrong & Only Boost correct & Only Linear correct \\
\midrule
35133 & 24919 & 10214 & 1320 & 1150 \\
\bottomrule
\end{tabular}

 (B)
\begin{tabular}{rrrrr}
\toprule
Both agreeing & Both correct & Both wrong & Only GNN correct & Only Boost correct \\
\midrule
34696 & 24824 & 9872 & 1492 & 1415 \\
\bottomrule
\end{tabular}

\caption{All statistical models predict similarly. (A) Predictions of gradient boosting (Boost) versus logistic regression (Linear) (B) Predictions of graph neural networks (GNN) versus gradient boosting (Boost) }
\label{tab:conf_matrix}
\end{table}

\begin{figure}[ht!]
    \centering
    \includegraphics[width=0.85\linewidth]{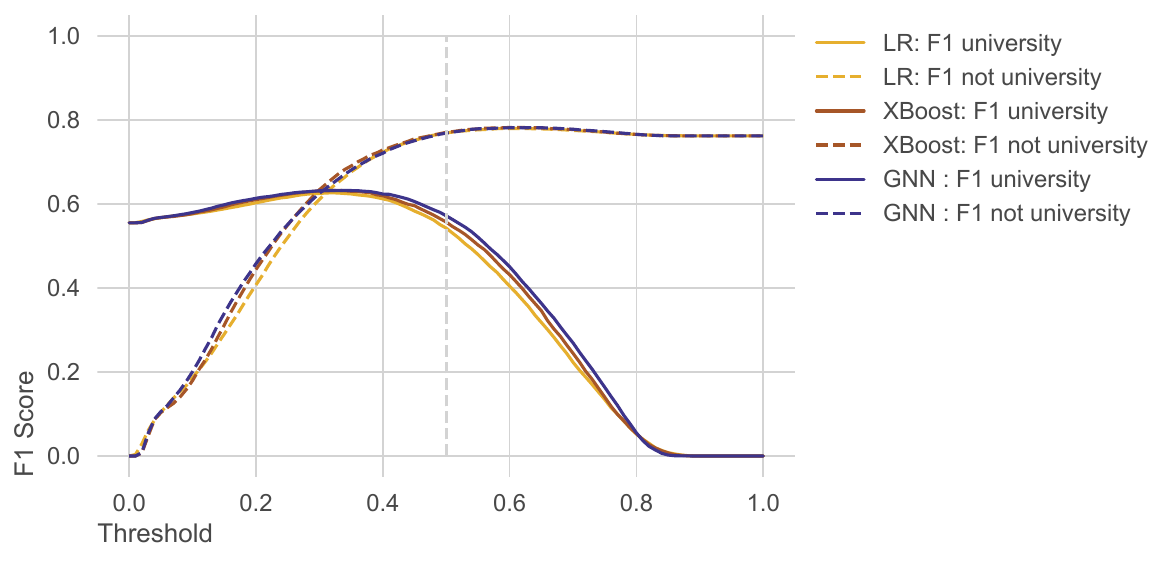}
    \caption{\textbf{F1 Score across classification thresholds.} Comparison of F1 scores for predicting university vs. non-university completion, across a range of probability thresholds. Gradient boosting (brown) and GNN (blue) consistently outperform logistic regression (gold), especially for the positive class (university completion). Dashed lines refer to F1 for non-completion; solid lines for completion.}
    \label{fig:roc_thresholds}
\end{figure}

However, these modest classification gains may understate the improvement in model quality. Complex models like Boost and GNNs generate probability scores that more accurately reflect underlying uncertainty. Figure~\ref{fig:roc_thresholds} shows that GNNs and boosting outperform logistic regression by achieving higher peak F1 scores and maintaining more stable performance across a broad range of thresholds—particularly for the minority class (university completion).

\FloatBarrier

\section{Correlation between the GNN embeddings and aggregated variables } \label{a:s:embeddings}

We conducted Principal Component Analysis (PCA) on the final-layer GNN embeddings of all students in the dataset. These embeddings represent each student's position in a high-dimensional feature space shaped by both individual attributes and social connections. The PCA was performed on the full embedding matrix (one row per student), after centering and scaling. We then computed Pearson correlations between the first three principal components (which jointly explain 89.1\% of the variance) and key input features and covariates. 

Table~\ref{tab:embedding_correlations} reports the correlations between the first three principal components of the GNN embeddings and the input features in the linear regression models. The outcome variable and the correlation with specific migration backgrounds are also shown, even when those variables were not included in the model. Each column corresponds to one PCA dimension, ordered by the proportion of variance it explains in the embedding space. To aid interpretation, we also show the correlation with the PCA dimension of the GNNs using only a subset of input layers. Specifically, \textit{PCA1 nuclear} refers to embeddings generated using only nuclear family features (e.g., parental education and income), and \textit{PCA1 school} refers to the first principal component of embeddings generated including school-related features.

\begin{table}[htbp]
  \centering
  \scriptsize
  \begin{minipage}[t]{0.31\textwidth}
    \raggedright
    \textbf{(A) First PCA dimension (variance explained 67.6\%)}\\[0.5em]
\begin{tabular}{p{3.5cm}p{1cm}}
\toprule
Variable & Corr. \\
\midrule
PCA1 school & -0.85 \\
PCA1 nuclear & 0.82 \\
Household income rank & -0.53 \\
Father univ. degree & -0.52 \\
Mother univ. degree & -0.51 \\
Mean education (n) & -0.46 \\
std. income rank (n) & -0.43 \\
std. education (n) & -0.43 \\
Birth year & -0.42 \\
Mean income rank (s) & -0.41 \\
Outcome & -0.38 \\
Father's income rank & -0.36 \\
Mother's income rank & -0.36 \\
Mean income rank (n) & -0.34 \\
Max grandparent income rank & -0.31 \\
WPO weight & 0.31 \\
\bottomrule
\end{tabular}

  \end{minipage}%
  \hfill
  \begin{minipage}[t]{0.31\textwidth}
    \raggedright
    \textbf{(B) Second PCA dimension (variance explained 15\%)}\\[0.5em]
\begin{tabular}{p{3.5cm}p{1cm}}
\toprule
Variable & Corr. \\
\midrule
PCA2 school & 0.77 \\
migration\_Dutch & -0.63 \\
Migration generation & 0.62 \\
Mean WPO weight in school & 0.46 \\
Log. school urbanicity & 0.46 \\
WPO weight & 0.39 \\
Known grandparents & -0.39 \\
std. WPO weight in school & 0.38 \\
PCA4 nuclear & 0.37 \\
Mean income rank (s) & -0.36 \\
Father's income rank & -0.36 \\
migration\_Morocco & 0.33 \\
Mean household income rank & -0.32 \\
migration\_Turkey & 0.31 \\
\bottomrule
\end{tabular}
  \end{minipage}%
  \hfill
  \begin{minipage}[t]{0.31\textwidth}
    \raggedright
    \textbf{(C) Third PCA dimension (variance explained 6.0\%)}\\[0.5em]
\begin{tabular}{p{3.5cm}p{1cm}}
\toprule
Variable & Corr. \\
\midrule
PCA2 nuclear & 0.68 \\
Special education (SBO) & 0.4 \\
PCA4 school & 0.35 \\
Gender (male) & -0.3 \\
\bottomrule
\end{tabular}

  \end{minipage}%

  \caption{Correlation between the first three PCA components of the embeddings and the covariates.}
  \label{tab:embedding_correlations}
\end{table}

\FloatBarrier

\section{Using AUC as evaluation metric} \label{a:s:auc}
In this section, we report the equivalent of Figure~\ref{fig:pred_gap}, using AUC-ROC instead of MCC as the evaluation metric. However, we do not provide equivalents of the subgroup analyses (Figures~\ref{fig:disaggregation}, \ref{fig:disaggregation_delta_a} and \ref{fig:disaggregation_delta_b}), since comparing predictability across subgroups is misleading: the proportion of individuals completing a degree varies substantially between subpopulations, which distorts comparisons of model performance. 

\begin{figure}[ht!]
    \centering
    \includegraphics[width=0.5\linewidth]{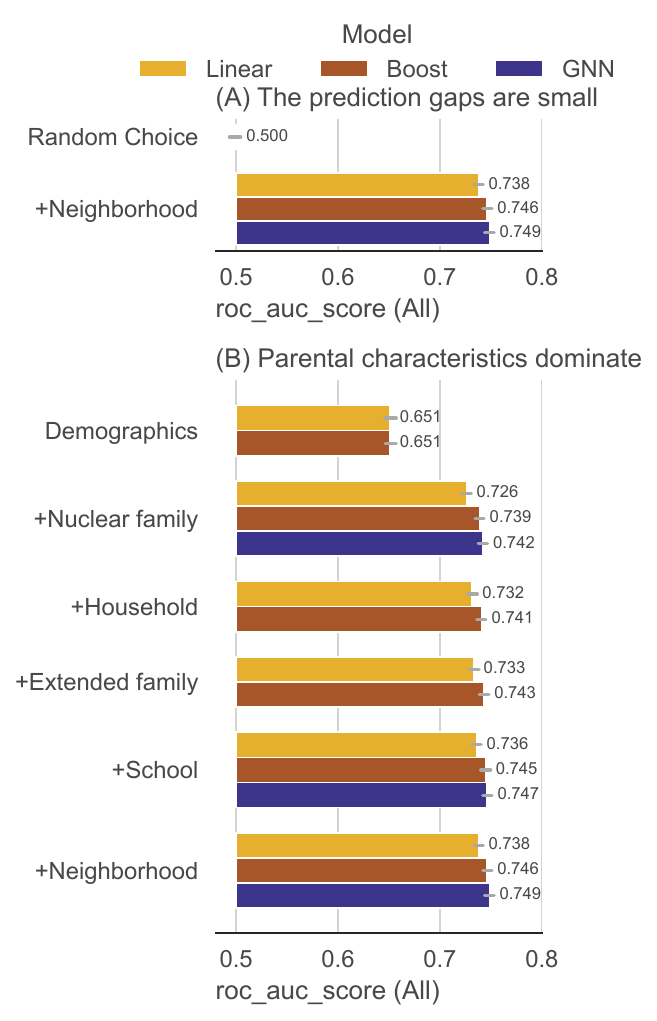}
    \caption{\textbf{Prediction gaps in university completion using AUC}. Out‐of‐sample predictive performance (AUC score) of three models of increasing flexibility: logistic regression (Linear), gradient boosting (Boost), and graph neural networks (GNN), alongside random guessing for (A) the complete model with all social contexts; (B) models in which social contexts are added sequentially. Note that the confidence intervals (95\% bootstrap confidence intervals) are not independent since they are based on the same data.}
    \label{fig:enter-label}
\end{figure}

\FloatBarrier

\section{Using log loss as evaluation metric}\label{a:s:log_loss}

\begin{figure}[ht!]
    \centering
    \includegraphics[width=0.5\linewidth]{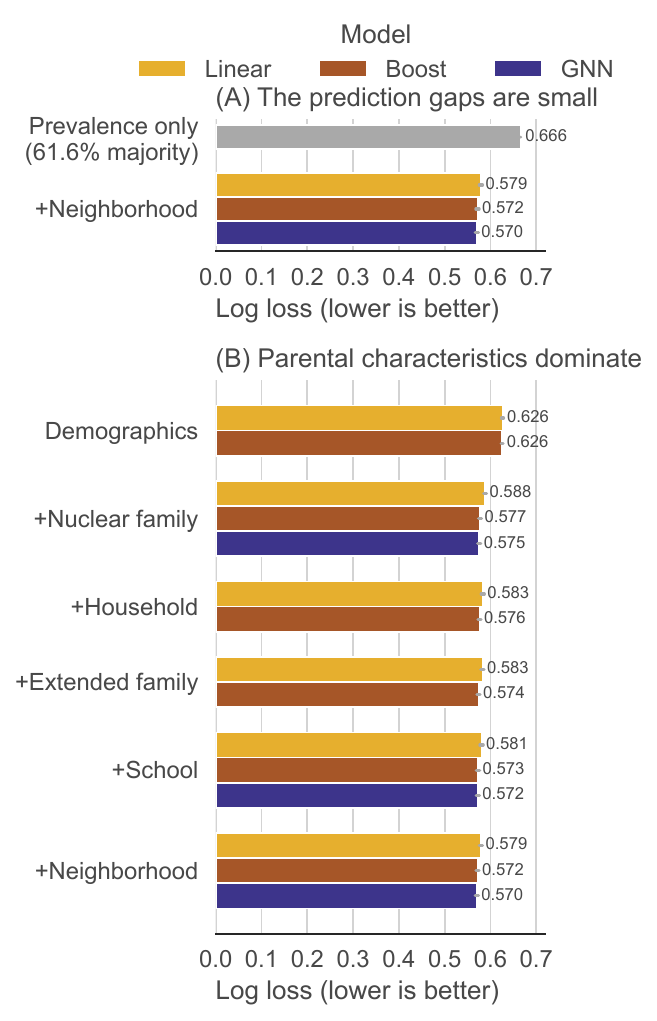}
    \caption{\textbf{Prediction gaps in university completion using log loss}. Held-out log loss for logistic regression (Linear), gradient boosting (Boost), and graph neural networks (GNN). Lower values indicate better predictions. Panel A compares the full-context models with a prevalence-only benchmark that assigns every student the observed class probabilities (61.6\% non-completion and 38.4\% completion); Panel B adds social contexts sequentially. Error bars show 95\% bootstrap confidence intervals, which are not independent because all models are evaluated on the same test observations.}
    \label{fig:pred_gap_log_loss}
\end{figure}

Log loss is the objective used to fit all three models and evaluates predicted probabilities without converting them into binary classifications. Lower values indicate better predictions. Figure~\ref{fig:pred_gap_log_loss} reports the equivalent of Figure~\ref{fig:pred_gap} using held-out log loss. A prevalence-only benchmark that assigns every student the observed shares of the outcome (61.6\% non-completion and 38.4\% completion) has a log loss of 0.666. For the full-context models, log loss is 0.579 for logistic regression, 0.572 for gradient boosting, and 0.570 for the GNN. The model ordering and substantive conclusion are unchanged: greater flexibility improves predicted probabilities, but the incremental gains are small.

\FloatBarrier

\section{Using accuracy as evaluation metric}\label{a:s:accuracy}

\begin{figure}[ht!]
    \centering
    \includegraphics[width=0.5\linewidth]{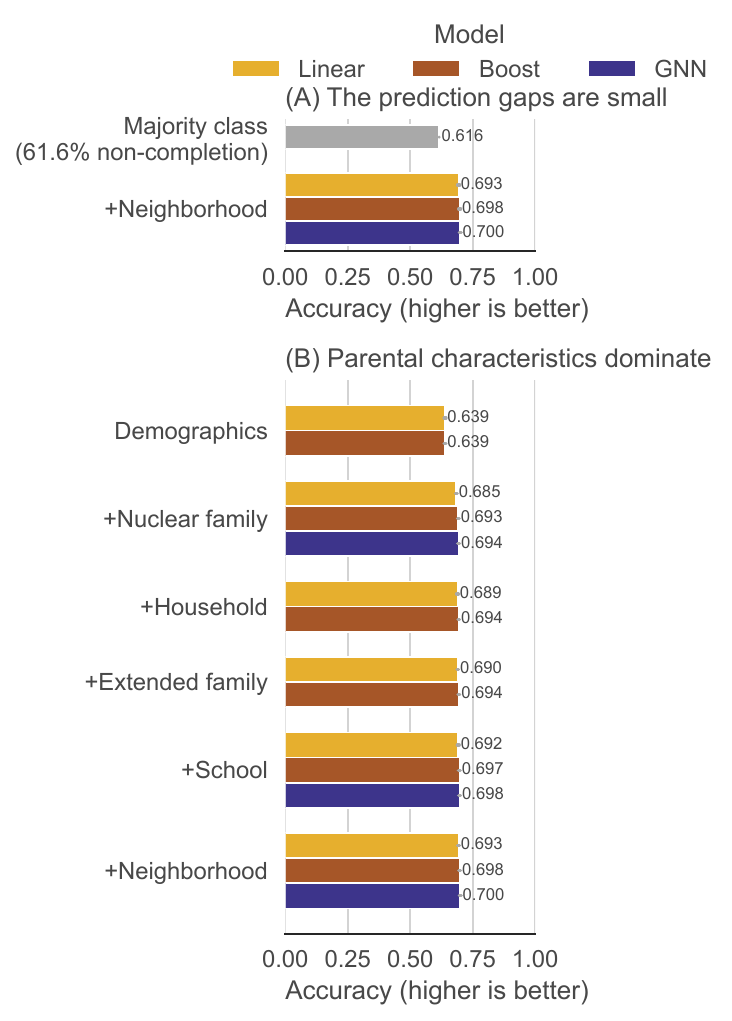}
    \caption{\textbf{Prediction gaps in university completion using accuracy}. Held-out accuracy for logistic regression (Linear), gradient boosting (Boost), and graph neural networks (GNN), using a common probability threshold of 0.5. Panel A compares the full-context models with a majority-class benchmark that predicts non-completion for every student; Panel B adds social contexts sequentially. Error bars show 95\% bootstrap confidence intervals, which are not independent because all models are evaluated on the same test observations.}
    \label{fig:pred_gap_accuracy}
\end{figure}

Accuracy is the share of held-out classifications that are correct at the common probability threshold of 0.5. Accuracy reports the fitted model's overall classification performance at this threshold, but it does not establish an upper bound on the predictability attainable with other data or models. Because 61.6\% of the held-out sample did not complete university, predicting non-completion for every student yields an accuracy of 0.616. The full-context models improve this benchmark to 0.693 for logistic regression, 0.698 for gradient boosting, and 0.700 for the GNN (Fig.~\ref{fig:pred_gap_accuracy}A). The sequential comparison again shows that parental characteristics account for most of the improvement, while later contexts and greater model flexibility add little (Fig.~\ref{fig:pred_gap_accuracy}B). Because accuracy gives greater weight to the more common outcome, we report it as an intuitive complement to MCC rather than as the principal evaluation metric.

\FloatBarrier

\section{Disaggregation of GNN-Linear gap into GNN-Boost and Boost-Linear} \label{a:s:disaggregation_gap}

\begin{figure}[ht!]
    \centering
    \includegraphics[width=1\linewidth]{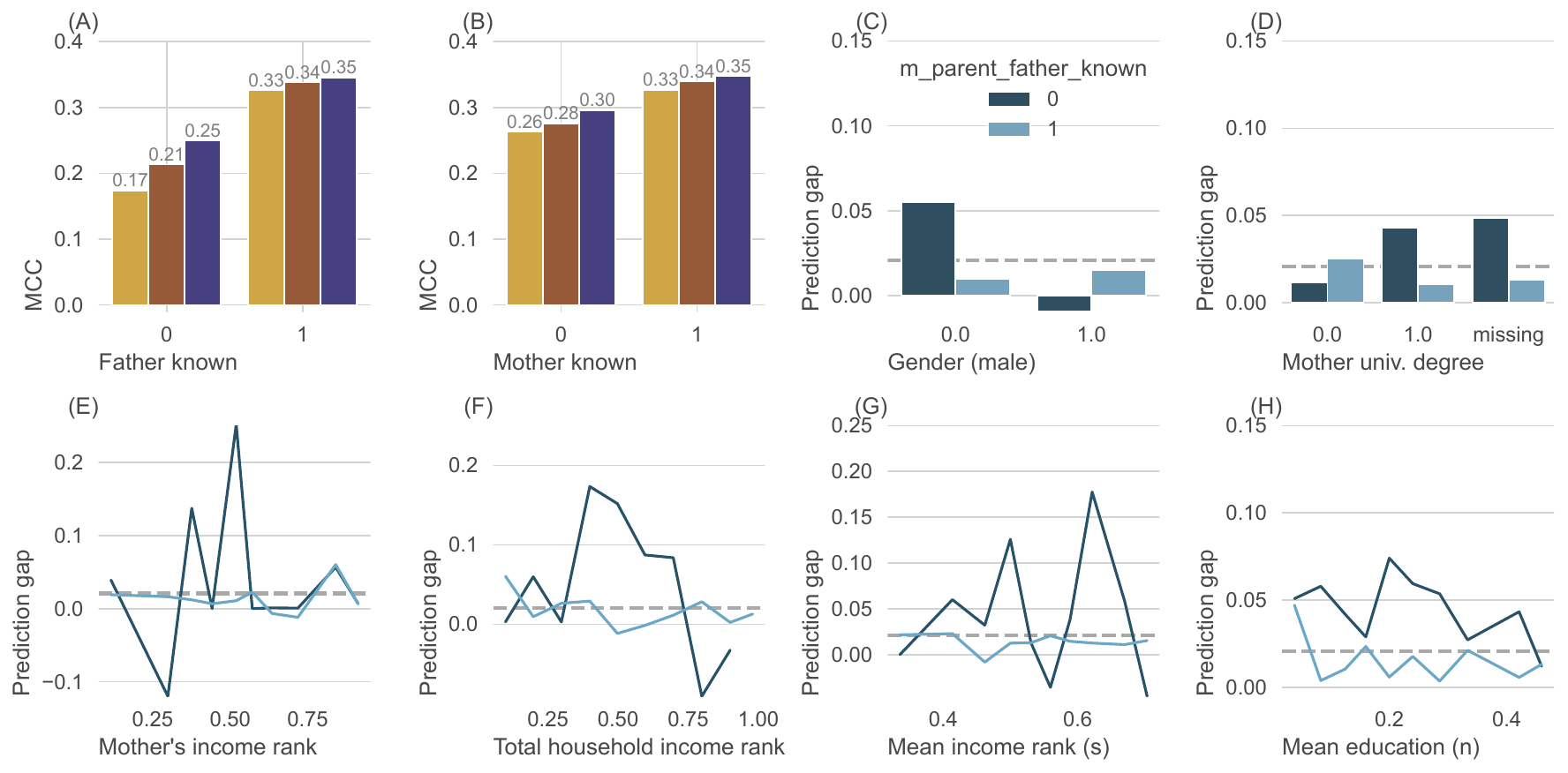}
    \caption{\textbf{Prediction gaps (Boost-Linear) are largest for girls without a registered father}. (A--B) Predictive performance (Matthews Correlation Coefficient, MCC) of the linear (gold), boosting (brown) and GNN (purple) models disaggregated by (A) father and (B) mother status. (C--H) Prediction gaps between gradient boosting and logistic regression disaggregated by father status (dark blue, father deceased or not known to the registry; light blue, father known) and (C) child's registered gender (D) mother educational attainment (E) mother's income rank (F) household's income rank (G) school peers' income rank (H) share of neighbors with a college degree. The dashed horizontal line in (C--H) indicates the average gain from using GNNs in the entire population.}
    \label{fig:disaggregation_delta_a}
\end{figure}

\begin{figure}[ht!]
    \centering
    \includegraphics[width=1\linewidth]{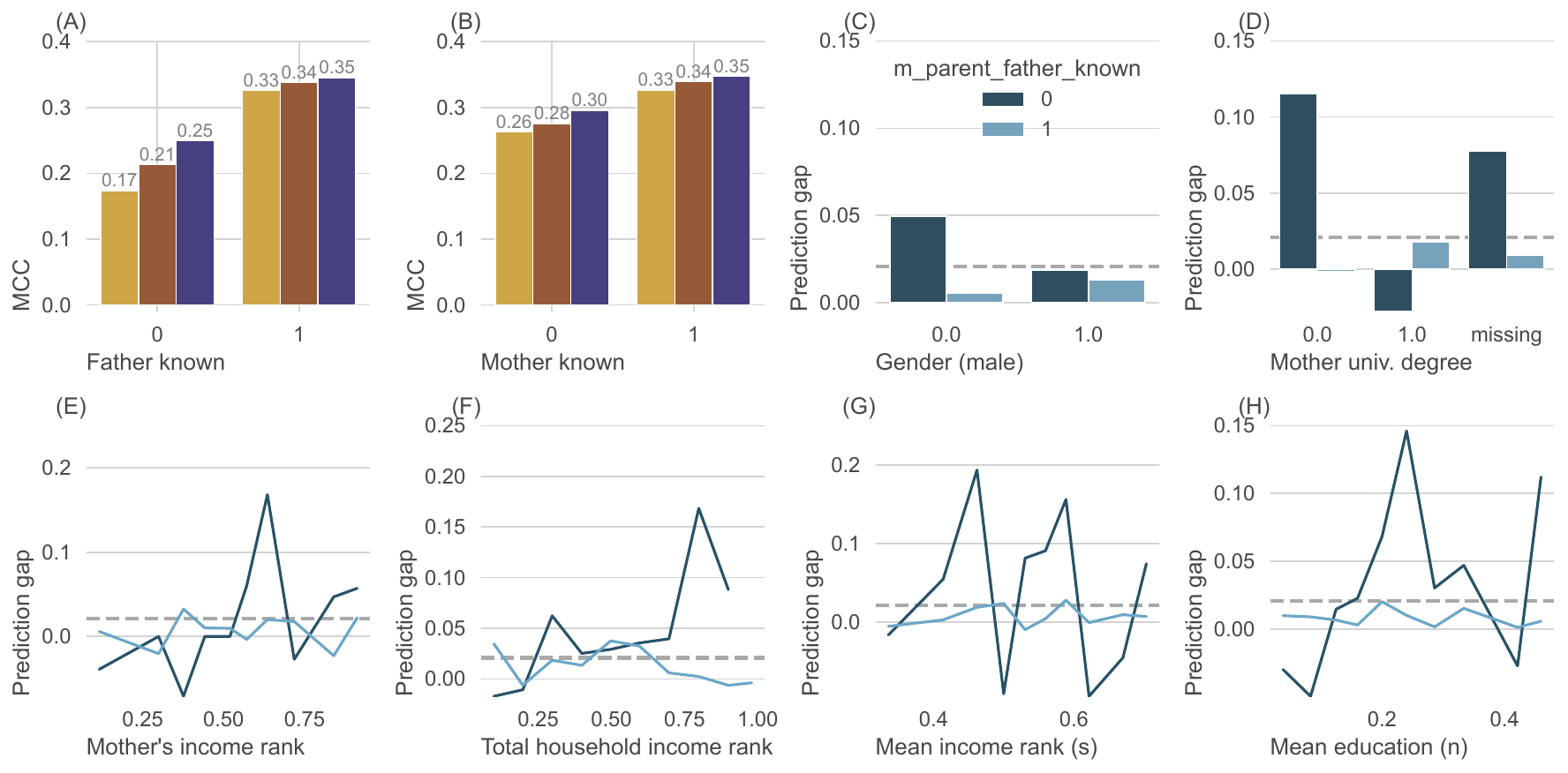}
    \caption{\textbf{Prediction gaps (GNN-Boost) are largest for girls without a registered father}. (A--B) Predictive performance (Matthews Correlation Coefficient, MCC) of the linear (gold), boosting (brown) and GNN (purple) models disaggregated by (A) father and (B) mother status. (C--H) Prediction gaps between GNNs and gradient boosting disaggregated by father status (dark blue, father deceased or not known to the registry; light blue, father known) and (C) child's registered gender (D) mother educational attainment (E) mother's income rank (F) household's income rank (G) school peers' income rank (H) share of neighbors with a college degree. The dashed horizontal line in (C--H) indicates the average gain from using GNNs in the entire population.}
    \label{fig:disaggregation_delta_b}
\end{figure}

Figures~\ref{fig:disaggregation_delta_a}--\ref{fig:disaggregation_delta_b} extend our main analysis (Fig.~\ref{fig:disaggregation}) by decomposing the full prediction gap between GNNs and logistic regression (GNN–Linear) into two components: the gain from moving to boosting (Boost–Linear) and the additional gain from moving to GNNs (GNN–Boost). This disaggregation reveals that the gains are evenly distributed in both steps. 

\FloatBarrier

\cleardoublepage

\section{Disaggregation (all variables)} \label{a:s:disaggregation_full}

\subsection{Nuclear}

\begin{figure}[h!]
    \centering
    \begin{minipage}[t]{0.5\textwidth}
        \centering
        \includegraphics[width=\linewidth]{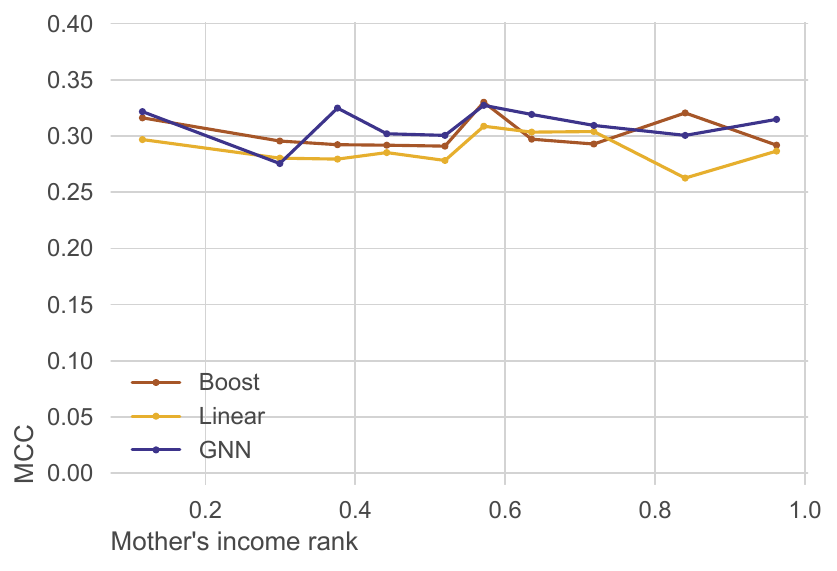}
        \caption{Disaggregation by Mother's income rank (Nuclear)}
        \label{fig:mother_income_rank}
    \end{minipage}%
    \hfill
    \begin{minipage}[t]{0.5\textwidth}
        \centering
        \includegraphics[width=\linewidth]{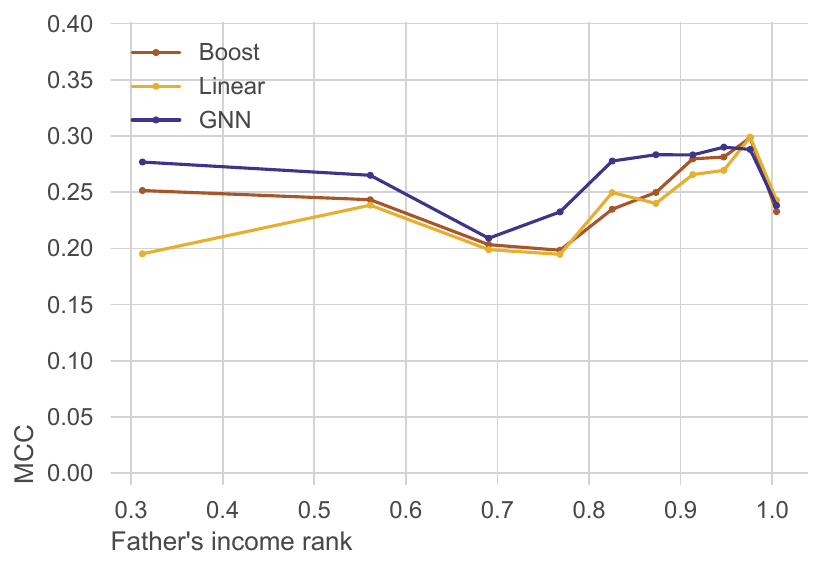}
        \caption{Disaggregation by Father's income rank (Nuclear)}
    \end{minipage}
\end{figure}

\begin{figure}[h!]
    \centering
    \begin{minipage}[t]{0.49\textwidth}
        \centering
        \includegraphics[width=\linewidth]{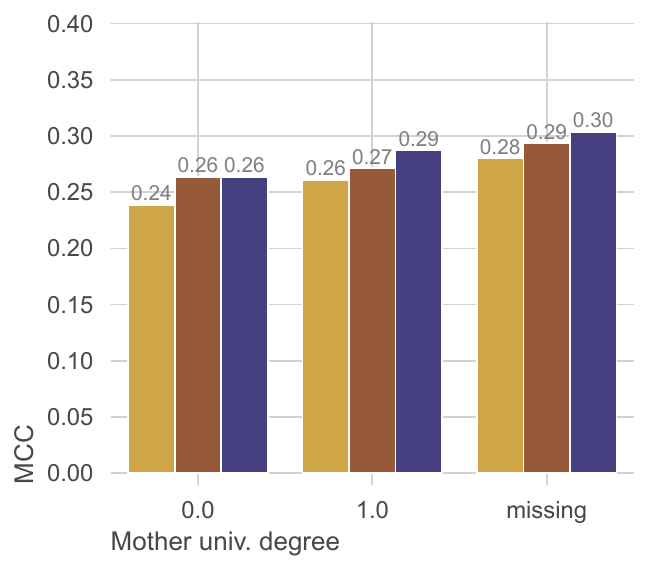}
        \caption{Disaggregation by Mother univ. degree (Nuclear)}
        \label{fig:mother_educ}
    \end{minipage}%
    \hfill
    \begin{minipage}[t]{0.49\textwidth}
        \centering
        \includegraphics[width=\linewidth]{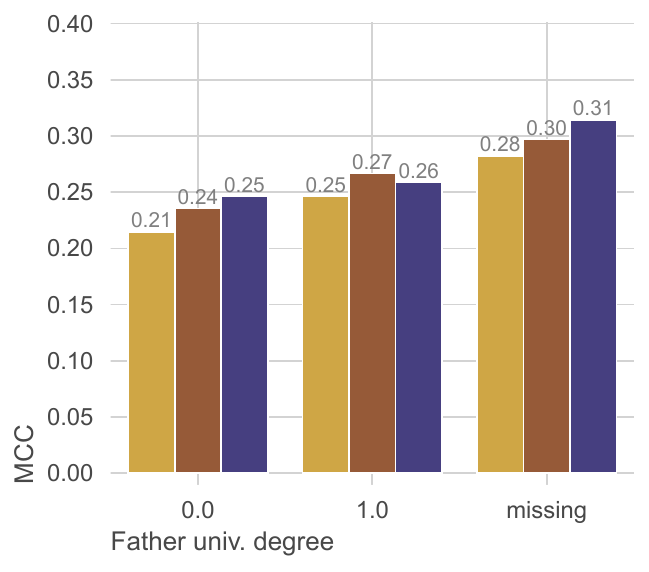}
        \caption{Disaggregation by Father univ. degree (Nuclear)}
        \label{fig:father_educ}
    \end{minipage}
\end{figure}

\begin{figure}[h!]
\centering
\includegraphics[width=0.6\linewidth]{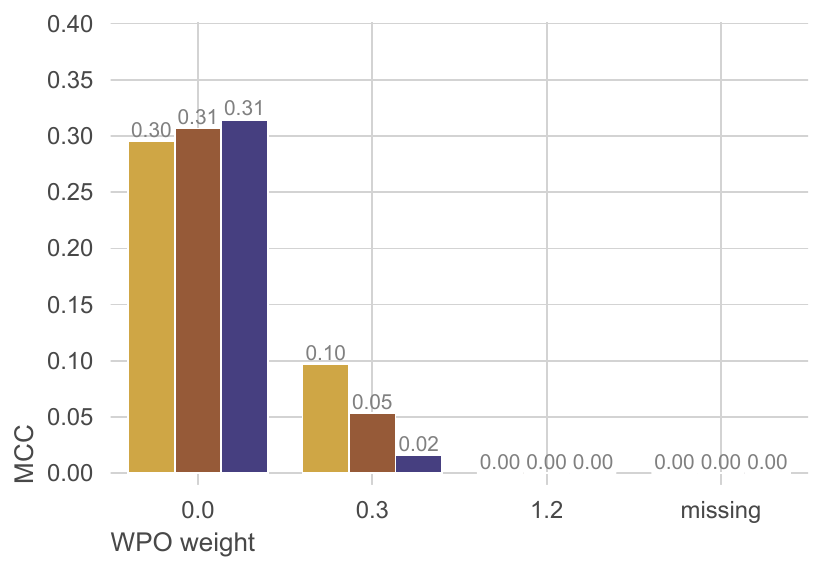}
\caption{Disaggregation by WPO weight (Nuclear)}
\end{figure}

\FloatBarrier
\newpage

\subsection{Extended}

\begin{figure}[h!]
    \centering
    \begin{minipage}[t]{0.49\textwidth}
        \centering
        \includegraphics[width=\linewidth]{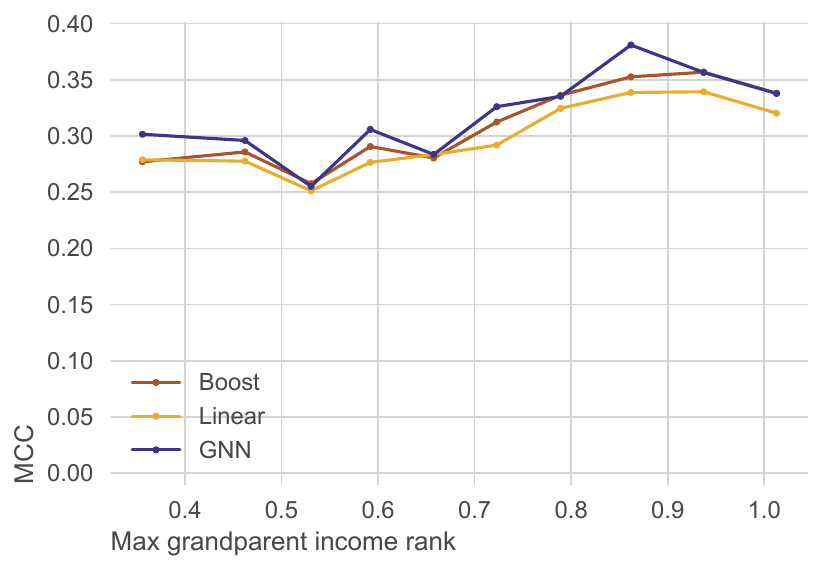}
        \caption{Disaggregation by Max grandparent income rank (Extended)}
        \label{fig:grandparent_max}
    \end{minipage}%
    \hfill
    \begin{minipage}[t]{0.49\textwidth}
        \centering
        \includegraphics[width=\linewidth]{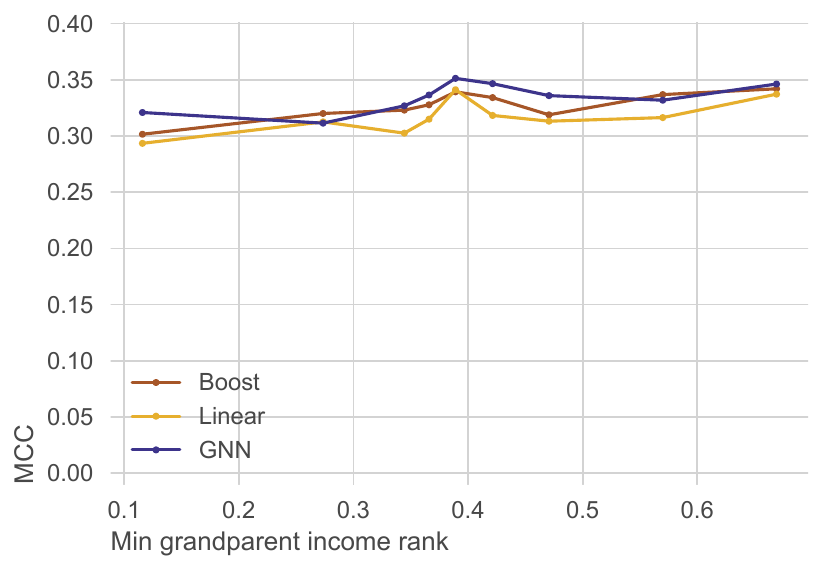}
        \caption{Disaggregation by Min grandparent income rank (Extended)}
        \label{fig:grandparent_min}
    \end{minipage}
\end{figure}

\begin{figure}[h!]
    \centering
    \begin{minipage}[t]{0.49\textwidth}
        \centering
        \includegraphics[width=\linewidth]{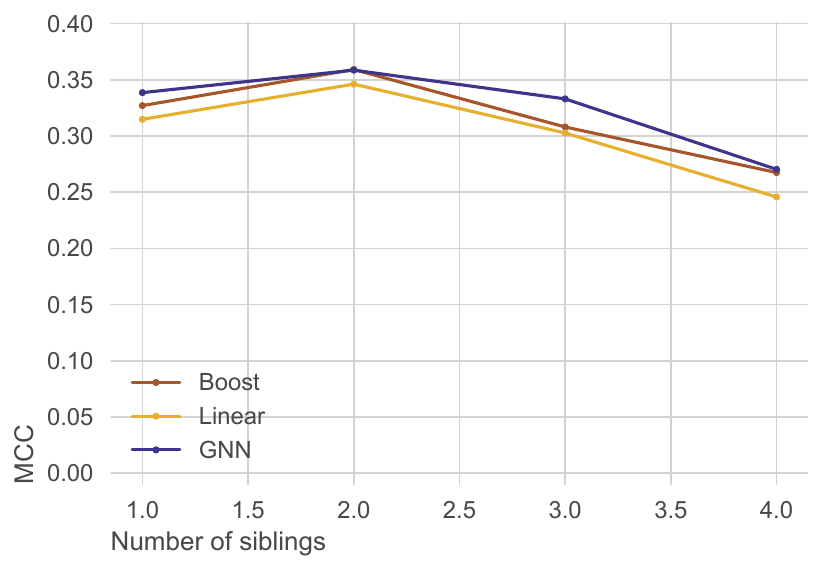}
        \caption{Disaggregation by Number of siblings (Extended)}
        \label{fig:siblings_n_known}
    \end{minipage}%
    \hfill
    \begin{minipage}[t]{0.49\textwidth}
        \centering
        \includegraphics[width=\linewidth]{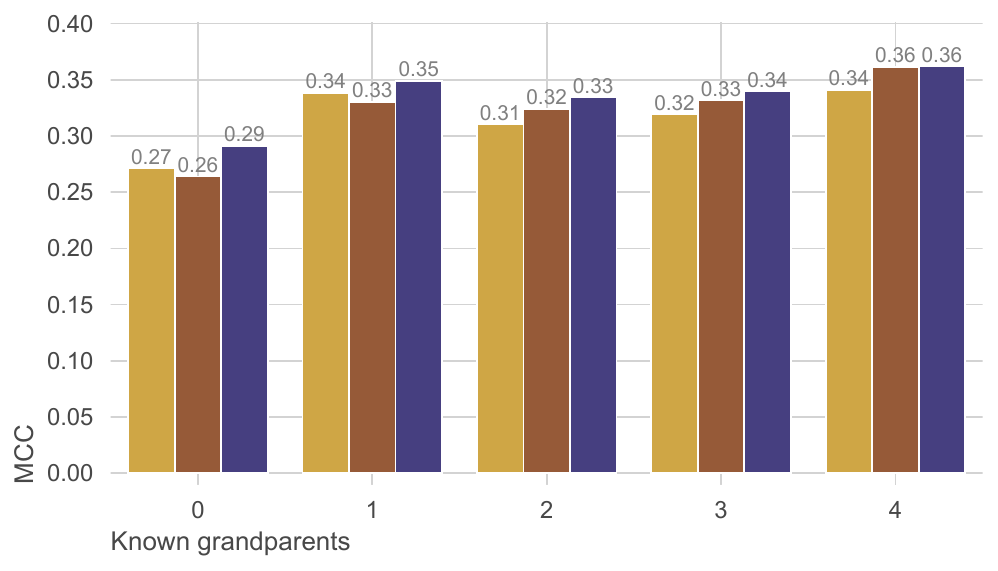}
        \caption{Disaggregation by Known grandparents (Extended)}
        \label{fig:grandparents_n_known}
    \end{minipage}
\end{figure}

\begin{figure}[h!]
\centering
\includegraphics[width=0.6\linewidth]{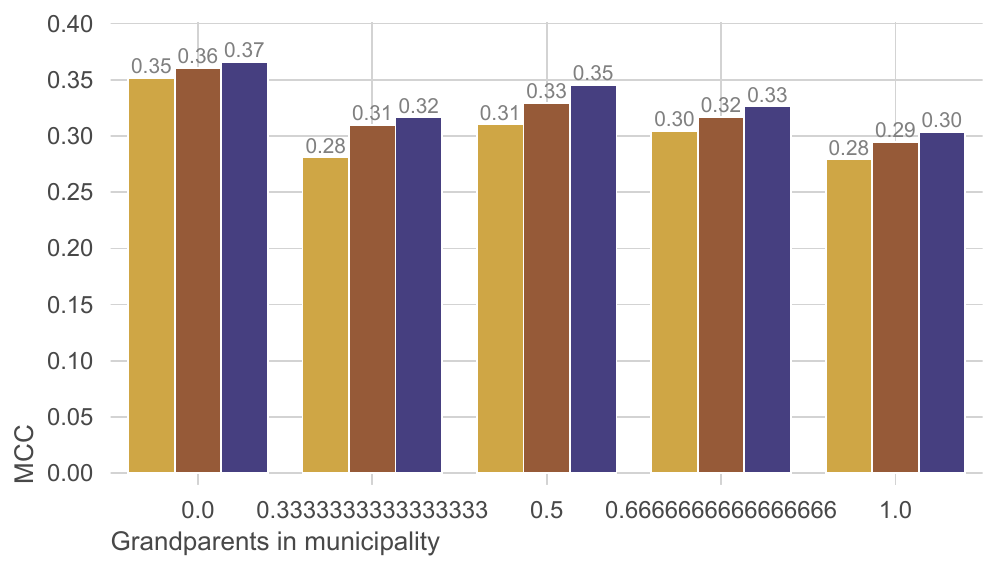}
\caption{Disaggregation by Grandparents in municipality (Extended)}
\end{figure}

\FloatBarrier
\newpage

\subsection{Household}

\begin{figure}[h!]
    \centering
    \begin{minipage}[t]{0.49\textwidth}
        \centering
        \includegraphics[width=\linewidth]{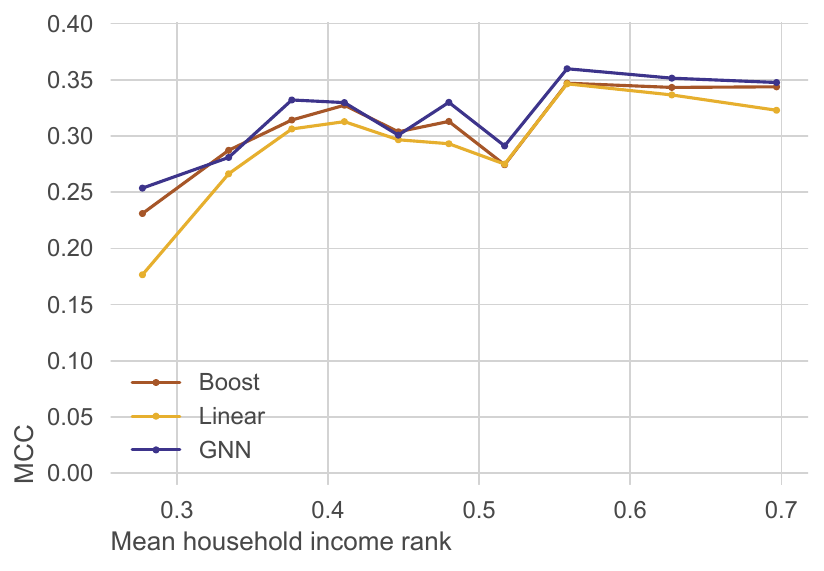}
        \caption{Disaggregation by Mean household income rank (Household)}
        \label{fig:house_mean_income}
    \end{minipage}%
    \hfill
    \begin{minipage}[t]{0.49\textwidth}
        \centering
        \includegraphics[width=\linewidth]{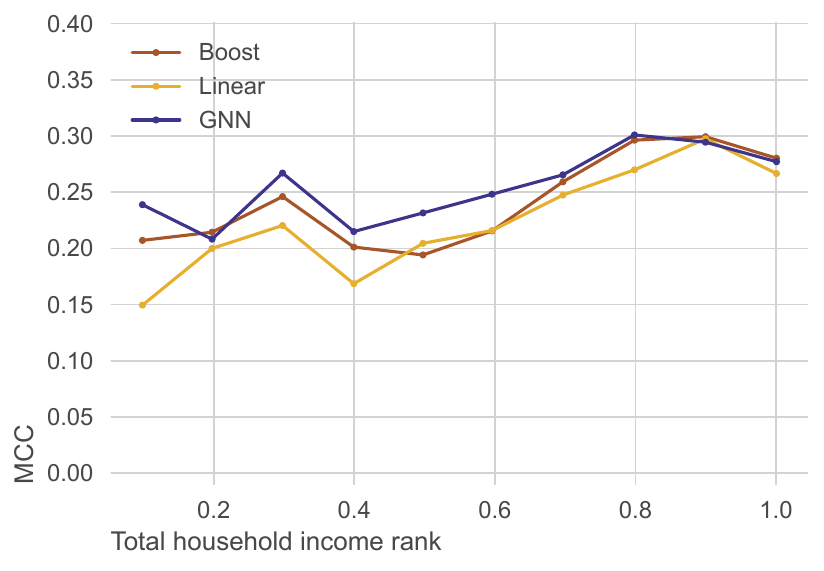}
        \caption{Disaggregation by Total household income rank (Household)}
        \label{fig:house_n_income}
    \end{minipage}
\end{figure}

\begin{figure}[h!]
\centering
\includegraphics[width=0.6\linewidth]{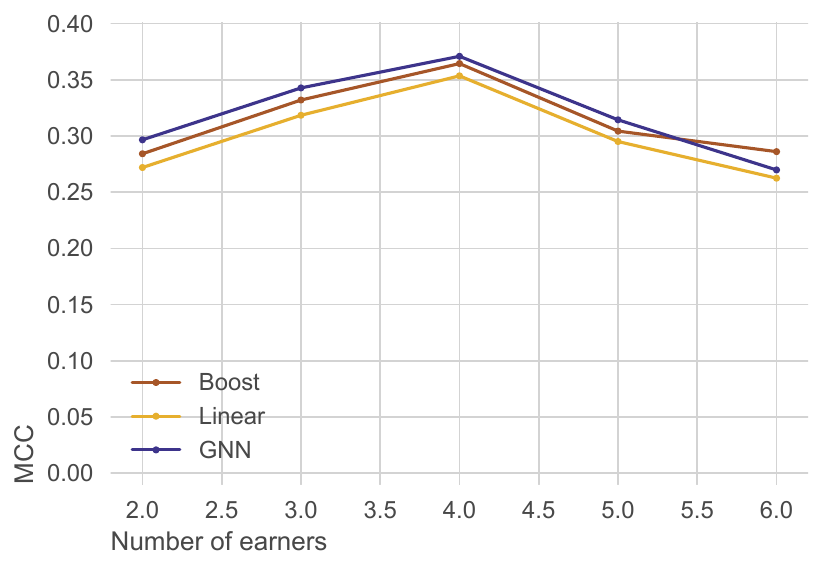}
\caption{Disaggregation by Number of earners (Household)}
\end{figure}

\FloatBarrier
\newpage

\subsection{School}

\begin{figure}[h!]
    \centering
    \begin{minipage}[t]{0.49\textwidth}
        \centering
        \includegraphics[width=\linewidth]{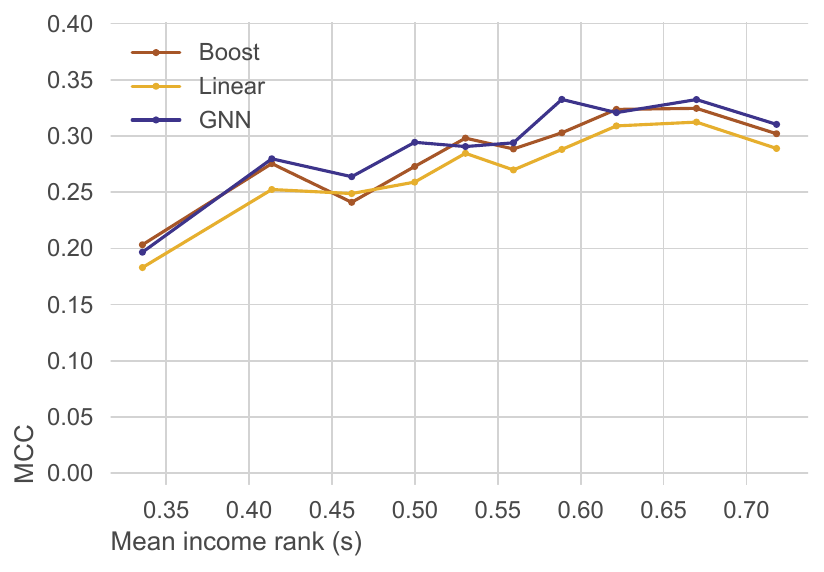}
        \caption{Disaggregation by Mean income rank (s) (School)}
        \label{fig:school_mean_income_rank}
    \end{minipage}%
    \hfill
    \begin{minipage}[t]{0.49\textwidth}
        \centering
        \includegraphics[width=0.8\linewidth]{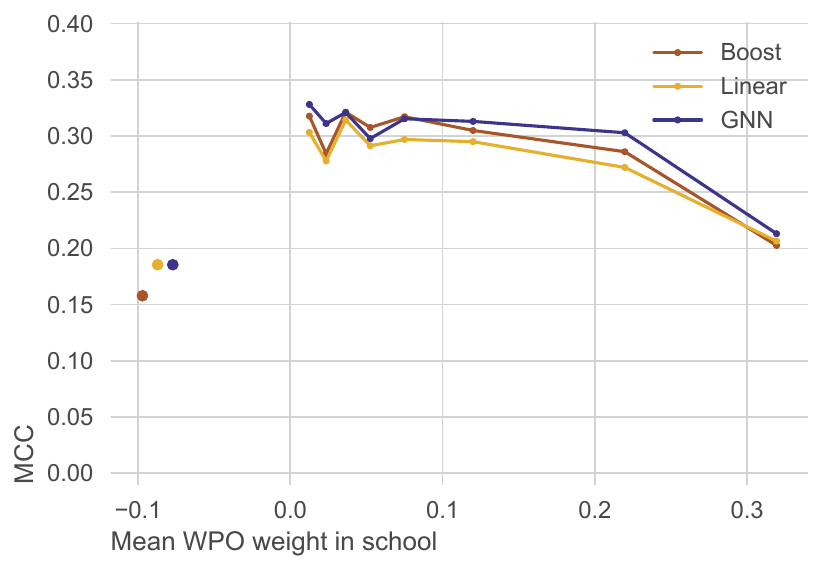}
        \caption{Disaggregation by Mean WPO weight in school (School)}
        \label{fig:school_income_reported}
    \end{minipage}
\end{figure}

\begin{figure}[h!]
    \centering
    \begin{minipage}[t]{0.49\textwidth}
        \centering
        \includegraphics[width=\linewidth]{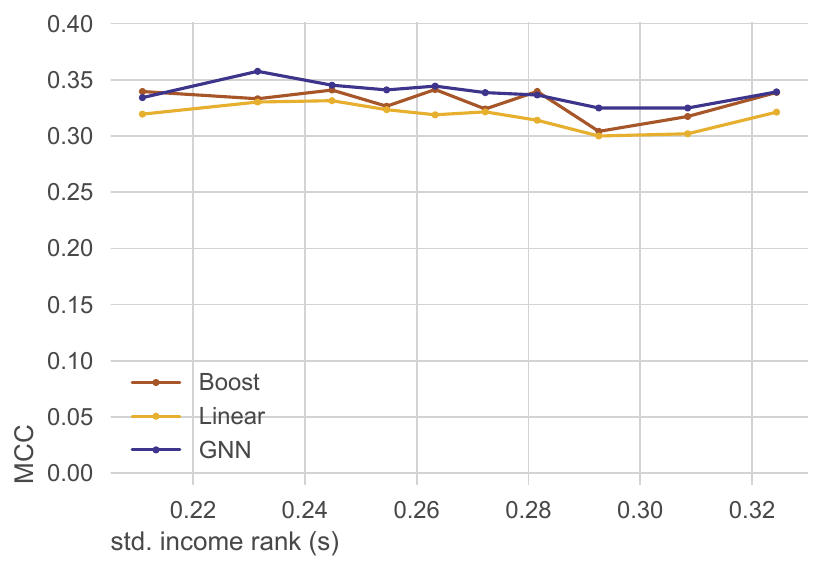}
        \caption{Disaggregation by std. income rank (s) (School)}
        \label{fig:school_std_income_rank}
    \end{minipage}%
    \hfill
    \begin{minipage}[t]{0.49\textwidth}
        \centering
        \includegraphics[width=\linewidth]{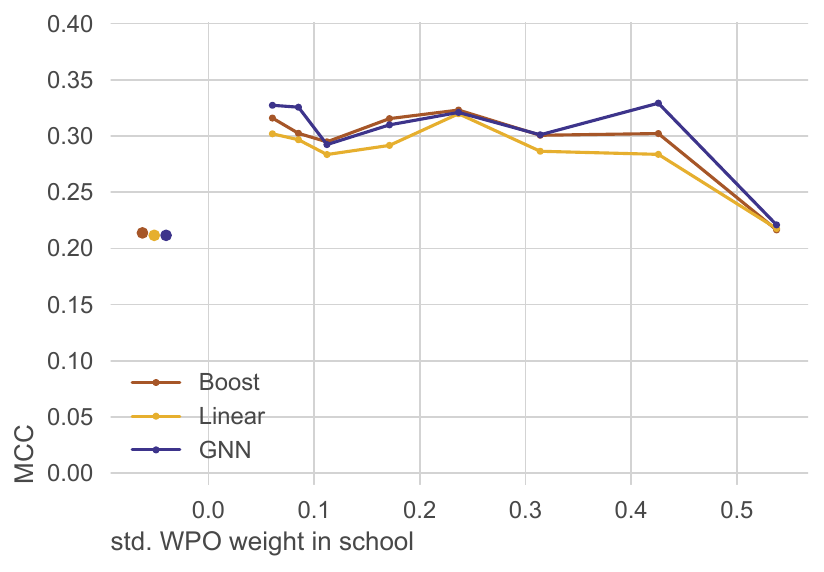}
        \caption{Disaggregation by std. WPO weight in school (School)}
        \label{fig:school_std_wpogewicht}
    \end{minipage}
\end{figure}

\begin{figure}[h!]
    \centering
    \begin{minipage}[t]{0.49\textwidth}
        \centering
        \includegraphics[width=\linewidth]{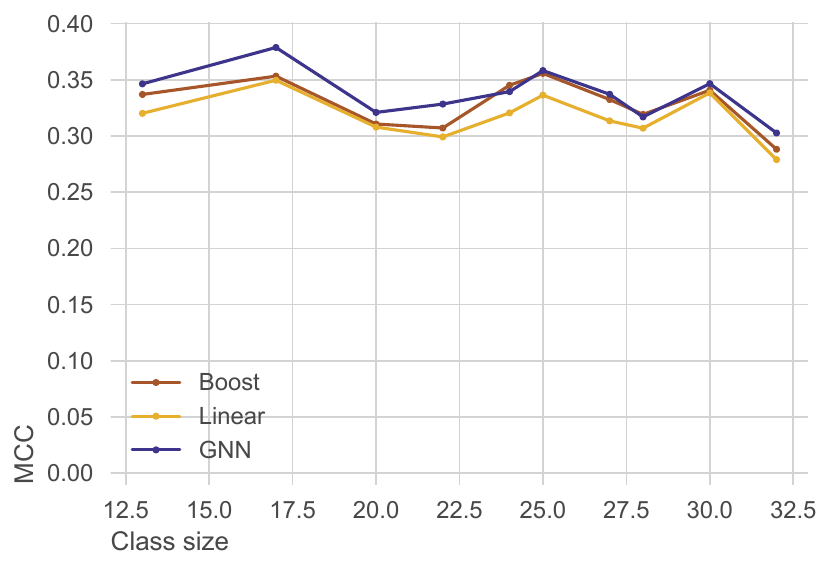}
        \caption{Disaggregation by Class size (School)}
        \label{fig:school_class_size}
    \end{minipage}%
    \hfill
    \begin{minipage}[t]{0.49\textwidth}
        \centering
        \includegraphics[width=\linewidth]{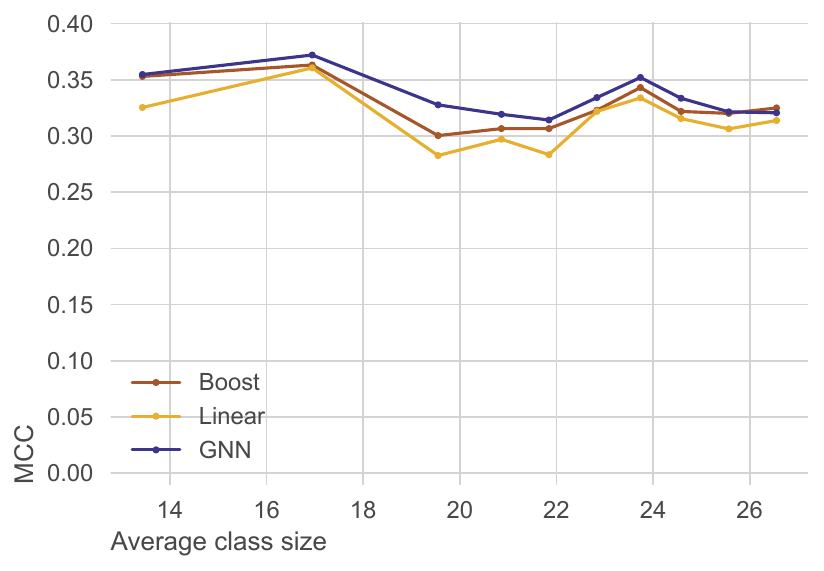}
        \caption{Disaggregation by Average class size (School)}
        \label{fig:school_mean_class_size}
    \end{minipage}
\end{figure}

\begin{figure}[h!]
    \centering
    \begin{minipage}[t]{0.49\textwidth}
        \centering
        \includegraphics[width=\linewidth]{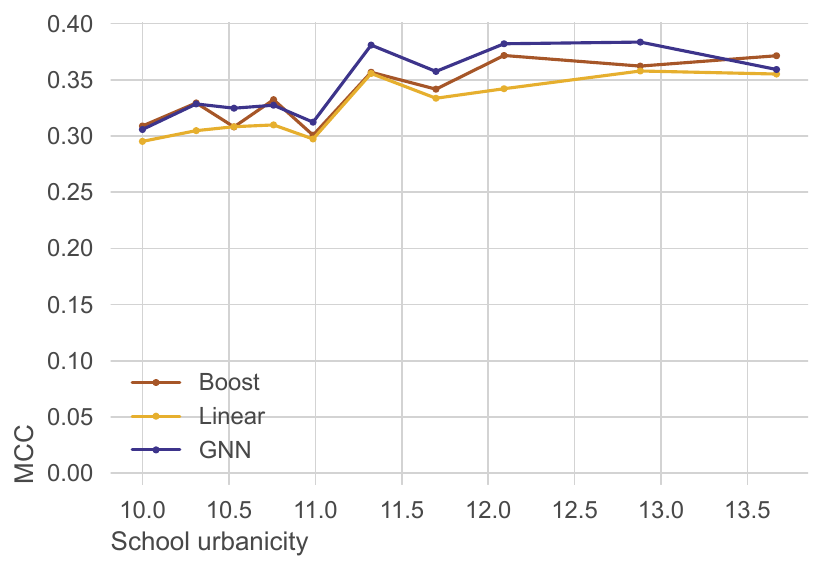}
        \caption{Disaggregation by School urbanicity (School)}
        \label{fig:school_urbanicity}
    \end{minipage}%
    \hfill
    \begin{minipage}[t]{0.49\textwidth}
        \centering
        \includegraphics[width=\linewidth]{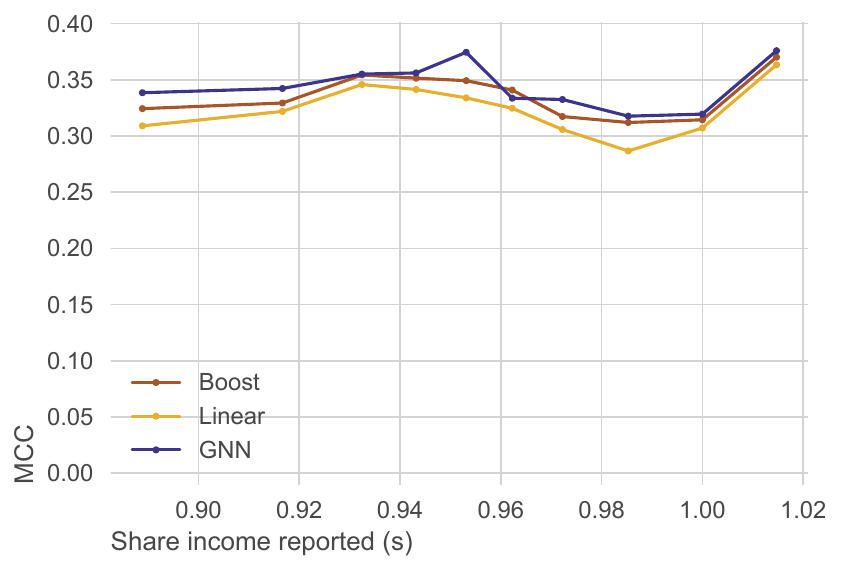}
        \caption{Disaggregation by Share income reported (s) (School)}
        \label{fig:school_income_share}
    \end{minipage}
\end{figure}

\begin{figure}[h!]
\centering
\includegraphics[width=0.6\linewidth]{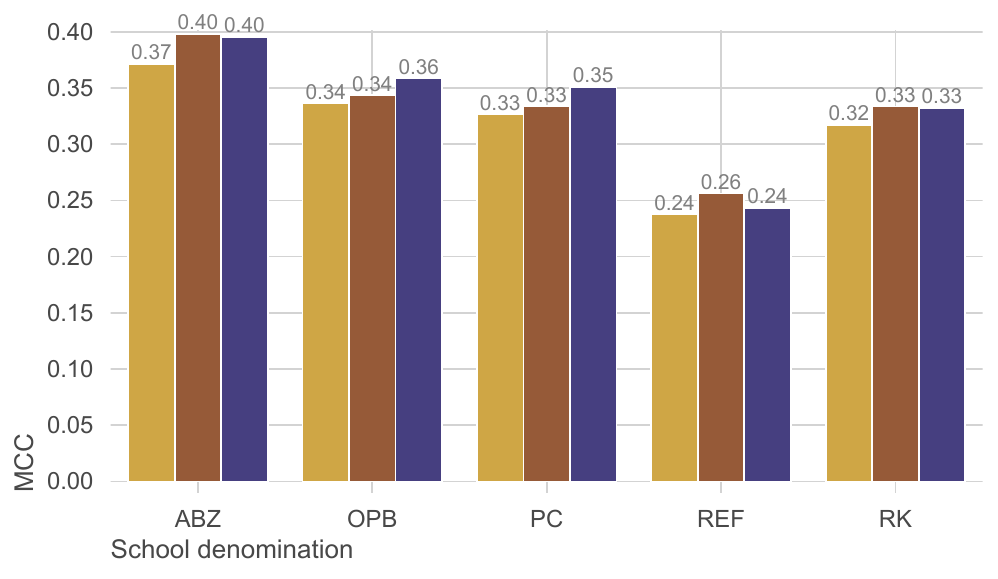}
\caption{Disaggregation by School denomination (School)}
\end{figure}

\FloatBarrier
\newpage

\subsection{Neighborhood}

\begin{figure}[h!]
    \centering
    \begin{minipage}[t]{0.49\textwidth}
        \centering
        \includegraphics[width=\linewidth]{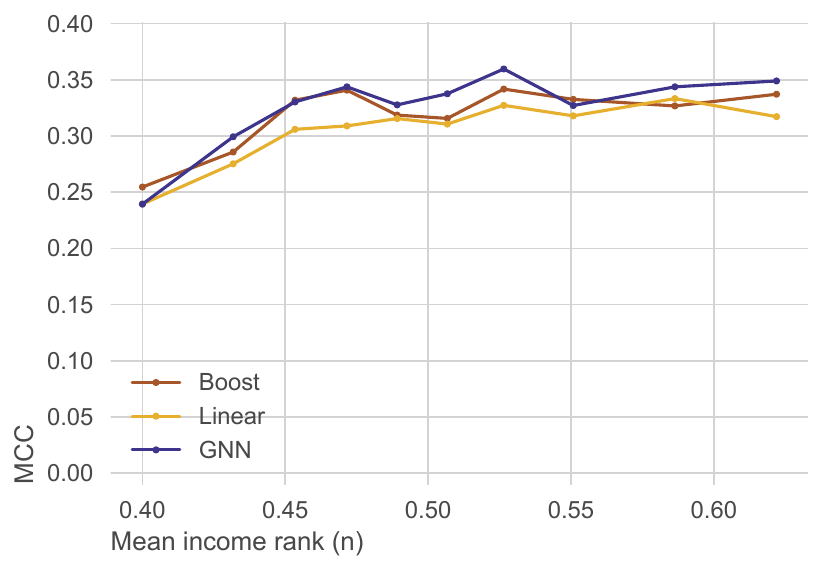}
        \caption{Disaggregation by Mean income rank (n) (Neighborhood)}
        \label{fig:neig_income_mean}
    \end{minipage}%
    \hfill
    \begin{minipage}[t]{0.49\textwidth}
        \centering
        \includegraphics[width=\linewidth]{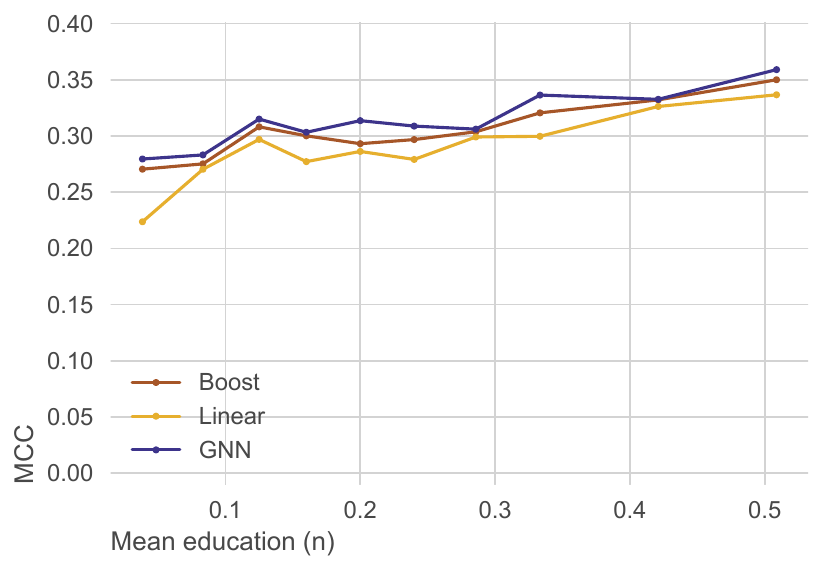}
        \caption{Disaggregation by Mean education (n) (Neighborhood)}
        \label{fig:neig_educ_mean}
    \end{minipage}
\end{figure}

\begin{figure}[h!]
\centering
\begin{minipage}[t]{0.48\textwidth}
    \centering
    \includegraphics[width=\linewidth]{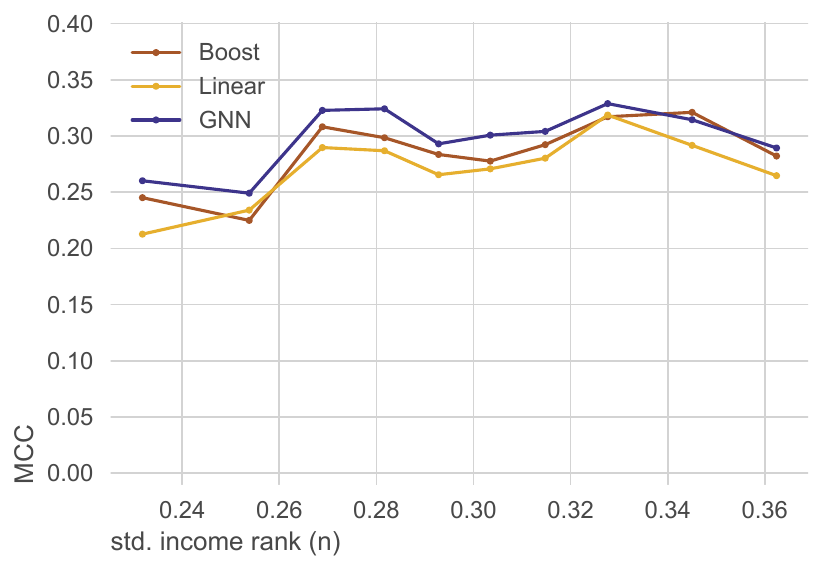}
    \caption{Disaggregation by std. income rank (n) (Neighborhood)}
\end{minipage}
\hfill
\begin{minipage}[t]{0.48\textwidth}
    \centering
    \includegraphics[width=\linewidth]{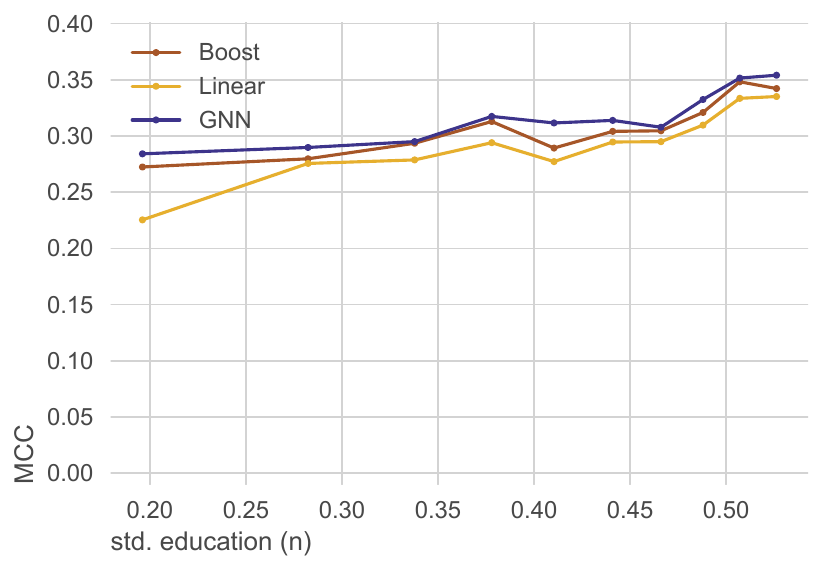}
    \caption{Disaggregation by std. education (n) (Neighborhood)}
\end{minipage}
\end{figure}

\FloatBarrier
\newpage

\subsection{Demographics}

\begin{figure}[h!]
\centering
\begin{minipage}[t]{0.32\textwidth}
    \centering
    \includegraphics[width=\linewidth]{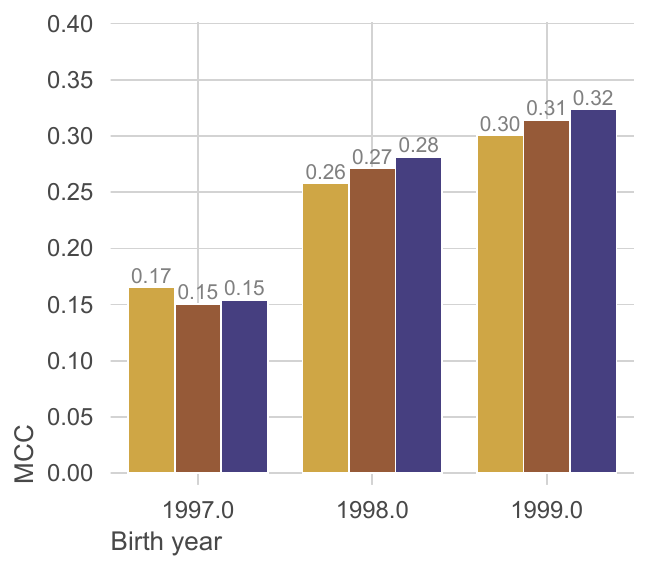}
    \caption{Disaggregation by Birth year (Demographics)}
\end{minipage}
\hfill
\begin{minipage}[t]{0.32\textwidth}
    \centering
    \includegraphics[width=\linewidth]{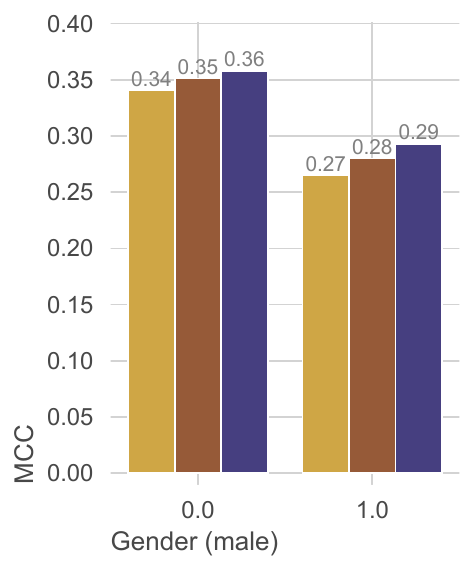}
    \caption{Disaggregation by Gender (male) (Demographics)}
\end{minipage}
\hfill
\end{figure}

\begin{figure}[h!]
\centering
\begin{minipage}[t]{0.54\textwidth}
    \centering
    \includegraphics[width=\linewidth]{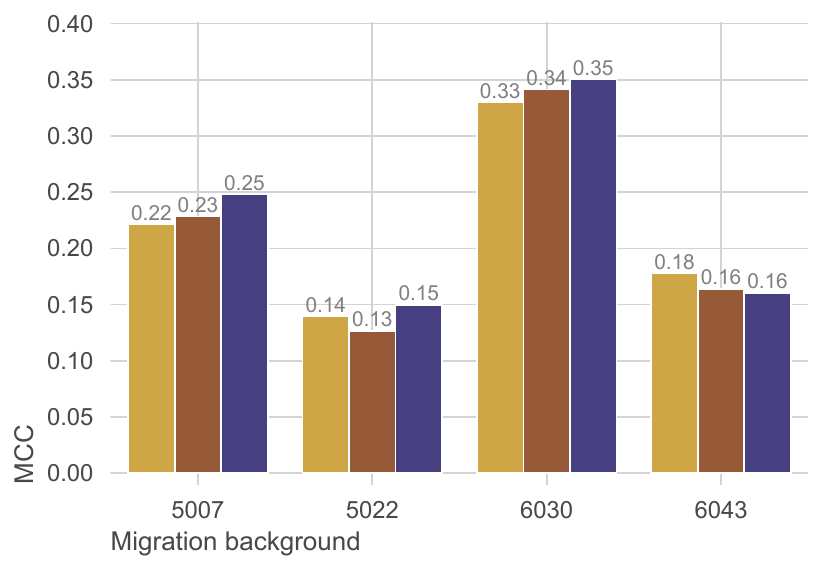}
    \caption{Disaggregation by Migration background. 6030: Dutch. 6043: Turkey. 5007: Suriname. 5022: Morocco (Demographics)}
\end{minipage}
\hfill
\begin{minipage}[t]{0.42\textwidth}
    \centering
    \includegraphics[width=\linewidth]{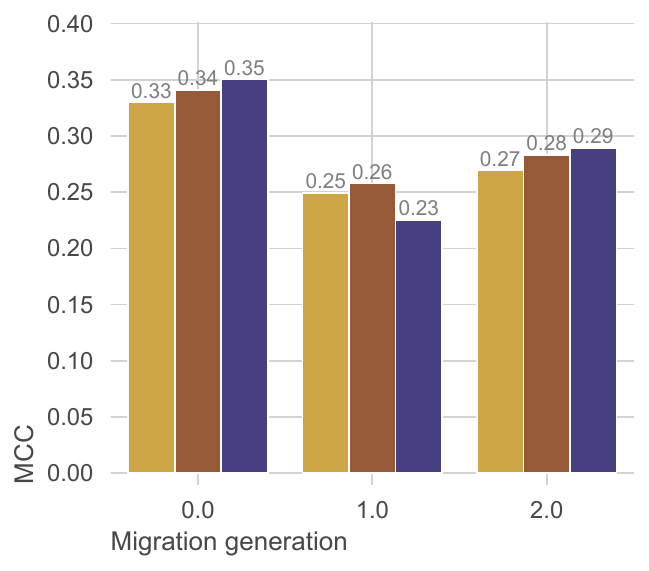}
    \caption{Disaggregation by Migration generation (Demographics)}
\end{minipage}
\end{figure}

\begin{figure}[h!]
\centering
\hfill
\begin{minipage}[t]{0.32\textwidth}
    \centering
    \includegraphics[width=\linewidth]{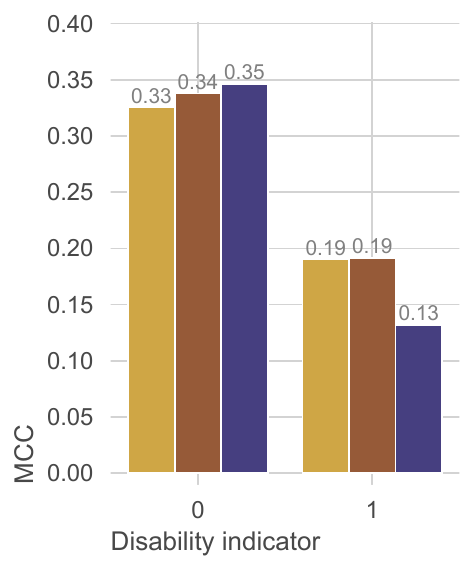}
    \caption{Disaggregation by Disability indicator (Demographics)}
\end{minipage}
\hfill
\begin{minipage}[t]{0.32\textwidth}
    \centering
    \includegraphics[width=\linewidth]{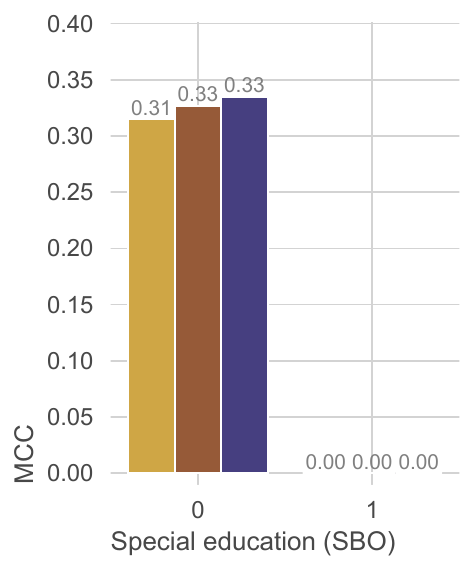}
    \caption{Disaggregation by Special education (SBO) (Demographics)}
\end{minipage}
\end{figure}

\FloatBarrier
\newpage

\subsection{Variables not included in the model}
\begin{figure}[h!]
\centering
\includegraphics[width=0.55\linewidth]{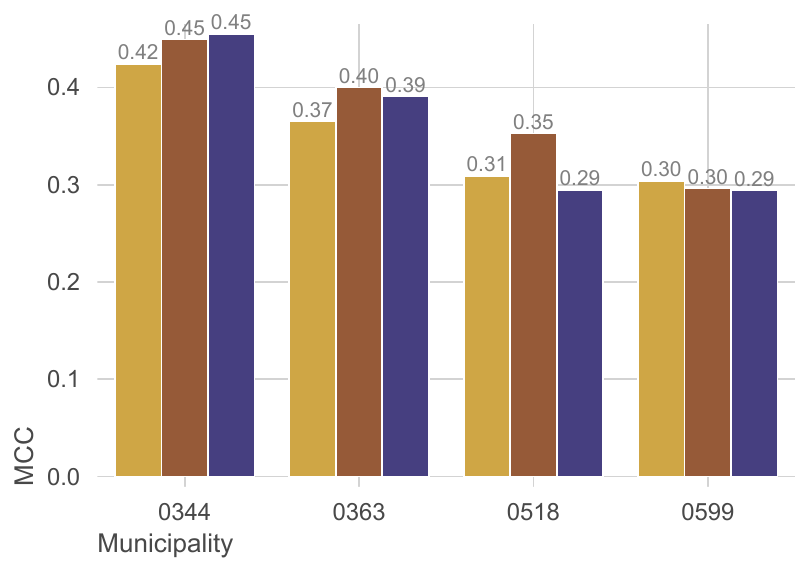}
\caption{Disaggregation by Municipality.  0344: Utrecht; 0363: Amsterdam; 0518: the Hague; 0599: Rotterdam}
\end{figure}

\begin{figure}[h!]
\centering
\includegraphics[width=0.6\linewidth]{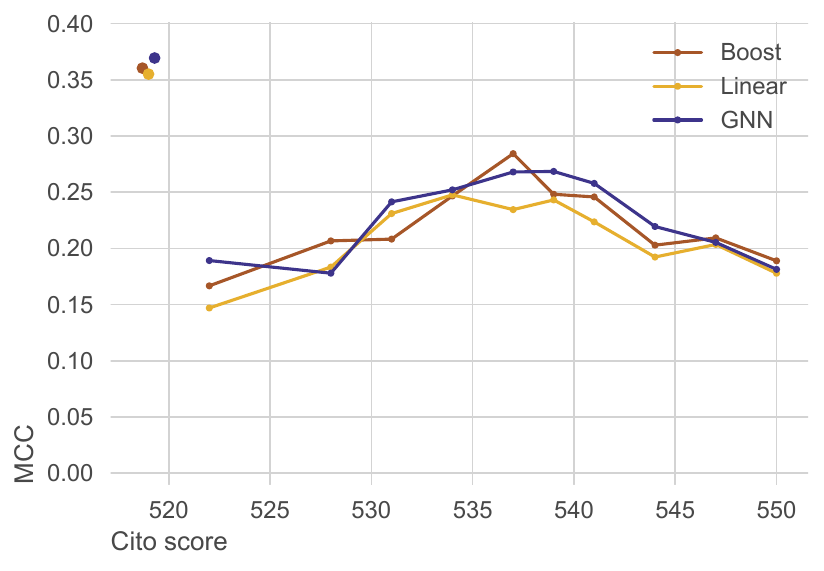}
\caption{Disaggregation by CITO score. The Cito score refers to the result of the national standardized final test (\emph{Eindtoets Basisonderwijs}) administered at the end of primary education in the Netherlands. It assesses core competencies in language, mathematics, and world knowledge, and is used to help determine the appropriate level of secondary education. }
\end{figure}

\begin{figure}[h!]
\centering
\includegraphics[width=1\linewidth]{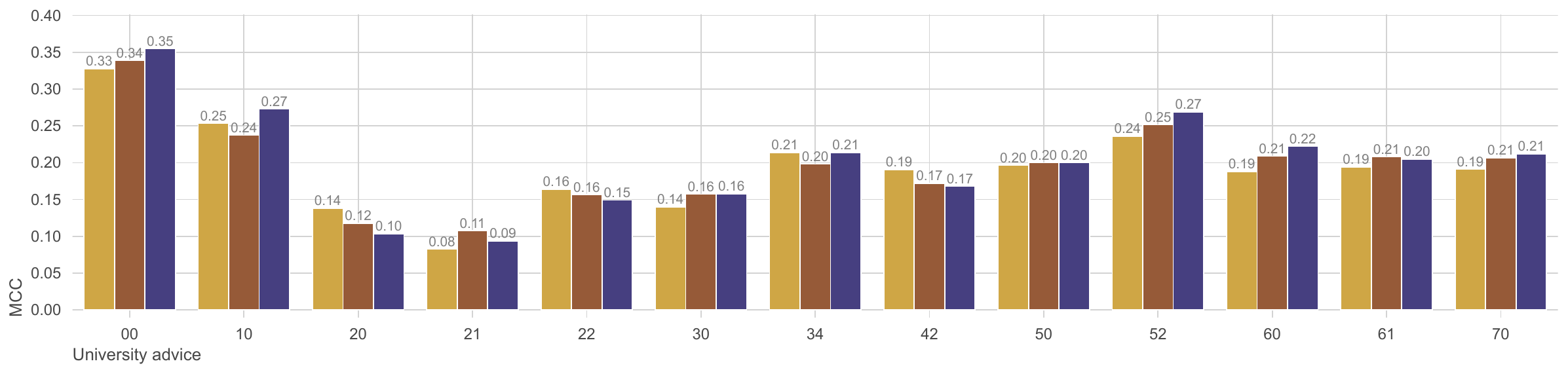}
\caption{Disaggregation by University track advice. The x-axis shows codes representing the highest level of secondary education a student was advised to pursue. 
00 = No advice, 
10 = Practical education (PrO), 
20 = VMBO basic vocational track (BL), 
21 = VMBO BL with learning support (LWOO), 
22 = VMBO BL to intermediate vocational (KL), 
30 = VMBO intermediate vocational track (KL), 
34 = VMBO intermediate to theoretical track (KL-TL), 
42 = VMBO general track (GL) to TL, 
50 = VMBO theoretical track (TL), 
52 = VMBO TL to HAVO, 
60 = HAVO (senior general secondary education), 
61 = HAVO to VWO (pre-university), 
70 = VWO (pre-university education).}
\end{figure}

\FloatBarrier

\printbibliography[heading=subbibliography,title={Appendix Bibliography}]

@article{torche_is_2011,
    title = {Is a {College} {Degree} {Still} the {Great} {Equalizer}? {Intergenerational} {Mobility} across {Levels} of {Schooling} in the {United} {States}},
    volume = {117},
    issn = {0002-9602},
    shorttitle = {Is a {College} {Degree} {Still} the {Great} {Equalizer}?},
    url = {https://www.journals.uchicago.edu/doi/abs/10.1086/661904},
    doi = {10.1086/661904},
    number = {3},
    urldate = {2024-01-03},
    journal = {American Journal of Sociology},
    author = {Torche, Florencia},
    month = nov,
    year = {2011},
    note = {Publisher: The University of Chicago Press},
    pages = {763--807},
}

@article{parolin_intergenerational_2024,
    title = {Intergenerational persistence of poverty in five high-income countries},
    url = {https://www.nature.com/articles/s41562-024-02029-w},
    urldate = {2024-11-26},
    journal = {Nature Human Behaviour},
    volume = {9},
    author = {Parolin, Zachary and Pintro-Schmitt, Rafael and Esping-Andersen, Gøsta and Fallesen, Peter},
    year = {2025},
    note = {Publisher: Nature Publishing Group UK London},
    pages = {254--267},
    number = {2},
    doi = {10.1038/s41562-024-02029-w},
}

@article{raghupathi_influence_2020,
    title = {The influence of education on health: an empirical assessment of {OECD} countries for the period 1995–2015},
    volume = {78},
    issn = {2049-3258},
    shorttitle = {The influence of education on health},
    url = {https://doi.org/10.1186/s13690-020-00402-5},
    doi = {10.1186/s13690-020-00402-5},
    number = {1},
    urldate = {2025-04-15},
    journal = {Archives of Public Health},
    author = {Raghupathi, Viju and Raghupathi, Wullianallur},
    month = apr,
    year = {2020},
    pages = {20},
}

@article{sivak2026births,
    title = {Are births predictable with linked survey and register data? Evidence from the Predicting Fertility data challenge},
    author = {Sivak, Elizaveta and Stulp, Gert},
    journal = {International Journal of Population Data Science},
    volume = {11},
    number = {5},
    eid = {3486},
    year = {2026},
    doi = {10.23889/ijpds.v11i5.3486},
}

@article{belsky_genetic_2018,
    title = {Genetic analysis of social-class mobility in five longitudinal studies},
    volume = {115},
    issn = {0027-8424, 1091-6490},
    url = {https://pnas.org/doi/full/10.1073/pnas.1801238115},
    doi = {10.1073/pnas.1801238115},
    language = {en},
    number = {31},
    pages = {E7275--E7284},
    urldate = {2024-01-26},
    journal = {Proceedings of the National Academy of Sciences},
    author = {Belsky, Daniel W. and Domingue, Benjamin W. and Wedow, Robbee and Arseneault, Louise and Boardman, Jason D. and Caspi, Avshalom and Conley, Dalton and Fletcher, Jason M. and Freese, Jeremy and Herd, Pamela and Moffitt, Terrie E. and Poulton, Richie and Sicinski, Kamil and Wertz, Jasmin and Harris, Kathleen Mullan},
    month = jul,
    year = {2018},
}

@report{Adreskwaliteit2024,
  author       = {{Centraal Bureau voor de Statistiek}},
  shortauthor  = {{CBS}},
  title        = {Adreskwaliteit 2024: Onderzoek naar de kwaliteit van adresgegevens in de {BRP}},
  institution  = {Centraal Bureau voor de Statistiek},
  date         = {2025-06-30},
  langid       = {dutch},
  url          = {https://dashboards.cbs.nl/longreads/Adreskwaliteit_2024/},
  urldate      = {2026-08-31}
}

@article{salganikIntroductionSpecialCollection2019,
  title = {Introduction to the {{Special Collection}} on the {{Fragile Families Challenge}}},
  author = {Salganik, Matthew J. and Lundberg, Ian and Kindel, Alexander T. and McLanahan, Sara},
  date = {2019-01-01},
  journaltitle = {Socius},
  volume = {5},
  pages = {2378023119871580},
  publisher = {SAGE Publications},
  issn = {2378-0231},
  doi = {10.1177/2378023119871580},
  url = {https://doi.org/10.1177/2378023119871580},
  urldate = {2023-01-22},
  langid = {english}
}

@article{sperrinMissingDataShould2020,
  title = {Missing Data Should Be Handled Differently for Prediction than for Description or Causal Explanation},
  author = {Sperrin, Matthew and Martin, Glen P. and Sisk, Rose and Peek, Niels},
  date = {2020-09-01},
  journaltitle = {Journal of Clinical Epidemiology},
  shortjournal = {J. Clin. Epidemiol.},
  volume = {125},
  eprint = {32540389},
  eprinttype = {pubmed},
  pages = {183--187},
  publisher = {Elsevier},
  issn = {0895-4356, 1878-5921},
  doi = {10.1016/j.jclinepi.2020.03.028},
  url = {https://www.jclinepi.com/article/S0895-4356(19)30766-8/abstract},
  urldate = {2026-08-31},
  langid = {english}
}

@book{frank_success_2016,
    title = {Success and Luck: Good Fortune and the Myth of Meritocracy},
    author = {Frank, Robert H.},
    publisher = {Princeton University Press},
    location = {Princeton, NJ},
    year = {2016},
    isbn = {978-1-4008-8027-0},
    doi = {10.1515/9781400880270},
}

@article{chettyImpactsNeighborhoodsIntergenerational2018,
    title = {The {Impacts} of {Neighborhoods} on {Intergenerational Mobility I}: {Childhood Exposure Effects}},
    volume = {133},
    number = {3},
    pages = {1107--1162},
    doi = {10.1093/qje/qjy007},
    url = {https://doi.org/10.1093/qje/qjy007},
    journal = {The Quarterly Journal of Economics},
    author = {Chetty, Raj and Hendren, Nathaniel},
    year = {2018},
}

@book{putnam_bowling_2000,
    title = {Bowling alone: {The} collapse and revival of {American} community},
    publisher = {Simon \& Schuster},
    author = {Putnam, Robert},
    year = {2000},
}

@book{bourdieu_reproduction_1990,
    title = {Reproduction in Education, Society and Culture},
    author = {Bourdieu, Pierre and Passeron, Jean-Claude},
    translator = {Nice, Richard},
    edition = {2},
    publisher = {Sage},
    location = {London},
    year = {1990},
}

@article{engzell_its_2020,
    title = {It's {All} about the {Parents}: {Inequality} {Transmission} across {Three} {Generations} in {Sweden}},
    volume = {7},
    issn = {23306696},
    shorttitle = {It's {All} about the {Parents}},
    url = {https://www.sociologicalscience.com/articles-v7-10-242/},
    doi = {10.15195/v7.a10},
    language = {en},
    urldate = {2024-01-03},
    journal = {Sociological Science},
    author = {Engzell, Per and Mood, Carina and Jonsson, Jan},
    year = {2020},
    pages = {242--267},
}

@article{hallsten_grand_2017,
    title = {Grand {Advantage}: {Family} {Wealth} and {Grandchildren}'s {Educational} {Achievement} in {Sweden}},
    volume = {82},
    issn = {0003-1224},
    shorttitle = {Grand {Advantage}},
    doi = {10.1177/0003122417695791},
    language = {eng},
    number = {2},
    journal = {American Sociological Review},
    author = {Hällsten, Martin and Pfeffer, Fabian T.},
    month = apr,
    year = {2017},
    pmid = {29200464},
    pmcid = {PMC5703428},
    pages = {328--360},
}

@incollection{erikson_how_2019,
    title = {How does education depend on social origin?},
    author = {Erikson, Robert},
    editor = {Becker, Rolf},
    booktitle = {Research Handbook on the Sociology of Education},
    publisher = {Edward Elgar Publishing},
    location = {Cheltenham},
    year = {2019},
    pages = {35--56},
    isbn = {978-1-78811-042-6},
    doi = {10.4337/9781788110426.00010},
}

@incollection{blanden_educational_2023,
    title = {Educational Inequality},
    author = {Blanden, Jo and Doepke, Matthias and Stuhler, Jan},
    editor = {Hanushek, Eric A. and Woessmann, Ludger and Machin, Stephen J.},
    booktitle = {Handbook of the Economics of Education},
    volume = {6},
    publisher = {Elsevier},
    year = {2023},
    pages = {405--497},
    doi = {10.1016/bs.hesedu.2022.11.003},
}

@report{coleman_equality_1966,
    title = {Equality of Educational Opportunity},
    author = {Coleman, James S. and Campbell, Ernest Q. and Hobson, Carol J. and McPartland, James and Mood, Alexander M. and Weinfeld, Frederic D. and York, Robert L.},
    institution = {U.S. Department of Health, Education, and Welfare, Office of Education},
    location = {Washington, DC},
    publisher = {U.S. Government Printing Office},
    year = {1966},
    url = {https://eric.ed.gov/?id=ED012275},
}

@article{owens_neighborhoods_2010,
    title = {Neighborhoods and {Schools} as {Competing} and {Reinforcing} {Contexts} for {Educational} {Attainment}},
    volume = {83},
    issn = {0038-0407},
    url = {https://doi.org/10.1177/0038040710383519},
    doi = {10.1177/0038040710383519},
    language = {en},
    number = {4},
    urldate = {2024-01-26},
    journal = {Sociology of Education},
    author = {Owens, Ann},
    month = oct,
    year = {2010},
    note = {Publisher: SAGE Publications Inc},
    pages = {287--311},
}

@article{hermansen_long-term_2020,
    title = {Long-{Term} {Trends} in {Adult} {Socio}-{Economic} {Resemblance} between {Former} {Schoolmates} and {Neighbouring} {Children}},
    volume = {36},
    issn = {0266-7215},
    url = {https://doi.org/10.1093/esr/jcz066},
    doi = {10.1093/esr/jcz066},
    number = {3},
    urldate = {2024-01-03},
    journal = {European Sociological Review},
    author = {Hermansen, Are Skeie and Borgen, Nicolai T and Mastekaasa, Arne},
    month = jun,
    year = {2020},
    pages = {366--380},
}

@article{massey_american_1990,
    title = {American {Apartheid}: {Segregation} and the {Making} of the {Underclass}},
    volume = {96},
    issn = {0002-9602, 1537-5390},
    shorttitle = {American {Apartheid}},
    url = {https://www.journals.uchicago.edu/doi/10.1086/229532},
    doi = {10.1086/229532},
    language = {en},
    number = {2},
    urldate = {2024-01-26},
    journal = {American Journal of Sociology},
    author = {Massey, Douglas S.},
    month = sep,
    year = {1990},
    pages = {329--357},
}

@article{ainsworth_why_2002,
    title = {Why does it take a village? {The} mediation of neighborhood effects on educational achievement},
    volume = {81},
    shorttitle = {Why does it take a village?},
    url = {https://academic.oup.com/sf/article-abstract/81/1/117/2234402},
    number = {1},
    urldate = {2024-01-26},
    journal = {Social forces},
    author = {Ainsworth, James W.},
    year = {2002},
    note = {Publisher: The University of North Carolina Press},
    pages = {117--152},
    doi = {10.1353/sof.2002.0038},
}

@article{sharkey_where_2014,
    title = {Where, {When}, {Why}, and {For} {Whom} {Do} {Residential} {Contexts} {Matter}? {Moving} {Away} from the {Dichotomous} {Understanding} of {Neighborhood} {Effects}},
    volume = {40},
    issn = {0360-0572, 1545-2115},
    shorttitle = {Where, {When}, {Why}, and {For} {Whom} {Do} {Residential} {Contexts} {Matter}?},
    url = {https://www.annualreviews.org/doi/10.1146/annurev-soc-071913-043350},
    doi = {10.1146/annurev-soc-071913-043350},
    language = {en},
    number = {1},
    urldate = {2024-01-03},
    journal = {Annual Review of Sociology},
    author = {Sharkey, Patrick and Faber, Jacob W.},
    month = jul,
    year = {2014},
    pages = {559--579},
}

@article{hedefalk_social_2020,
    title = {The social context of nearest neighbors shapes educational attainment regardless of class origin},
    volume = {117},
    url = {https://www.pnas.org/doi/abs/10.1073/pnas.1922532117},
    doi = {10.1073/pnas.1922532117},
    number = {26},
    urldate = {2023-12-15},
    journal = {Proceedings of the National Academy of Sciences},
    author = {Hedefalk, Finn and Dribe, Martin},
    month = jun,
    year = {2020},
    note = {Publisher: Proceedings of the National Academy of Sciences},
    pages = {14918--14925},
}

@article{lee_success_2014,
    title = {The {Success} {Frame} and {Achievement} {Paradox}: {The} {Costs} and {Consequences} for {Asian} {Americans}},
    volume = {6},
    issn = {1867-1756},
    shorttitle = {The {Success} {Frame} and {Achievement} {Paradox}},
    url = {https://doi.org/10.1007/s12552-014-9112-7},
    doi = {10.1007/s12552-014-9112-7},
    language = {en},
    number = {1},
    urldate = {2024-01-26},
    journal = {Race and Social Problems},
    author = {Lee, Jennifer and Zhou, Min},
    month = mar,
    year = {2014},
    pages = {38--55},
}

@article{solon_correlations_2000,
    title = {Correlations between neighboring children in their subsequent educational attainment},
    volume = {82},
    url = {https://direct.mit.edu/rest/article-abstract/82/3/383/57192},
    number = {3},
    urldate = {2024-01-26},
    journal = {Review of Economics and Statistics},
    author = {Solon, Gary and Page, Marianne E. and Duncan, Greg J.},
    year = {2000},
    note = {Publisher: MIT Press 238 Main St., Suite 500, Cambridge, MA 02142-1046, USA journals …},
    pages = {383--392},
    doi = {10.1162/003465300558885},
}

@article{page_correlations_2003,
    title = {Correlations between {Brothers} and {Neighboring} {Boys} in {Their} {Adult} {Earnings}: {The} {Importance} of {Being} {Urban}},
    volume = {21},
    issn = {0734-306X, 1537-5307},
    shorttitle = {Correlations between {Brothers} and {Neighboring} {Boys} in {Their} {Adult} {Earnings}},
    url = {https://www.journals.uchicago.edu/doi/10.1086/377021},
    doi = {10.1086/377021},
    language = {en},
    number = {4},
    urldate = {2024-01-26},
    journal = {Journal of Labor Economics},
    author = {Page, Marianne E. and Solon, Gary},
    month = oct,
    year = {2003},
    pages = {831--855},
}

@incollection{altonji_role_2011,
    title = {The role of family, school, and community characteristics in inequality in education and labor-market outcomes},
    author = {Altonji, Joseph G. and Mansfield, Richard K.},
    editor = {Duncan, Greg J. and Murnane, Richard J.},
    booktitle = {Whither Opportunity? Rising Inequality, Schools, and Children's Life Chances},
    publisher = {Russell Sage Foundation},
    location = {New York},
    year = {2011},
    pages = {339--358},
}

@article{kuyvenhoven_neighbourhood_2021,
    title = {Neighbourhood and school effects on educational inequalities in the transition from primary to secondary education in {Amsterdam}},
    volume = {58},
    issn = {0042-0980},
    url = {https://doi.org/10.1177/0042098020959011},
    doi = {10.1177/0042098020959011},
    language = {EN},
    number = {13},
    urldate = {2025-04-02},
    journal = {Urban Studies},
    author = {Kuyvenhoven, Joeke and Boterman, Willem R.},
    month = oct,
    year = {2021},
    note = {Publisher: SAGE Publications Ltd},
    pages = {2660--2682},
}

@article{chung_peers_2020,
    title = {Peers’ parents and educational attainment: {The} exposure effect},
    volume = {64},
    issn = {09275371},
    shorttitle = {Peers’ parents and educational attainment},
    url = {https://linkinghub.elsevier.com/retrieve/pii/S092753712030018X},
    doi = {10.1016/j.labeco.2020.101812},
    language = {en},
    urldate = {2024-01-29},
    journal = {Labour Economics},
    author = {Chung, Bobby W.},
    month = jun,
    year = {2020},
    pages = {101812},
}

@article{bernardi_understanding_2016,
    title = {Understanding heterogeneity in the effects of parental separation on educational attainment in {Britain}: {Do} children from lower educational backgrounds have less to lose?},
    volume = {32},
    shorttitle = {Understanding heterogeneity in the effects of parental separation on educational attainment in {Britain}},
    url = {https://academic.oup.com/esr/article-abstract/32/6/807/2525511},
    number = {6},
    urldate = {2024-01-29},
    journal = {European Sociological Review},
    author = {Bernardi, Fabrizio and Boertien, Diederik},
    year = {2016},
    note = {Publisher: Oxford University Press},
    pages = {807--819},
    doi = {10.1093/esr/jcw036},
}

@article{gratz_when_2015,
    title = {When {Growing} {Up} {Without} a {Parent} {Does} {Not} {Hurt}: {Parental} {Separation} and the {Compensatory} {Effect} of {Social} {Origin}},
    volume = {31},
    issn = {0266-7215},
    shorttitle = {When {Growing} {Up} {Without} a {Parent} {Does} {Not} {Hurt}},
    url = {https://doi.org/10.1093/esr/jcv057},
    doi = {10.1093/esr/jcv057},
    number = {5},
    urldate = {2024-01-29},
    journal = {European Sociological Review},
    author = {Grätz, Michael},
    month = oct,
    year = {2015},
    pages = {546--557},
}

@article{lang_does_2001,
    title = {Does growing up with a parent absent really hurt?},
    url = {https://www.jstor.org/stable/3069659?casa_token=-ZXzfslYglcAAAAA:vTS0XF_pAFjSTOxm_S-j0rOUY--JSvRzEKTg2nY0NpaGVMwPGMRDRtapwBfA3QbipipRwqaze6vCutRO6pWjZBAd1zmuw5Fnz0p6ABIbL4kAsdIxp2Xj},
    urldate = {2024-01-29},
    volume = {36},
    number = {2},
    journal = {Journal of Human Resources},
    author = {Lang, Kevin and Zagorsky, Jay L.},
    year = {2001},
    note = {Publisher: JSTOR},
    pages = {253--273},
    doi = {10.2307/3069659},
}

@article{monserud_household_2011,
    title = {Household {Structure} and {Children}'s {Educational} {Attainment}: {A} {Perspective} on {Coresidence} with {Grandparents}},
    volume = {73},
    copyright = {Copyright © National Council on Family Relations, 2011},
    issn = {1741-3737},
    shorttitle = {Household {Structure} and {Children}'s {Educational} {Attainment}},
    url = {https://onlinelibrary.wiley.com/doi/abs/10.1111/j.1741-3737.2011.00858.x},
    doi = {10.1111/j.1741-3737.2011.00858.x},
    language = {en},
    number = {5},
    urldate = {2024-01-29},
    journal = {Journal of Marriage and Family},
    author = {Monserud, Maria A. and Elder Jr., Glen H.},
    year = {2011},
    note = {\_eprint: https://onlinelibrary.wiley.com/doi/pdf/10.1111/j.1741-3737.2011.00858.x},
    pages = {981--1000},
}

@article{hindman_building_2015,
    title = {Building {Better} {Models}: {Prediction}, {Replication}, and {Machine} {Learning} in the {Social} {Sciences}},
    volume = {659},
    issn = {0002-7162},
    shorttitle = {Building {Better} {Models}},
    url = {https://doi.org/10.1177/0002716215570279},
    doi = {10.1177/0002716215570279},
    language = {en},
    number = {1},
    urldate = {2023-01-21},
    journal = {The ANNALS of the American Academy of Political and Social Science},
    author = {Hindman, Matthew},
    month = may,
    year = {2015},
    note = {Publisher: SAGE Publications Inc},
    pages = {48--62},
}

@article{brandt_abductive_2021,
    title = {Abductive {Logic} of {Inquiry} for {Quantitative} {Research} in the {Digital} {Age}},
    volume = {8},
    issn = {2330-6696},
    url = {https://sociologicalscience.com/articles-v8-10-191/},
    doi = {10.15195/v8.a10},
    language = {en-US},
    urldate = {2025-04-02},
    journal = {Sociological Science},
    author = {Brandt, Philipp and Timmermans, Stefan},
    month = jun,
    year = {2021},
    pages = {191--210},
}

@article{knigge_delayed_2022,
    title = {Delayed tracking and inequality of opportunity: {Gene}-environment interactions in educational attainment},
    volume = {7},
    copyright = {2022 The Author(s)},
    issn = {2056-7936},
    shorttitle = {Delayed tracking and inequality of opportunity},
    url = {https://www.nature.com/articles/s41539-022-00122-1},
    doi = {10.1038/s41539-022-00122-1},
    language = {en},
    number = {1},
    urldate = {2025-04-15},
    journal = {npj Science of Learning},
    author = {Knigge, Antonie and Maas, Ineke and Stienstra, Kim and de Zeeuw, Eveline L. and Boomsma, Dorret I.},
    month = may,
    year = {2022},
    note = {Publisher: Nature Publishing Group},
    pages = {6},
}

@article{atav_persistent_2026,
    title = {The persistent relationship between family background and educational attainment in the {Netherlands}},
    author = {Atav, Tilbe and Rietveld, Cornelius A. and van Kippersluis, Hans},
    journal = {The Journal of Economic Inequality},
    year = {2026},
    doi = {10.1007/s10888-026-09725-9},
    url = {https://doi.org/10.1007/s10888-026-09725-9},
}

@article{keskiner_is_2015,
    title = {“{Is} it {Merit} or {Cultural} {Capital}?” {The} role of parents during early tracking in {Amsterdam} and {Strasbourg} among descendants of immigrants from {Turkey}},
    volume = {3},
    issn = {2214-594X},
    shorttitle = {“{Is} it {Merit} or {Cultural} {Capital}?},
    url = {https://doi.org/10.1186/s40878-015-0014-7},
    doi = {10.1186/s40878-015-0014-7},
    number = {1},
    urldate = {2025-04-02},
    journal = {Comparative Migration Studies},
    author = {Keskiner, Elif},
    month = oct,
    year = {2015},
    pages = {9},
}

@article{van_de_werfhorst_ethnicity_2007,
    title = {Ethnicity, schooling, and merit in the {Netherlands}},
    volume = {7},
    copyright = {https://journals.sagepub.com/page/policies/text-and-data-mining-license},
    issn = {1468-7968, 1741-2706},
    url = {https://journals.sagepub.com/doi/10.1177/1468796807080236},
    doi = {10.1177/1468796807080236},
    language = {en},
    number = {3},
    urldate = {2025-04-22},
    journal = {Ethnicities},
    author = {Van De Werfhorst, Herman G. and Van Tubergen, Frank},
    month = sep,
    year = {2007},
    pages = {416--444},
}

@article{troost_neighbourhood_2023,
    title = {Neighbourhood effects on educational attainment. {What} matters more: {Exposure} to poverty or exposure to affluence?},
    volume = {18},
    issn = {1932-6203},
    shorttitle = {Neighbourhood effects on educational attainment. {What} matters more},
    url = {https://www.ncbi.nlm.nih.gov/pmc/articles/PMC9994736/},
    doi = {10.1371/journal.pone.0281928},
    number = {3},
    urldate = {2025-04-02},
    journal = {PLOS ONE},
    author = {Troost, Agata A. and van Ham, Maarten and Manley, David J.},
    month = mar,
    year = {2023},
    pmid = {36888593},
    pmcid = {PMC9994736},
    pages = {e0281928},
}

@article{pankowska2024potential,
    title = {The potential of benchmark challenges in the social sciences},
    volume = {63},
    number = {4},
    journal = {Social Science Information},
    author = {Pankowska, Paulina and Mendrik, Adrienne and Emery, Tom and Garcia-Bernardo, Javier},
    year = {2024},
    note = {Publisher: SAGE Publications Sage UK: London, England},
    pages = {498--519},
    doi = {10.1177/05390184241297742},
}

@article{engzell_understanding_2023,
    title = {Understanding {Patterns} and {Trends} in {Income} {Mobility} through {Multiverse} {Analysis}},
    volume = {88},
    issn = {0003-1224},
    url = {https://doi.org/10.1177/00031224231180607},
    doi = {10.1177/00031224231180607},
    language = {EN},
    number = {4},
    urldate = {2025-04-02},
    journal = {American Sociological Review},
    author = {Engzell, Per and Mood, Carina},
    month = aug,
    year = {2023},
    note = {Publisher: SAGE Publications Inc},
    pages = {600--626},
}

@article{verhagen_incorporating_2024,
    title = {Incorporating {Machine} {Learning} into {Sociological} {Model}-{Building}},
    issn = {0081-1750, 1467-9531},
    url = {http://journals.sagepub.com/doi/10.1177/00811750231217734},
    doi = {10.1177/00811750231217734},
    language = {en},
    urldate = {2024-05-22},
    journal = {Sociological Methodology},
    author = {Verhagen, Mark D.},
    month = jan,
    year = {2024},
    pages = {217--268},
    volume = {54},
    number = {2}
}

@article{scarselli_graph_2009,
    title = {The graph neural network model},
    volume = {20},
    number = {1},
    journal = {IEEE transactions on neural networks},
    author = {Scarselli, Franco and Gori, Marco and Tsoi, Ah Chung and Hagenbuchner, Markus and Monfardini, Gabriele},
    year = {2009},
    note = {Publisher: IEEE},
    pages = {61--80},
    doi = {10.1109/TNN.2008.2005605},
}

@article{portes_new_1993,
    title = {The {New} {Second} {Generation}: {Segmented} {Assimilation} and its {Variants}},
    volume = {530},
    copyright = {https://journals.sagepub.com/page/policies/text-and-data-mining-license},
    issn = {0002-7162, 1552-3349},
    shorttitle = {The {New} {Second} {Generation}},
    url = {https://journals.sagepub.com/doi/10.1177/0002716293530001006},
    doi = {10.1177/0002716293530001006},
    language = {en},
    number = {1},
    urldate = {2025-06-22},
    journal = {The ANNALS of the American Academy of Political and Social Science},
    author = {Portes, Alejandro and Zhou, Min},
    month = nov,
    year = {1993},
    pages = {74--96},
}

@article{van_der_laan_whole_2022,
    title = {A whole population network and its application for the social sciences},
    issn = {0266-7215},
    url = {https://doi.org/10.1093/esr/jcac026},
    doi = {10.1093/esr/jcac026},
    urldate = {2023-01-12},
    journal = {European Sociological Review},
    author = {van der Laan, Jan and de Jonge, Edwin and Das, Marjolijn and Te Riele, Saskia and Emery, Tom},
    month = jun,
    year = {2023},
    volume = {39},
    number = {1},
    pages = {145--160},
}

@article{bokanyi_anatomy_2023,
    title = {The anatomy of a population-scale social network},
    volume = {13},
    number = {1},
    journal = {Scientific Reports},
    author = {Bokányi, Eszter and Heemskerk, Eelke M. and Takes, Frank W.},
    year = {2023},
    pages = {9209},
    doi = {10.1038/s41598-023-36324-9},
}

@software{garcia-bernardo_netcbs_2024,
    title = {{netCBS}: Package to efficiently create network measures using {CBS} networks in the {RA}},
    author = {Garcia-Bernardo, Javier},
    publisher = {Zenodo},
    year = {2024},
    doi = {10.5281/zenodo.13908121},
    url = {https://zenodo.org/records/13908121},
}

@article{pedregosa_scikit-learn_2011,
    title = {Scikit-learn: Machine Learning in {Python}},
    author = {Pedregosa, Fabian and Varoquaux, Ga{\"e}l and Gramfort, Alexandre and Michel, Vincent and Thirion, Bertrand and Grisel, Olivier and Blondel, Mathieu and Prettenhofer, Peter and Weiss, Ron and Dubourg, Vincent and Vanderplas, Jake and Passos, Alexandre and Cournapeau, David and Brucher, Matthieu and Perrot, Matthieu and Duchesnay, {\'E}douard},
    journal = {Journal of Machine Learning Research},
    volume = {12},
    number = {85},
    pages = {2825--2830},
    year = {2011},
    url = {https://jmlr.org/papers/v12/pedregosa11a.html},
    eprint = {1201.0490},
    eprinttype = {arxiv},
    doi = {10.48550/arXiv.1201.0490},
}

@inproceedings{hamilton_inductive_2017,
    title = {Inductive Representation Learning on Large Graphs},
    author = {Hamilton, Will and Ying, Zhitao and Leskovec, Jure},
    booktitle = {Advances in Neural Information Processing Systems},
    volume = {30},
    year = {2017},
    url = {https://proceedings.neurips.cc/paper/2017/hash/5dd9db5e033da9c6fb5ba83c7a7ebea9-Abstract.html},
    eprint = {1706.02216},
    eprinttype = {arxiv},
    doi = {10.48550/arXiv.1706.02216},
}

@article{savcisens_using_2024,
    title = {Using sequences of life-events to predict human lives},
    volume = {4},
    copyright = {2023 The Author(s), under exclusive licence to Springer Nature America, Inc.},
    issn = {2662-8457},
    url = {https://www.nature.com/articles/s43588-023-00573-5},
    doi = {10.1038/s43588-023-00573-5},
    language = {en},
    number = {1},
    urldate = {2025-04-15},
    journal = {Nature Computational Science},
    author = {Savcisens, Germans and Eliassi-Rad, Tina and Hansen, Lars Kai and Mortensen, Laust Hvas and Lilleholt, Lau and Rogers, Anna and Zettler, Ingo and Lehmann, Sune},
    month = jan,
    year = {2024},
    note = {Publisher: Nature Publishing Group},
    pages = {43--56},
}

@article{chicco_advantages_2020,
    title = {The advantages of the {Matthews} correlation coefficient ({MCC}) over {F1} score and accuracy in binary classification evaluation},
    volume = {21},
    issn = {1471-2164},
    url = {https://doi.org/10.1186/s12864-019-6413-7},
    doi = {10.1186/s12864-019-6413-7},
    number = {1},
    urldate = {2025-04-15},
    journal = {BMC Genomics},
    author = {Chicco, Davide and Jurman, Giuseppe},
    month = jan,
    year = {2020},
    pages = {6},
}

@techreport{lundberg_father_2017,
    title = {Father Absence and the Educational Gender Gap},
    author = {Lundberg, Shelly},
    institution = {IZA Institute of Labor Economics},
    type = {IZA Discussion Paper},
    number = {10814},
    year = {2017},
    month = may,
    doi = {10.2139/ssrn.2979961},
    url = {https://docs.iza.org/dp10814.pdf},
}

@article{boertien_gendered_2022,
    title = {Gendered {Diverging} {Destinies}: {Changing} {Family} {Structures} and the {Reproduction} of {Educational} {Inequalities} {Among} {Sons} and {Daughters} in the {United} {States}},
    volume = {59},
    issn = {0070-3370},
    shorttitle = {Gendered {Diverging} {Destinies}},
    url = {https://doi.org/10.1215/00703370-9612710},
    doi = {10.1215/00703370-9612710},
    number = {1},
    urldate = {2025-04-25},
    journal = {Demography},
    author = {Boertien, Diederik and Bernardi, Fabrizio},
    month = feb,
    year = {2022},
    pages = {111--136},
}

@misc{cbs_persoonskenmerken_2022,
    title = {Persoonskenmerken van alle in de {Gemeentelijke} {Basisadministratie} {Persoonsgegevens} ({GBA}) ingeschreven personen, niet-gecoördineerd},
    copyright = {https://portal.odissei.nl/api/datasets/:persistentId/versions/1.0/customlicense?persistentId=doi:10.57934/0b01e4108071daf2},
    url = {https://portal.odissei.nl/dataset.xhtml?persistentId=doi:10.57934/0b01e4108071daf2},
    doi = {10.57934/0b01e4108071daf2},
    language = {en},
    urldate = {2025-05-26},
    publisher = {ODISSEI Portal},
    author = {{Centraal Bureau voor de Statistiek}},
    month = feb,
    year = {2022},
}

@misc{cbs_hoogst_2019,
    title = {Hoogst behaald en hoogst gevolgd opleidingsniveau en opleidingsrichting van de bevolking in {Nederland}},
    copyright = {https://portal.odissei.nl/api/datasets/:persistentId/versions/1.0/customlicense?persistentId=doi:10.57934/0b01e410806c9615},
    url = {https://portal.odissei.nl/dataset.xhtml?persistentId=doi:10.57934/0b01e410806c9615},
    doi = {10.57934/0b01e410806c9615},
    language = {en},
    urldate = {2025-05-26},
    publisher = {ODISSEI Portal},
    author = {{Centraal Bureau voor de Statistiek}},
    month = oct,
    year = {2019},
}

@misc{cbs_inkomen_2011,
    title = {Inkomen van personen},
    copyright = {https://portal.odissei.nl/api/datasets/:persistentId/versions/1.0/customlicense?persistentId=doi:10.57934/0b01e41080372fbd},
    url = {https://portal.odissei.nl/dataset.xhtml?persistentId=doi:10.57934/0b01e41080372fbd},
    doi = {10.57934/0b01e41080372fbd},
    language = {en},
    urldate = {2025-05-26},
    publisher = {ODISSEI Portal},
    author = {{Centraal Bureau voor de Statistiek}},
    month = jan,
    year = {2011},
}

@misc{cbs_inschrijvingen_2014,
    title = {Inschrijvingen van leerlingen in het basisonderwijs},
    copyright = {https://portal.odissei.nl/api/datasets/:persistentId/versions/1.0/customlicense?persistentId=doi:10.57934/0b01e410802d7a90},
    url = {https://portal.odissei.nl/dataset.xhtml?persistentId=doi:10.57934/0b01e410802d7a90},
    doi = {10.57934/0b01e410802d7a90},
    language = {en},
    urldate = {2025-05-26},
    publisher = {ODISSEI Portal},
    author = {{Centraal Bureau voor de Statistiek}},
    month = sep,
    year = {2014},
}

@misc{cbs_familienetwerk_2009,
    title = {Familienetwerk},
    copyright = {https://portal.odissei.nl/api/datasets/:persistentId/versions/1.0/customlicense?persistentId=doi:10.57934/0b01e41080760802},
    url = {https://portal.odissei.nl/dataset.xhtml?persistentId=doi:10.57934/0b01e41080760802},
    doi = {10.57934/0b01e41080760802},
    language = {en},
    urldate = {2025-05-26},
    publisher = {ODISSEI Portal},
    author = {{Centraal Bureau voor de Statistiek}},
    month = jan,
    year = {2009},
}

@misc{cbs_burennetwerk_2009,
    title = {Burennetwerk},
    copyright = {https://portal.odissei.nl/api/datasets/:persistentId/versions/1.0/customlicense?persistentId=doi:10.57934/0b01e410807607b7},
    url = {https://portal.odissei.nl/dataset.xhtml?persistentId=doi:10.57934/0b01e410807607b7},
    doi = {10.57934/0b01e410807607b7},
    language = {en},
    urldate = {2025-05-26},
    publisher = {ODISSEI Portal},
    author = {{Centraal Bureau voor de Statistiek}},
    month = jan,
    year = {2009},
}

@misc{cbs_huisgenotennetwerk_2009,
    title = {Huisgenotennetwerk},
    copyright = {https://portal.odissei.nl/api/datasets/:persistentId/versions/1.0/customlicense?persistentId=doi:10.57934/0b01e4108076077c},
    url = {https://portal.odissei.nl/dataset.xhtml?persistentId=doi:10.57934/0b01e4108076077c},
    doi = {10.57934/0b01e4108076077c},
    language = {en},
    urldate = {2025-05-26},
    publisher = {ODISSEI Portal},
    author = {{Centraal Bureau voor de Statistiek}},
    month = jan,
    year = {2009},
}

@misc{cbs_klasgenotennetwerk_2009,
    title = {Klasgenotennetwerk},
    copyright = {https://portal.odissei.nl/api/datasets/:persistentId/versions/1.0/customlicense?persistentId=doi:10.57934/0b01e4108075f996},
    url = {https://portal.odissei.nl/dataset.xhtml?persistentId=doi:10.57934/0b01e4108075f996},
    doi = {10.57934/0b01e4108075f996},
    language = {en},
    urldate = {2025-05-26},
    publisher = {ODISSEI Portal},
    author = {{Centraal Bureau voor de Statistiek}},
    month = jan,
    year = {2009},
}

@misc{cbs_gemeente-_2017,
    title = {Gemeente-, wijk- en buurtcodes van een verblijfsobject},
    copyright = {https://portal.odissei.nl/api/datasets/:persistentId/versions/1.0/customlicense?persistentId=doi:10.57934/0b01e41080383bfc},
    url = {https://portal.odissei.nl/dataset.xhtml?persistentId=doi:10.57934/0b01e41080383bfc},
    doi = {10.57934/0b01e41080383bfc},
    language = {en},
    urldate = {2025-05-26},
    publisher = {ODISSEI Portal},
    author = {{Centraal Bureau voor de Statistiek}},
    month = jan,
    year = {2017},
}

@article{thaning_end_2020,
    title = {The end of dominance? {Evaluating} measures of socio-economic background in stratification research},
    volume = {36},
    shorttitle = {The end of dominance?},
    url = {https://academic.oup.com/esr/article-abstract/36/4/533/5841732},
    number = {4},
    urldate = {2025-06-18},
    journal = {European Sociological Review},
    author = {Thaning, Max and Hällsten, Martin},
    year = {2020},
    note = {Publisher: Oxford University Press},
    pages = {533--547},
    doi = {10.1093/esr/jcaa009},
}

@inproceedings{akiba_optuna_2019,
    address = {New York, NY, USA},
    series = {{KDD} '19},
    title = {Optuna: {A} {Next}-generation {Hyperparameter} {Optimization} {Framework}},
    isbn = {978-1-4503-6201-6},
    shorttitle = {Optuna},
    url = {https://dl.acm.org/doi/10.1145/3292500.3330701},
    doi = {10.1145/3292500.3330701},
    urldate = {2025-05-05},
    booktitle = {Proceedings of the 25th {ACM} {SIGKDD} {International} {Conference} on {Knowledge} {Discovery} \& {Data} {Mining}},
    publisher = {Association for Computing Machinery},
    author = {Akiba, Takuya and Sano, Shotaro and Yanase, Toshihiko and Ohta, Takeru and Koyama, Masanori},
    month = jul,
    year = {2019},
    pages = {2623--2631},
}

@article{arseniev-koehler_machine_2022,
    title = {Machine {Learning} as a {Model} for {Cultural} {Learning}: {Teaching} an {Algorithm} {What} it {Means} to be {Fat}},
    volume = {51},
    issn = {0049-1241, 1552-8294},
    shorttitle = {Machine {Learning} as a {Model} for {Cultural} {Learning}},
    url = {https://journals.sagepub.com/doi/10.1177/00491241221122603},
    doi = {10.1177/00491241221122603},
    language = {en},
    number = {4},
    urldate = {2025-09-22},
    journal = {Sociological Methods \& Research},
    author = {Arseniev-Koehler, Alina and Foster, Jacob G.},
    month = nov,
    year = {2022},
    pages = {1484--1539},
}

@article{peterson_using_2021,
    title = {Using large-scale experiments and machine learning to discover theories of human decision-making},
    volume = {372},
    url = {https://www.science.org/doi/abs/10.1126/science.abe2629},
    doi = {10.1126/science.abe2629},
    number = {6547},
    urldate = {2023-01-20},
    journal = {Science},
    author = {Peterson, Joshua C. and Bourgin, David D. and Agrawal, Mayank and Reichman, Daniel and Griffiths, Thomas L.},
    month = jun,
    year = {2021},
    note = {Publisher: American Association for the Advancement of Science},
    pages = {1209--1214},
}

@article{chettySocialCapitalMeasurement2022,
  title = {Social Capital {{I}}: Measurement and Associations with Economic Mobility},
  shorttitle = {Social Capital {{I}}},
  author = {Chetty, Raj and Jackson, Matthew O. and Kuchler, Theresa and Stroebel, Johannes and Hendren, Nathaniel and Fluegge, Robert B. and Gong, Sara and Gonzalez, Federico and Grondin, Armelle and Jacob, Matthew and Johnston, Drew and Koenen, Martin and Laguna-Muggenburg, Eduardo and Mudekereza, Florian and Rutter, Tom and Thor, Nicolaj and Townsend, Wilbur and Zhang, Ruby and Bailey, Mike and Barberá, Pablo and Bhole, Monica and Wernerfelt, Nils},
  date = {2022-08-04},
  journaltitle = {Nature},
  shortjournal = {Nature},
  volume = {608},
  number = {7921},
  pages = {108--121},
  issn = {0028-0836, 1476-4687},
  doi = {10.1038/s41586-022-04996-4},
  url = {https://www.nature.com/articles/s41586-022-04996-4},
  urldate = {2022-11-28},
  langid = {english}
}

@book{rosenthal_pygmalion_1968,
    title = {Pygmalion in the Classroom: Teacher Expectation and Pupils' Intellectual Development},
    author = {Rosenthal, Robert and Jacobson, Lenore},
    publisher = {Holt, Rinehart and Winston},
    location = {New York},
    year = {1968},
}

@article{chettyMeasuringImpactsTeachers2014,
  title = {Measuring the {{Impacts}} of {{Teachers II}}: {{Teacher Value-Added}} and {{Student Outcomes}} in {{Adulthood}}},
  shorttitle = {Measuring the {{Impacts}} of {{Teachers II}}},
  author = {Chetty, Raj and Friedman, John N. and Rockoff, Jonah E.},
  date = {2014-09},
  journaltitle = {American Economic Review},
  shortjournal = {Am. Econ. Rev.},
  volume = {104},
  number = {9},
  pages = {2633--2679},
  issn = {0002-8282},
  doi = {10.1257/aer.104.9.2633},
  url = {https://www.aeaweb.org/articles?id=10.1257/aer.104.9.2633},
  urldate = {2026-10-02},
  langid = {english}
}

@article{hamreCanInstructionalEmotional2005,
  title = {Can Instructional and Emotional Support in the First-Grade Classroom Make a Difference for Children at Risk of School Failure?},
  author = {Hamre, Bridget K. and Pianta, Robert C.},
  date = {2005},
  journaltitle = {Child development},
  shortjournal = {Child Dev.},
  volume = {76},
  number = {5},
  pages = {949--967},
  url = {https://academic.oup.com/chidev/article-abstract/76/5/949/8275336},
  urldate = {2026-10-02},
    doi = {10.1111/j.1467-8624.2005.00889.x},
}

@article{jacksonEffectsSchoolSpending2016,
  title = {The {{Effects}} of {{School Spending}} on {{Educational}} and {{Economic Outcomes}}: {{Evidence}} from {{School Finance Reforms}}},
  shorttitle = {The {{Effects}} of {{School Spending}} on {{Educational}} and {{Economic Outcomes}}},
  author = {Jackson, C. Kirabo and Johnson, Rucker C. and Persico, Claudia},
  date = {2016-02-01},
  journaltitle = {The Quarterly Journal of Economics},
  shortjournal = {Q. J. Econ.},
  volume = {131},
  number = {1},
  pages = {157--218},
  issn = {0033-5533},
  doi = {10.1093/qje/qjv036},
  url = {https://doi.org/10.1093/qje/qjv036},
  urldate = {2026-10-02}
}

@article{jaegerExtendedFamilyChildren2012,
  title = {The {{Extended Family}} and {{Children}}’s {{Educational Success}}},
  author = {Jæger, Mads Meier},
  date = {2012-12-01},
  journaltitle = {American Sociological Review},
  shortjournal = {Am. Sociol. Rev.},
  volume = {77},
  number = {6},
  pages = {903--922},
  publisher = {SAGE Publications Inc},
  issn = {0003-1224},
  doi = {10.1177/0003122412464040},
  url = {https://doi.org/10.1177/0003122412464040},
  urldate = {2026-10-02},
  langid = {english}
}

@article{jussimTeacherExpectationsSelfFulfilling2005,
  title = {Teacher {{Expectations}} and {{Self-Fulfilling Prophecies}}: {{Knowns}} and {{Unknowns}}, {{Resolved}} and {{Unresolved Controversies}}},
  shorttitle = {Teacher {{Expectations}} and {{Self-Fulfilling Prophecies}}},
  author = {Jussim, Lee and Harber, Kent D.},
  date = {2005-05-01},
  journaltitle = {Personality and Social Psychology Review},
  shortjournal = {Personal. Soc. Psychol. Rev.},
  volume = {9},
  number = {2},
  pages = {131--155},
  publisher = {SAGE Publications Inc},
  issn = {1088-8683},
  doi = {10.1207/s15327957pspr0902_3},
  url = {https://doi.org/10.1207/s15327957pspr0902_3},
  urldate = {2026-10-02},
  langid = {english}
}

@article{leventhalNeighborhoodsTheyLive2000,
  title = {The Neighborhoods They Live in: The Effects of Neighborhood Residence on Child and Adolescent Outcomes.},
  shorttitle = {The Neighborhoods They Live In},
  author = {Leventhal, Tama and Brooks-Gunn, Jeanne},
  date = {2000},
  journaltitle = {Psychological bulletin},
  shortjournal = {Psychol. Bull.},
  volume = {126},
  number = {2},
  pages = {309--337},
  publisher = {American Psychological Association},
  url = {https://psycnet.apa.org/record/2000-03444-006},
  urldate = {2026-10-02},
    doi = {10.1037/0033-2909.126.2.309},
}

@book{linSocialCapitalTheory2001,
    title = {Social Capital: A Theory of Social Structure and Action},
    author = {Lin, Nan},
    publisher = {Cambridge University Press},
    location = {Cambridge},
    year = {2001},
    doi = {10.1017/CBO9780511815447},
}

@article{mareMultigenerationalViewInequality2011,
  title = {A {{Multigenerational View}} of {{Inequality}}},
  author = {Mare, Robert D.},
  date = {2011-01-27},
  journaltitle = {Demography},
  volume = {48},
  number = {1},
  pages = {1--23},
  issn = {0070-3370},
  doi = {10.1007/s13524-011-0014-7},
  url = {https://doi.org/10.1007/s13524-011-0014-7},
  urldate = {2026-10-02}
}

@book{wilson_truly_1987,
    title = {The Truly Disadvantaged: The Inner City, the Underclass, and Public Policy},
    author = {Wilson, William Julius},
    publisher = {University of Chicago Press},
    location = {Chicago},
    year = {1987},
    isbn = {9780226901268},
}

@incollection{bourdieu1986forms,
    title = {The Forms of Capital},
    author = {Bourdieu, Pierre},
    editor = {Richardson, John G.},
    booktitle = {Handbook of Theory and Research for the Sociology of Education},
    publisher = {Greenwood},
    location = {New York},
    year = {1986},
    pages = {241--258},
}

@article{portesSocialCapitalIts1998,
  title = {Social {{Capital}}: {{Its Origins}} and {{Applications}} in {{Modern Sociology}}},
  shorttitle = {Social {{Capital}}},
  author = {Portes, Alejandro},
  date = {1998},
  journaltitle = {Annual Review of Sociology},
  shortjournal = {Annu. Rev. Sociol.},
  volume = {24},
  pages = {1--24},
  publisher = {Annual Reviews},
  issn = {1545-2115},
  doi = {10.1146/annurev.soc.24.1.1},
  url = {https://www.annualreviews.org/content/journals/10.1146/annurev.soc.24.1.1},
  urldate = {2026-10-02}
}

@article{rivkinTeachersSchoolsAcademic2005,
  title = {Teachers, {{Schools}}, and {{Academic Achievement}}},
  author = {Rivkin, Steven G. and Hanushek, Eric A. and Kain, John F.},
  date = {2005-03-01},
  journaltitle = {Econometrica},
  volume = {73},
  number = {2},
  pages = {417--458},
  publisher = {John Wiley \& Sons, Ltd},
  issn = {1468-0262},
  doi = {10.1111/j.1468-0262.2005.00584.x},
  url = {https://onlinelibrary.wiley.com/doi/10.1111/j.1468-0262.2005.00584.x},
  urldate = {2026-10-02},
  langid = {english}
}

@article{sampsonNeighborhoodsViolentCrime1997,
  title = {Neighborhoods and {{Violent Crime}}: {{A Multilevel Study}} of {{Collective Efficacy}}},
  shorttitle = {Neighborhoods and {{Violent Crime}}},
  author = {Sampson, Robert J. and Raudenbush, Stephen W. and Earls, Felton},
  date = {1997-08-15},
  journaltitle = {Science},
  volume = {277},
  number = {5328},
  pages = {918--924},
  publisher = {American Association for the Advancement of Science},
  doi = {10.1126/science.277.5328.918},
  url = {https://www.science.org/doi/10.1126/science.277.5328.918},
  urldate = {2026-10-02}
}

@book{smallUnanticipatedGainsOrigins2009,
  title = {Unanticipated {{Gains}}: {{Origins}} of {{Network Inequality}} in {{Everyday Life}}},
  shorttitle = {Unanticipated {{Gains}}},
  author = {Small, Mario Luis},
  date = {2009-08-01},
  publisher = {Oxford University Press},
  doi = {10.1093/acprof:oso/9780195384352.001.0001},
  url = {https://doi.org/10.1093/acprof:oso/9780195384352.001.0001},
  urldate = {2026-10-02},
  isbn = {978-0-19-538435-2}
}

@article{colemanSocialCapitalCreation1988,
    title = {Social Capital in the Creation of Human Capital},
    author = {Coleman, James S.},
    journal = {American Journal of Sociology},
    volume = {94},
    pages = {S95--S120},
    year = {1988},
    doi = {10.1086/228943},
}
\end{refsection}

\end{document}